\documentclass[apj,twocolumn]{openjournal}

\pdfoutput=1

\usepackage[utf8]{inputenc}
\usepackage{graphicx}
\usepackage{hyperref}
\usepackage{comment}
\usepackage{amsmath}
\usepackage{ amssymb }
\usepackage{float}
\usepackage{cleveref}
\usepackage[caption=false]{subfig}
\usepackage{placeins}
\usepackage{xcolor}
\usepackage[normalem]{ulem}

\usepackage{orcidlink}

\begin{document}

\title{Network Theory in Galaxy Distributions: The Coma supercluster Neighborhood}     

\author{Evelise Gausmann\orcidlink{0000-0001-5509-6108} }
\author{Fabricio Ferrari\orcidlink{0000-0002-0056-1970} }

\affiliation{Instituto de Matem\'atica Estat\'istica e F\'isica -- IMEF,   Universidade Federal do Rio Grande -- FURG,
Rio Grande, Rio Grande do Sul, 96.201-900, Brazil. }
\email{egausmann@gmail.com} 
\email{fabricio.ferrari@furg.br}

\shortauthors{Gausmann and Ferrari}


\begin{abstract}
In this work, we use the theory of spatial networks to analyze  galaxy
distributions. The aim is to develop  new approaches to study the spatial galaxy
environment properties by means of the network parameters. We investigate
how each of  the network parameters (degree, closeness and betweeness
centrality; diameter; giant component; transitivity) map  the cluster structure
and properties. We measure the network parameters of galaxy samples comprising
the Coma Supercluster and 4 regions in their neighborhood ($z<0.0674$) using the
catalog produced by \citet{tempel2014flux}. For comparison we repeat the same
procedures for Random Geometric Graphs and Segment Cox process, generated with
the same dimensions and mean density of nodes. We found that there is a strong
correlation between degree centrality and the normalized environmental density.  Also, at high degrees  there are  more
elliptical than spiral galaxies, which confirms the density-morphology relation.
The mean degree as a function of the connection radius is an estimator of the
count-of-spheres and consequently provides the correlation dimension as a
function of the connection radius. The correlation dimension indicates high
clustering at scales indicated by the network diameter. Further, at this scales,
high values of betweeness centrality characterize galaxy bridges connecting
dense regions,  tracing very well the filamentary structures. Then, since
galaxies with the highest closeness centrality belongs to the largest components
of the network, associated to supercluster regions, we can produce a catalog of
superclusters only by extracting the largest connected components of the
network. Establishing the correlation between the well-studied normalized
environmental densities and the parameters of the network theory allows us to
develop alternative tools to the study of the large-scale structures.
\end{abstract}

\maketitle

\section{Introduction}

The network theory is a multidisciplinary area involving graph theory,
statistical mechanics and data analysis to compile and to extract information of
a massive quantity of data. This characteristic makes the network theory a
powerful tool for the study of the structures in the galaxy distribution and its
environment -- \textit{the cosmic web}.    The  cosmic web \citep{bond1996filaments}  is a highly complex and compound structure composed by
clusters, walls, filaments and voids of galaxies that drive the matter
distribution in the Universe \citep{da1988southern,da1994complete}. The first
studies of graph theory were applied to the cosmic web  for detecting groups
of galaxies \citep{press1982identify} and trace filaments \citep{barrow1985minimal}.

One of the most used methods to detect groups of galaxies is the
friends-of-friends (FoF) group finder algorithm based on finding the closest
neighbor of a galaxy and then the neighbor of the neighbor and so on, limited by
a maximum linking length between the neighbors \citep{davis1985evolution}. The
result of this method is a network with nodes connected only to the closest
neighbors. Another analogous network is obtained by the minimal spanning tree
(MST) algorithms that, by definition, is a single and shortest path that
connects all nodes of the network without circuits or loops (in this way, MST is
a particular solution of the FoF method \citep{colberg2007quantifying}.) MST
algorithms are used for tracing filaments  when applied to the center
of groups of galaxies. Frequently, both methods are applied, the FoF algorithms
to detect the groups of galaxies, and then the MST algorithms to trace the
filaments \citep{colberg2007quantifying,alpaslan2014galaxy}. 

A recurring problem in both methods is to define the best linking length between
the structures. One of the approaches is to calculate the mean separation of the
structures, take several fractions of this value and search for one that best
represents the data set (see~\citet{martinez2002clustering} for a discussion
of several other methods, including finding the maximum length between the MST
connections.) In a random geometric graph (RGG), the maximum linking length of a
MST define the connectivity regime, which  means  that  all the nodes belongs
to a single component~\citep{penrose2003random}. The best linking length is
related to the notion of percolation phase; 
at the percolation phase 
the small structures 
connect to others forming larger structures. The percolation
method detects the connectedness in the galaxy distribution that lower orders of
correlation functions do not~\citep{radicchi2015predicting}. 

Network theory itself was applied by~\citet{hong2015network} to measure three
network centrality measurements: degree, closeness, and betweeness centrality
using the Cosmological Evolution Survey (COSMOS) catalog \citep{scoville2013evolution} in the two-dimensional projective-space. They used
these measurements to distinguish different structures and concluded that the
degree centrality is an indicator of voids, clusters, and walls. They used
betweeness centrality to distinguish galaxies in regions defined as
\textit{Main Branch}  and \textit{Dangling Leaf} and closeness centrality to
define regions of \textit{Kernel, Backbone}, and \textit{Fracture}. However,
it was not established if the betweeness and closeness centrality differs
for a random spatial network.  Another problem,
 also present in the FoF and MST methods, is
how to determine the best linking length to define the
network. \citet{hong2015network} did a statistical analysis based on the mean
density of a sample with a Poisson degree distribution to define the linking
length. However, they visually adjusted the mean density  according to the
sample. This linking length needs visual inspection and does not always reflect
the better choice of this parameter, as is shown by~\citet{hong2016discriminating} using the maximum network diameter. In a recent
work, \citet{hong2020constraining} studied the topological structure of
gravitational clustering in five different universes produced by cosmological
N-body simulations. They compared the N-body graph simulations with the expected
values for the same network parameters  of RGG
samples. The number of particles used in each simulation is $N_{p} = 2048^{3}$ in a
box size of  $L=1024h^{-1}$Mpc in the comoving scale and represents all the
observable universe.  They found that the average vertex degree, the
transitivity and the cumulative number density of subcomponents with connected
component size larger than 5 can distinguish five models of universes.

In this work, we apply the network theory to three-dimensional
data of galaxy distribution from the catalog of \citet{tempel2014flux}, 
which was created to study methods to develop
catalogs of superclusters and filaments \citep{liivamagi2012sdss,libeskind2015filaments}. 
The catalog is  Finger-of-God effect ~\citep{tully1978proc} corrected by using FoF group finder. 


We use the \citet{tempel2014flux} catalog  to study the statistical proprieties of the three-dimensional spatial network associated with the connection radius $r$ that
maximizes the network diameter of each particular data set at different distance
scales indicated by the network diameter as a function of $r$. Our purpose is to
test the network theory as a tool for studying the complex spatial structure of
the galaxy distribution and which network measures can be used to find regions
of superclusters and filaments. We also compare the spatial network of the data samples with the RGGs and the Segment Cox Process \citep{stoyan1995statistic}
generated with the same dimensions and same mean density of nodes. The RGG is the appropriate network to compare with a real spatial network because is a network embedded in a metric
space.

This work is structured as follows: In Section \ref{sec:network}, we present the
basic network theory. In Section \ref{sec:data}, we give the details of the data
sample. In section \ref{sec:diamAPL}, we find the connection
radii used to generate the networks. In section \ref{sec:Meandeg}, we
investigate clustering measures as a function of the connection radius. In
Section \ref{sec:results}, we present and discuss the results. Finally, in Section  \ref{sec:conclusions}, we summarize the relevant
points.

\section{Network Theory}
\label{sec:network}
Here we briefly present some definitions  from network theory. The basic
reference  is  \citet{newman2010networks} and \citet{barabasi2016network}. We
define our galaxy network by using the same definitions of network measures used
by ~\citet{hong2015network} to compare our results with the ones obtained  by ~\citet{hong2015network,hong2016discriminating,de2017network}. We developed our
algorithms with  \texttt{Igraph-python} \citep{csardi2006igraph}, Numpy \citep{oliphant2006guide}, SciPy \citep{oliphant2007python,millman2011python}
and Matplotlib \citep{hunter2007matplotlib} within the Python language ecosystem \citep{jones2001scipy}.

The \textbf{adjacency matrix} gives the complete description of the spatial
network for a given data set and we can express several measures by this matrix.
The element $A_{ij}$ of the adjacency matrix describes the relationship between
the objects $i$ and $j$:  it is one if  they are  connected   and zero
otherwise:
\begin{align}
\label{eq:adj}
A_{ij}=
   \begin{cases}
     1   & \quad \text{if } i\neq j \text{ and } i  \text{ is connected to } j\\
     0   & \quad \text{otherwise}. \\
   \end{cases}
\end{align}
The dimension of $A_{ij}$ is $n \times n$, where $n$ is the number of elements
in the sample.  For an undirected network, this matrix is symmetric, $A_{ij} =
A_{ji}$. The diagonal elements $A_{ii}=0$, that represents a connection of one
point  with itself and  are zero because there are no loops in our network. 

Several measurements can be used to characterize the network. The \textbf{degree
centrality} is  an important measure of the local relevance of each  node on the
network; for the node  $i$ the degree is
\begin{align}
\label{eq:deg}
k_{i}=\sum_{j} A_{ij},
\end{align}
and it states  the number of connections of the element
$i$. In our galaxy network, the galaxy with a higher degree centrality has more
neighbors at a given distance; the galaxies with lower degree centrality are
more isolated: thus the degree centrality is a local measure of the relationship
between galaxies. Its value is related to the local density of galaxies \citep{hong2015network} (see Section \ref{sec:results}).

In a network, a \textbf{path} between the node $i$ and $j$  is an ordered list
of $m$ connections;  the length of the path is $m$. Among all possible paths,
the  \textbf{shortest path}  $d_{ij}$  between $i$ and $j$ corresponds to the
path with the lowest value of $m$. The network diameter $diam$ is the  longest shortest path
between all nodes paths. In a spatial network, the network diameter does not
represent a physical distance. In the same way, the \textbf{average path length}
(APL), $\langle d \rangle$, is the average of the shortest paths between all
pairs of nodes \citep{barabasi2016network}. The \textbf{giant component} of the network is the magnitude of the largest
connected component, which means the number of nodes at this component. The
\textbf{betweeness centrality} $x_i$ for the node $i$ is a measure of how many
shortest path of all pairs pass through a certain node. It is defined as:
\begin{align}
\label{eq:bet}
x_{i}=\sum_{s,t} \frac{n^{i}_{st}}{g_{st}},
\end{align}
where $g_{st}$ is the number of shortest path  between the nodes $s$ and $t$,
and $n^{i}_{st}$ is the number of them that pass through the node $i$. Like the
diameter, the betweeness centrality   is a global measure: it quantifies the
importance of the node on joining one region and another. High values of
betweeness centrality mean that the node works efficiently as a bridge joining
two dense regions \citep{hong2015network}. 

The \textbf{closeness centrality} $c_{i}$ of a node $i$ measure how easily a
node can be to reach by other nodes. The closeness centrality  is inversely
proportional to the sum of lengths of all shortest paths from a node $i$ to all
other nodes, i.e. 
\begin{align}
\label{eq:close}
c_{i}= \left [ \frac{1}{n-1 }\sum_{j(\neq i)} d_{ij}  \right ] ^ {-1}.
\end{align}
One node with higher closeness centrality has a smaller average path to all
other pairs. It is also a global measure.
 
The \textbf{ local clustering
coefficient}  $C_i$ measures how connected  the neighbors of a node $i$ are. It
is defined as 
\begin{align}
\label{eq:clucoef}
C_{i}= \frac{2 L_i}{k_i (k_i -1) },
\end{align}
where $L_i$  is the number of connections between $k_i$
neighbors of $i$. The $C_i = 0$ means that no neighbor of $i$ is connected, and
$C_i = 1$ means that all neighbors of $i$ are connected. The mean local
clustering coefficient for each degree $C(k)$ as a function of the degree
centrality can indicate the cluster organization. The decrease of $C(k)$ with
the increase of $k$ indicates a hierarchical organization. The invariance of
$C(k)$ with $k$ represents the lack of hierarchical organization \citep{ravasz2003hierarchical,newman2010networks}. The mean clustering
coefficient is the average value of all local clustering coefficients,
\begin{align}
\label{eq:meanclucoef}
C=\frac{1}{n} \sum_{i=1}^{n} C_{i}.
\end{align}

The \textbf{transitivity} or \textbf{global clustering coefficient} $T$ is the
ratio between the number of closed triplets (three nodes connected by three
connections, forming a triangle) and  all possibles triplets (three
nodes connected by three or two connections, forming a closed or open triangle).
In general, these measures give an indicator of the clustering of the network
related to its ability to produce triangles.

\section{Data Sample}
\label{sec:data}

To test our algorithm, we use the catalog developed by~\citet{tempel2014flux} and
published in the CosmoDB Database of cosmology-related catalogs at the Tartu
Observatory\footnote{\url{http://cosmodb.to.ee/}}. \citet{tempel2014flux} published a
three-dimensional real space galaxy catalog, with 588\,193 entries corrected in
redshift for  the motion to the cosmic microwave background (CMB)
and FoG-effect, based on  Sloan Digital Sky Survey (SDSS) data release 10. This
catalog was produced to be used to find galaxy groups and clusters for
flux-limited galaxy surveys. The effects of the selection are very strong in the
flux-limited samples, the most important are the decrease of the group density
and its richness with increasing distance from the observer \citep{tago2010groups}. However, \citet{tago2010groups} has shown that by
calibrating the group properties with the radial distance is possible to
minimize the effects of the selection by choosing a variable linking length on
the FoF group finder. To properly scale the linking length with distance, they
calculated the mean distance to the nearest galaxy in the plane of the sky,
producing a transversal linking length. The group members were sought within a
cylindrical volume limited by the ratio of the radial and transversal linking
lengths equal to ten \citep{tempel2014flux}. This correction is statistical and was used for large-scale structures  studies such as superclusters detection \citep{liivamagi2012sdss}
and filaments detection \citep{tempel2014detecting}.

For linear dimensions, \citet{tempel2014flux} used comoving distances in
$h^{-1}$Mpc units \citep{martinez2001statistics}. The standard cosmological
parameters are: $H_0 = 100 \ h $ km s$^{-1}$  Mpc$^{-1}$, the matter density
$\Omega _{m} = 0.27$, and the dark energy density $\Omega _{\Lambda} = 0.73$.
The catalog in \citet{tempel2014flux} also provided a morphological
classification based on Galaxy Zoo \citep{lintott2008galaxy}, classifying
galaxies as a spiral ($1$), elliptical ($2$), and unclear ($0$).

To define the region of the supercluster samples, we use the SDSS DR8
supercluster catalog \citep{liivamagi2012sdss}. They delineated superclusters by
using the luminosity density field with densities over a selected threshold.
They have created two types of catalogs: one with an adaptive local threshold
and another with different global thresholds. The catalog with an adaptive local
threshold has the advantage of minimizing the selection effects \citep{liivamagi2012sdss}.  

\begin{figure}
\centering
\includegraphics[width=\columnwidth]{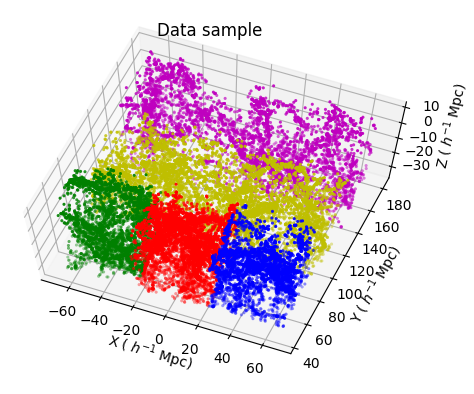}
\caption{ Data sample composed by Super Coma Region in red points; Region 1 in blue points; Region 2 in green points; Region 3 in yellow points and Region 4 in magenta points.}
\label{fig:samples}
\end{figure}

In this work, we analyze the region of the Coma Supercluster and its  neighborhood. We
call the Coma Region, Super Coma Region, and Region 1, 2, 3, and 4, the regions
for the Coma Cluster, Coma Supercluster, and their neighborhood, respectively.
Figure \ref{fig:samples} shows the disposition of the data samples, in comoving
coordinates. Regarding the Coma Supercluster selected in \citet{liivamagi2012sdss}, for this work we select  \citet{tempel2014flux}
Super Coma Region containing 7802 galaxies with two much richer clusters - the
Leo Cluster  (A1367) and the Coma Cluster (A1656). Super Coma Region, Region 1,
and Region 2 have dimensions of $(45 \times 45 \times 45) h^{-3}$ Mpc$^3$, and
Region 3 and Region 4 of $(135 \times 45 \times 45) h^{-3}$ Mpc$^3$. We select
from the Super Coma Region, the Coma Region with $N_{Coma}=1467$ galaxies
distributed in a cube of dimensions $(16 \times 16 \times 16) h^{-3}$ Mpc$^3$.
The coordinates of the regions and the respective number of galaxies are shown
in Table \ref{tab:my_label}.

\begin{table*}[]
    \centering
    \begin{tabular}{lccccccc} \hline \\
                    &   \multicolumn{2}{c}{RA} & \multicolumn{2}{c}{DEC} & \multicolumn{2}{c}{$z$} & $N_{\rm gal}$\\ 
                    \hline \\
        Super Coma & $157.26^\circ{}$ & $216.51^\circ{}$ &  $-0.04^\circ{}$ & $42.71^\circ{}$ & $0.0134$ & $0.0320$ & 7802 \\
        Region 1  &  $123.53^\circ{}$  &  $169.72^\circ{}$  &  $-0.012^\circ{}$      &     $38.45^\circ{}$     & $0.0165$ & $0.0377$ & 4806 \\
        Region 2  &  $199.49^\circ{}$  &  $245.69^\circ{}$  & $-0.032^\circ{}$ & $38.44^\circ{}$ &  $0.0144$ & $0.0379$ &  5257 \\
        Region 3 &  $144.02^\circ{}$  &  $224.17^\circ{}$ &   $11.039^\circ{}$ & $38.35^\circ{}$ & $0.0276$ & $0.0502$ & 7928 \\ 
        Region 4 &  $153.50^\circ{}$  & $215.39^\circ{}$ &   $16.47^\circ{}$ & $36.19^\circ{}$ & $0.0419$ & $0.0674$ & 5985 \\
        \hline
    \end{tabular}
    \caption{The coordinates of the regions in the data samples and the respective number of galaxies}
    \label{tab:my_label}
\end{table*}

We also perform the analysis with  simulations of a Cox
process. A Cox process, also known as a doubly stochastic Poisson process since it is an extension of the Poisson process,  is a type of stochastic process where the intensity rate (or the rate parameter) itself is a random process. This adds an extra layer of randomness to the model \citep{stoyan1995statistic}. A particular case of a Cox process is the Segment Cox
process where $N_{\rm seg}$ segments of length $l$ are randomly distributed in a
cube of volume $V$ with mean density $\lambda_s = N_{\rm seg}/V$. The points  are distributed on the segments, $N_{\rm point}$ points in
each segment with mean intensity function of points per segment given by
$\lambda_l =N_{\rm point} l$. The intensity function is the product of the mean
density of segments, the mean density of points per segment and the length of
the segment $\lambda=\lambda_{s}\lambda_{l}l= N_{\rm points} N_{\rm seg}/V $ \citep{martinez2001statistics}. In the simulations, we use an average total
number of points $N_{\rm point} N_{\rm seg}= 4800$ in two configurations:
Segment Cox process 1 with $N_{\rm point}=3$, $ N_{\rm seg}=1600$ and Segment
Cox process 2 with $N_{\rm point}=8$, $ N_{\rm seg}=600$; both with $l=4.5\
h^{-1}$ Mpc. The dimensions and average number of points in this point process
is the  same of Region 1 for better comparison. 

To compare the spatial galaxy network and  the Segment Cox
process network with a random spatial network, we create a  three-dimensional
RGG sample by placing $N$ nodes in the interval $[0,L_{sample}]^3$, where
$L_{sample}$  is the side of the polyhedron representing the sample. We
distributed the nodes following a given distribution probability function $P(x)$,
where $x$ is the position coordinate and connected them by a connection
probability function $H(r')$, where $r'$ is the Euclidean distance between the
nodes. In this study, the node distribution of the undirected RGG network
follows a random uniform probability distribution and the connection
probability is $H(r')=1$ for $r' \leq r$ and $H(r')=0$ otherwise \citep{dettmann2016random}. There is a  connection between a node and all other nodes if their  distance is less  or equal to $r'$. We define the undirected spatial galaxy network following the same connection probability. We generate RGG samples with the same dimensions and number of points that we have in the data samples and follow the same procedure to construct the spatial network. We called Random Region 1 the RGG sample with the dimensions and mean density of
Region 1. We followed this procedure  for all samples.

\section{The choice of the connection radius}
\label{sec:diamAPL}

We are interested in finding the connection radius $r$ that better represents the
spatial distribution of galaxies. A small $r$ produces a
disconnected network with several small components, whereas a large $r$ produces
a component with all galaxies in one single connected component (in this case,
any galaxy is connected to every other.) In
spatial networks, the choice of $r$ is associated with the studies in bond
percolation \citep{barthelemy2011spatial,radicchi2015predicting} 
that quantify the phase transition between a non-percolating and a
percolating phase. 

The bond percolation is  the emergence of the giant component
in the network \citep{radicchi2015predicting} and the maximum network diameter
is related to the inflection point of the growth curve of the giant component \citep{hong2016discriminating}. \citet{einasto1984structure} showed that samples
with different structures present different percolation thresholds even if the
samples have the same mean density.  We will define the $r$ by associating this
length to the maximum network diameter, just after the emergence of the giant
component.

\begin{figure*}
\centering
\includegraphics[scale=.32]{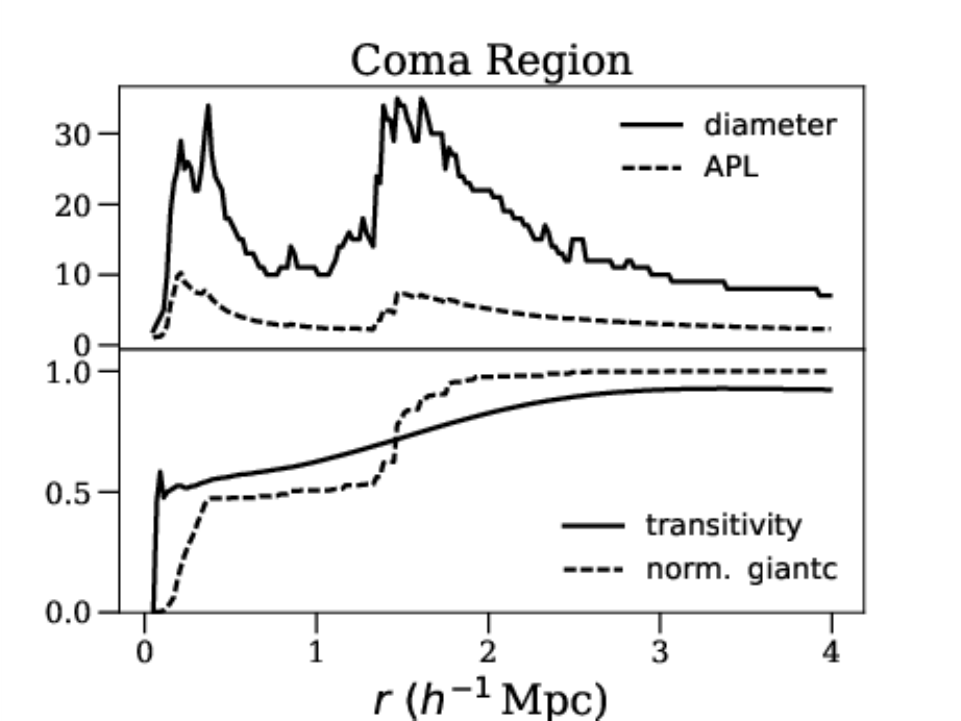} 
\includegraphics[scale=.32]{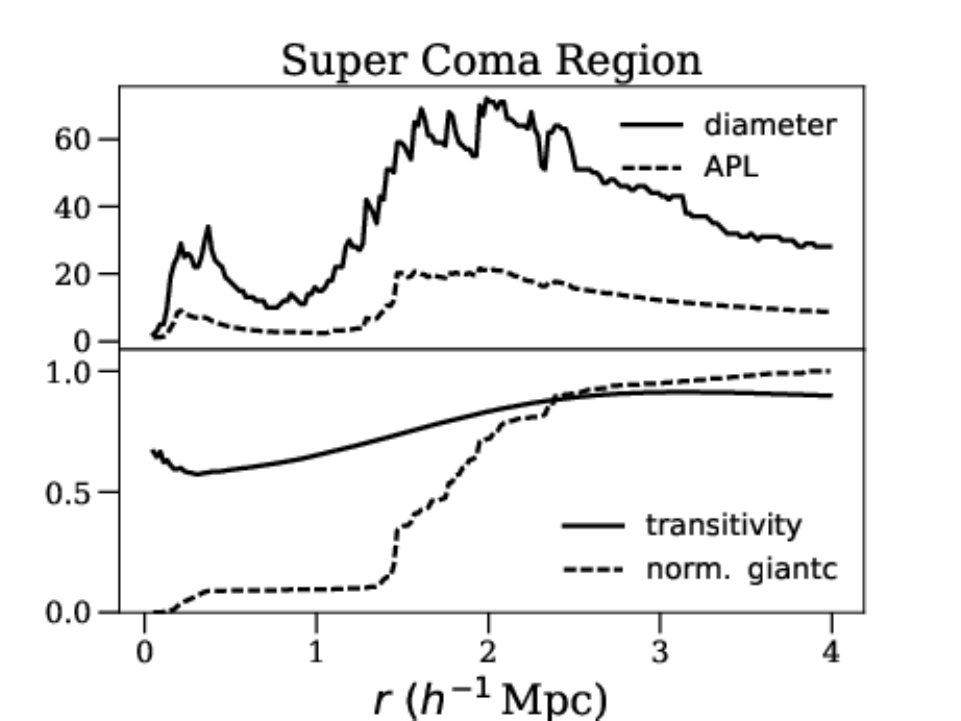}
\includegraphics[scale=.32]{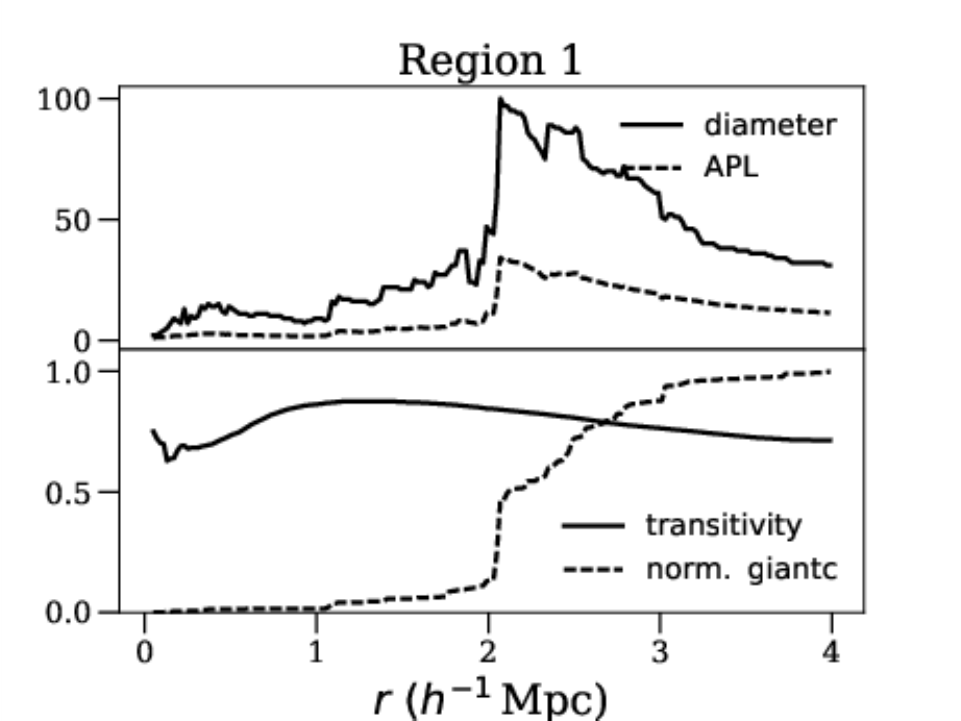}
 \includegraphics[scale=.32]{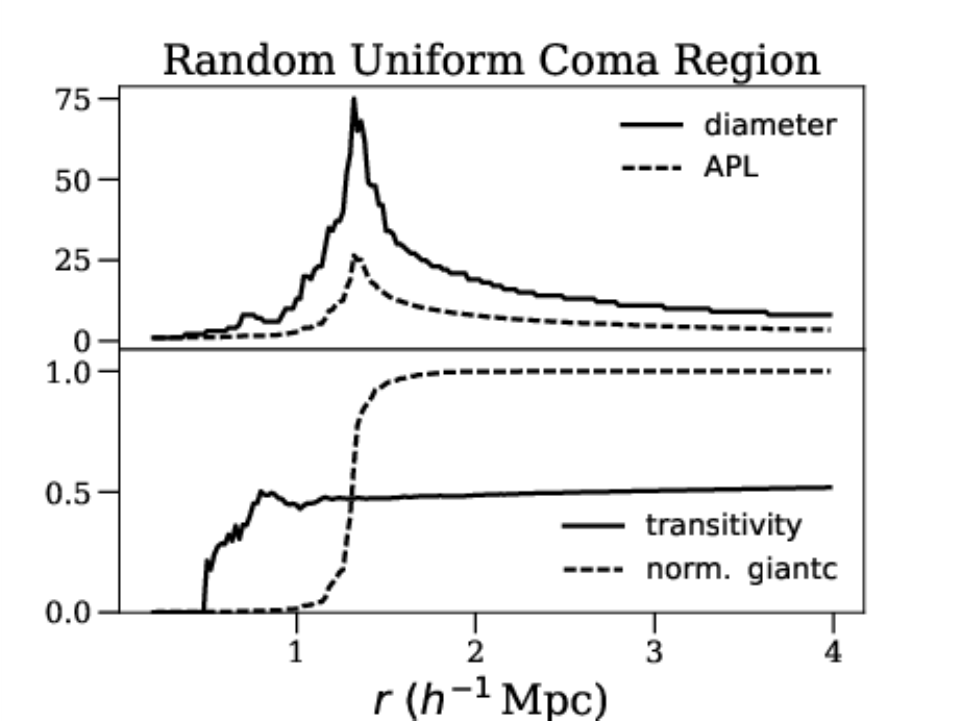}
 \includegraphics[scale=.32]{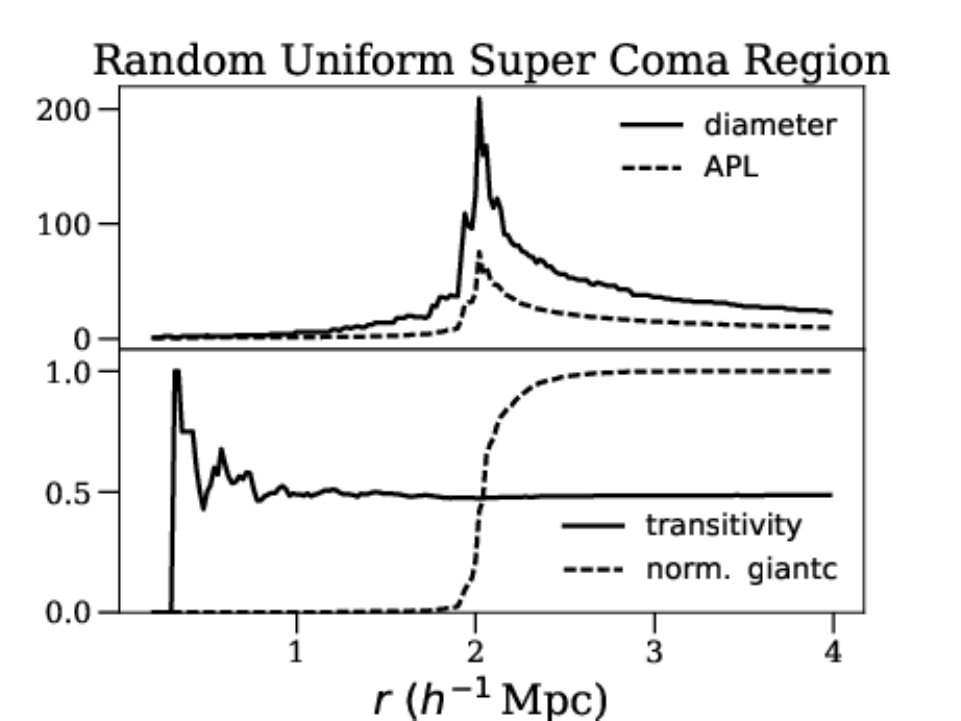}
\includegraphics[scale=.32]{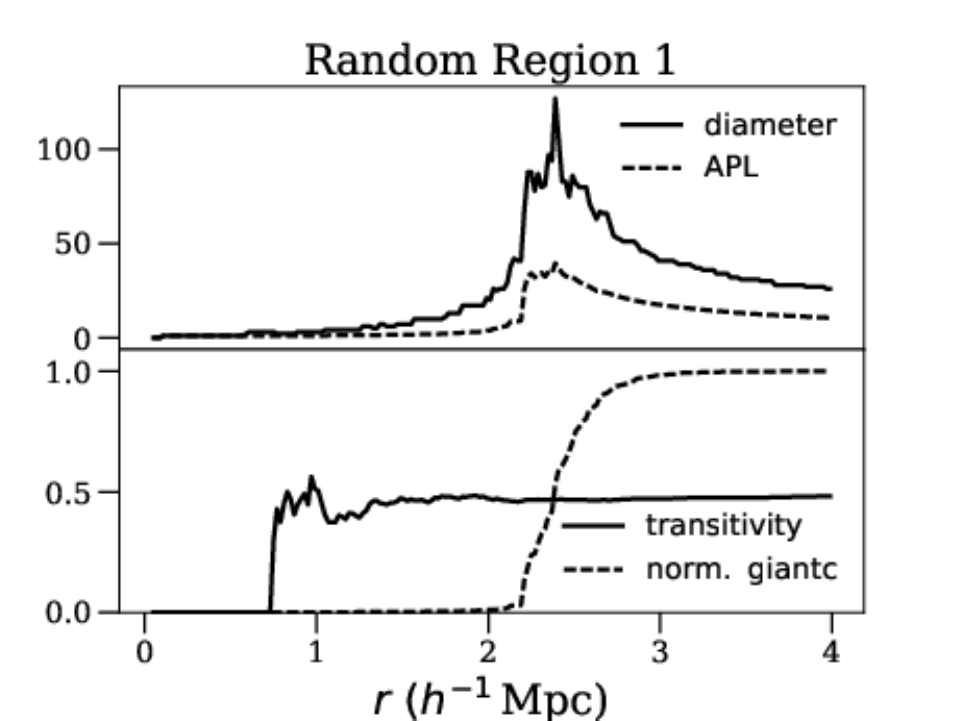}
\includegraphics[scale=.32]{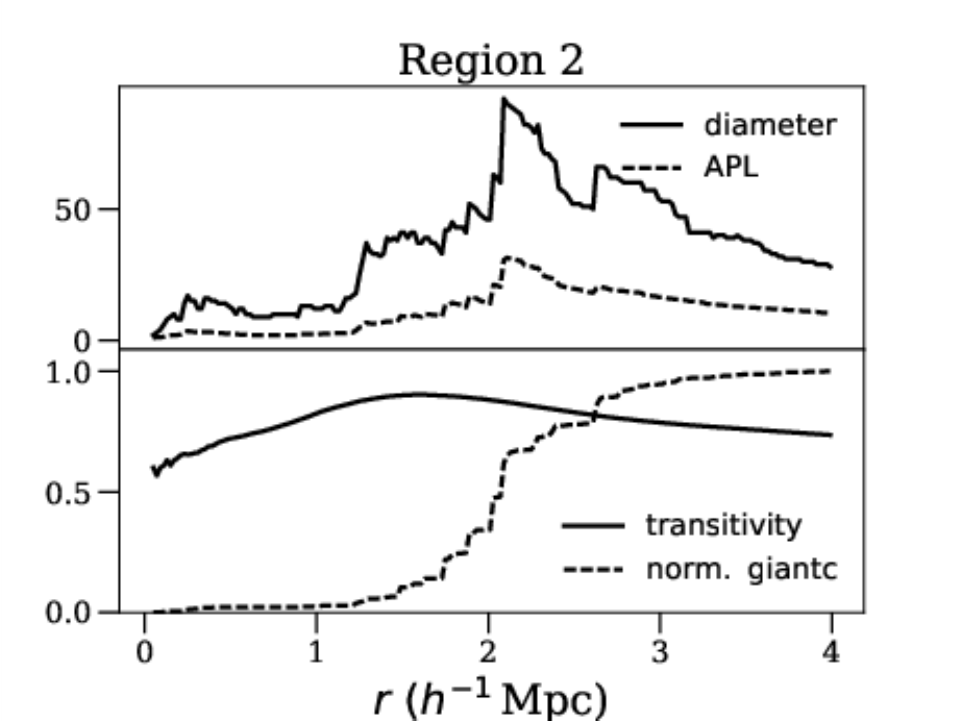} 
\includegraphics[scale=.32]{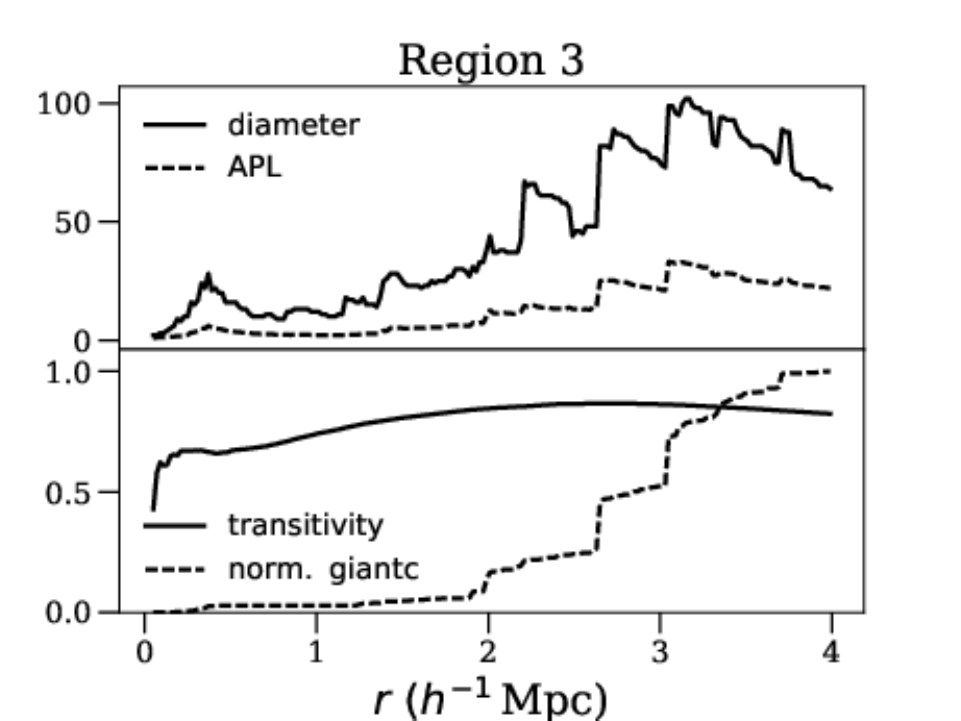}
\includegraphics[scale=.32]{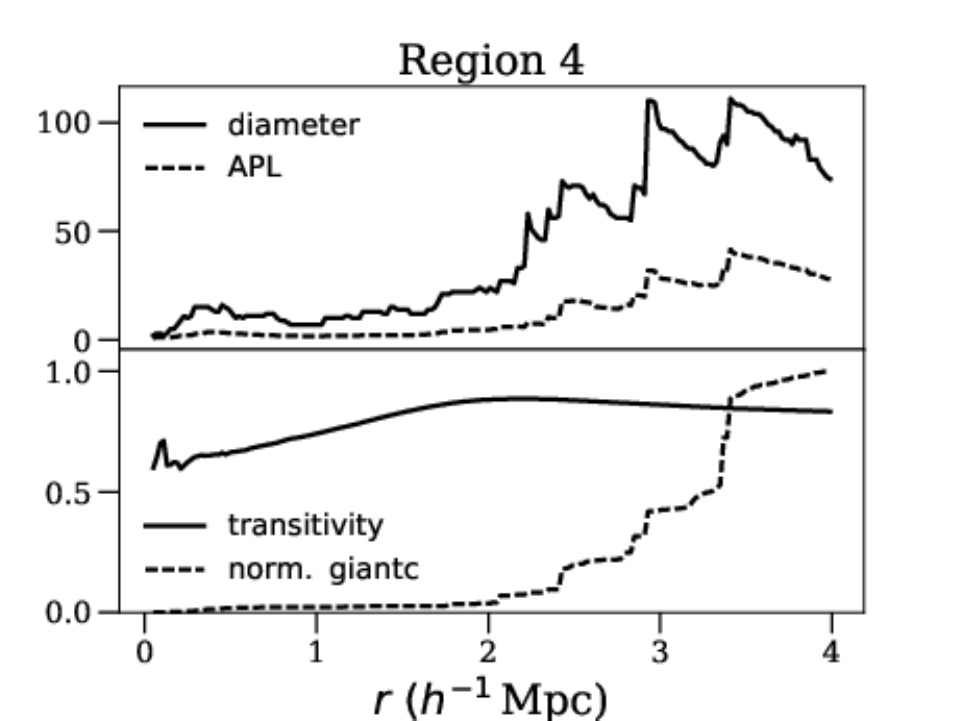}
\includegraphics[scale=.32]{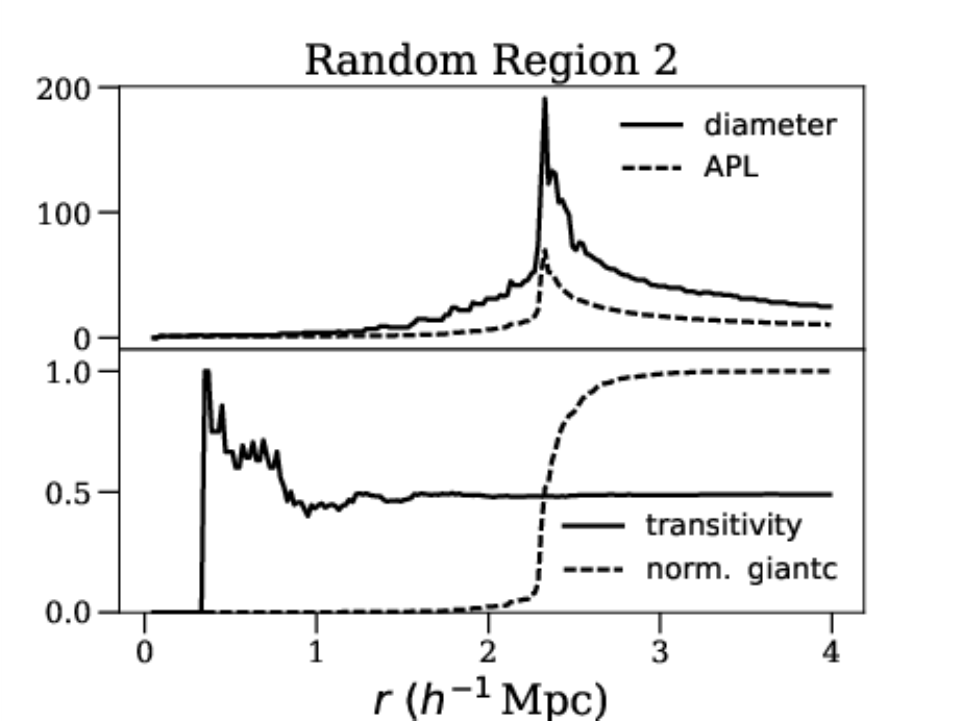}
\includegraphics[scale=.32]{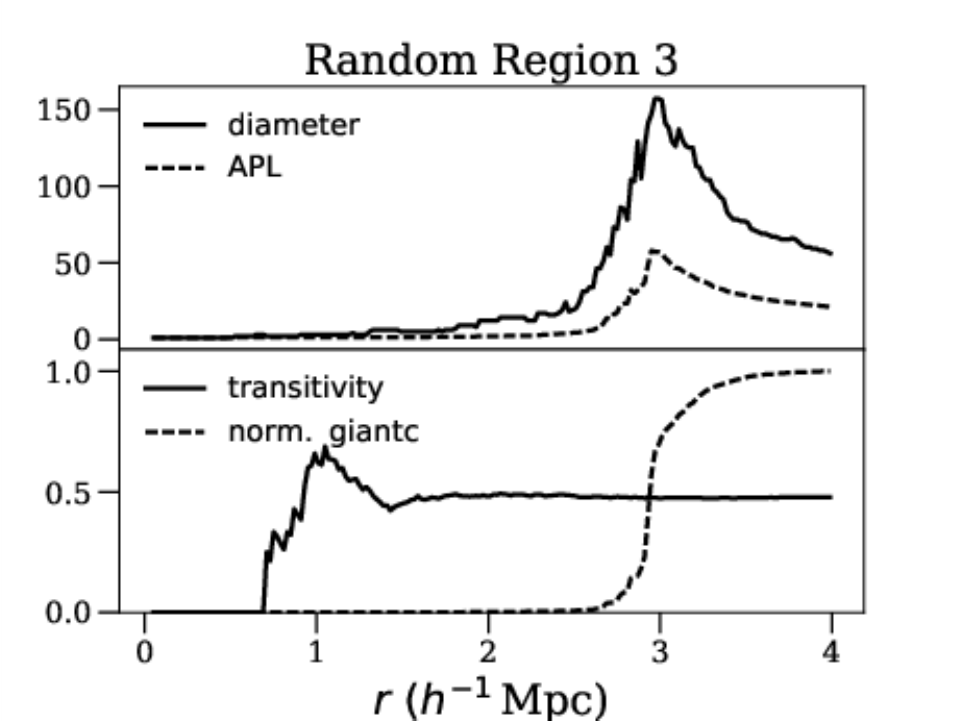}
\includegraphics[scale=.32]{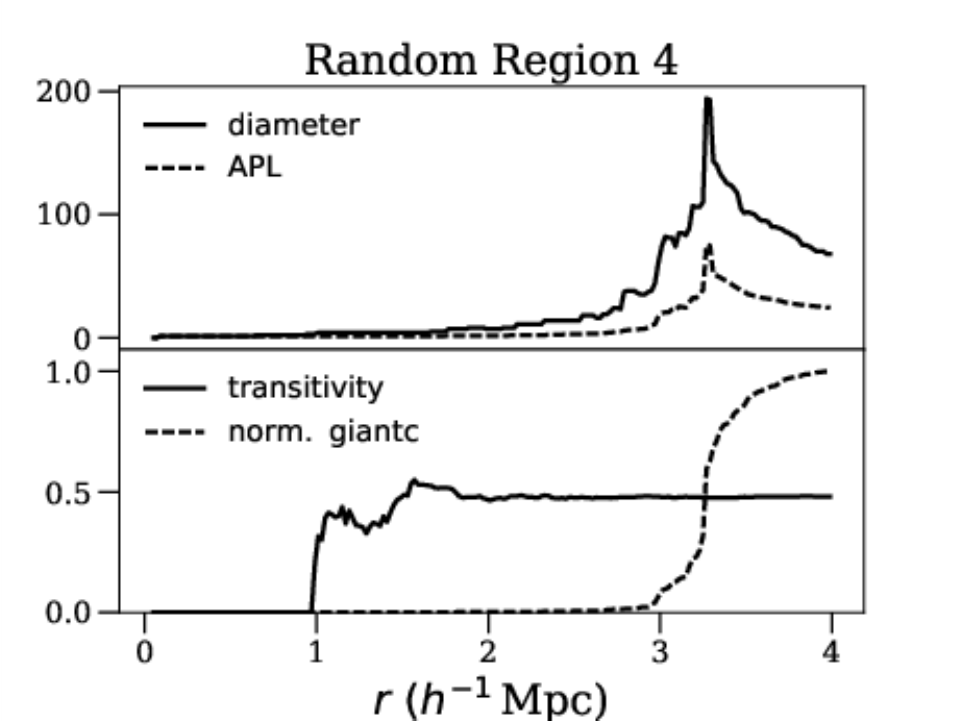}

\caption{Network diameter, as well as APL, normalized giant component and
transitivity, \textit{versus} $r$ for Coma, Super Coma, Regions 1, 2, 3, 4 and
the respective RGG sample (for each dataset -- rows 1 and 3 -- the corresponding random dataset is presented below -- rows 2 and 4.) The network diameter increases to a maximum value,
then  the graph tends to a complete graph and the network diameter tends to one \citep{hong2016discriminating}. The APL follows a similar behavior. The giant
component emerges on steps due to density variations on the sample.} 
\label{fig:diam}
\end{figure*}

To explore the behavior of the network diameter as a function of $r$,  we first
create an adjacency matrix for various $r$, starting at $r = 0.05 \ h^{-1} $Mpc
and increasing by $\Delta r = 0.02 \ h^{-1}$Mpc. Then, we calculate the network
diameter for each adjacency matrix.  In the same way, we construct  graphs for
the average  path length (APL), the giant component, and the transitivity. We
show the network parameters as a function of $r$ for all the samples in Figure
\ref{fig:diam}.

\begin{figure*}
\centering
  \includegraphics[scale=.32]{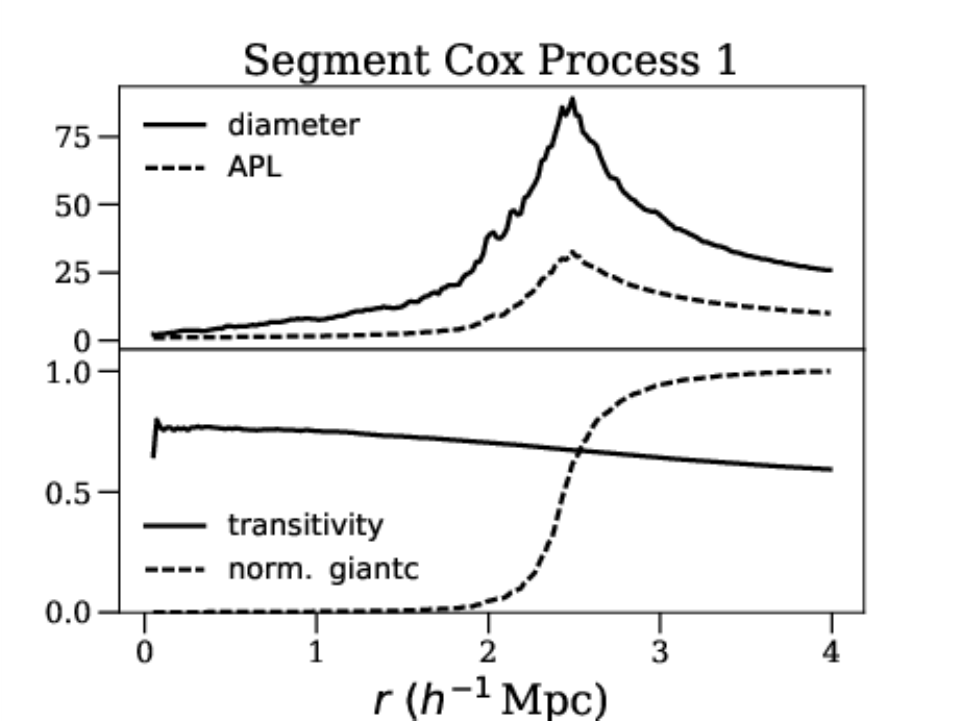}
  \includegraphics[scale=.32]{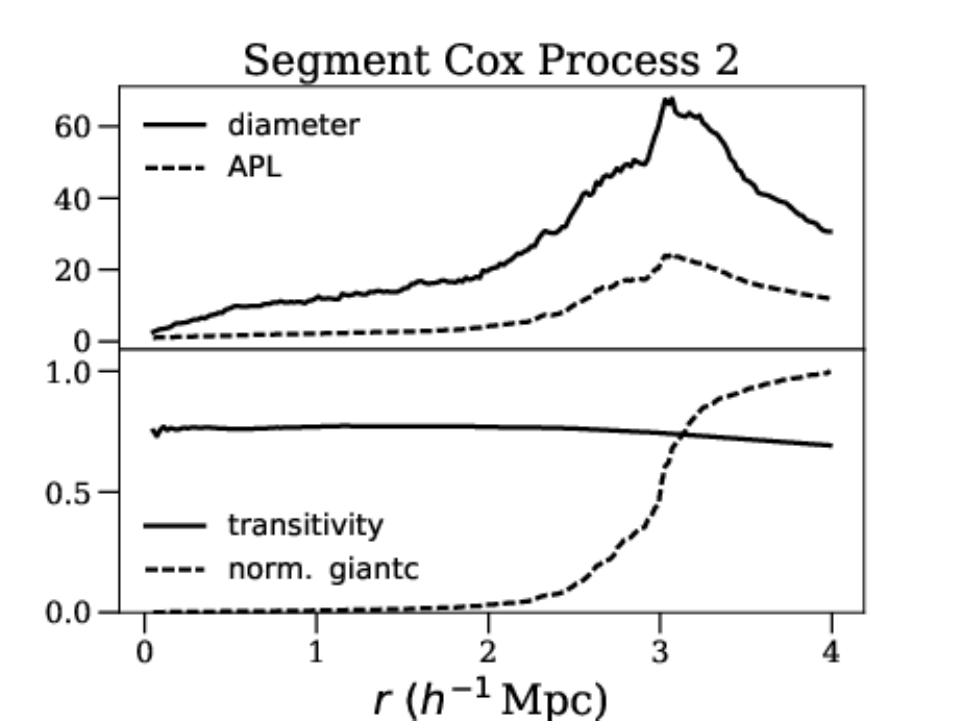}
  \includegraphics[scale=.32]{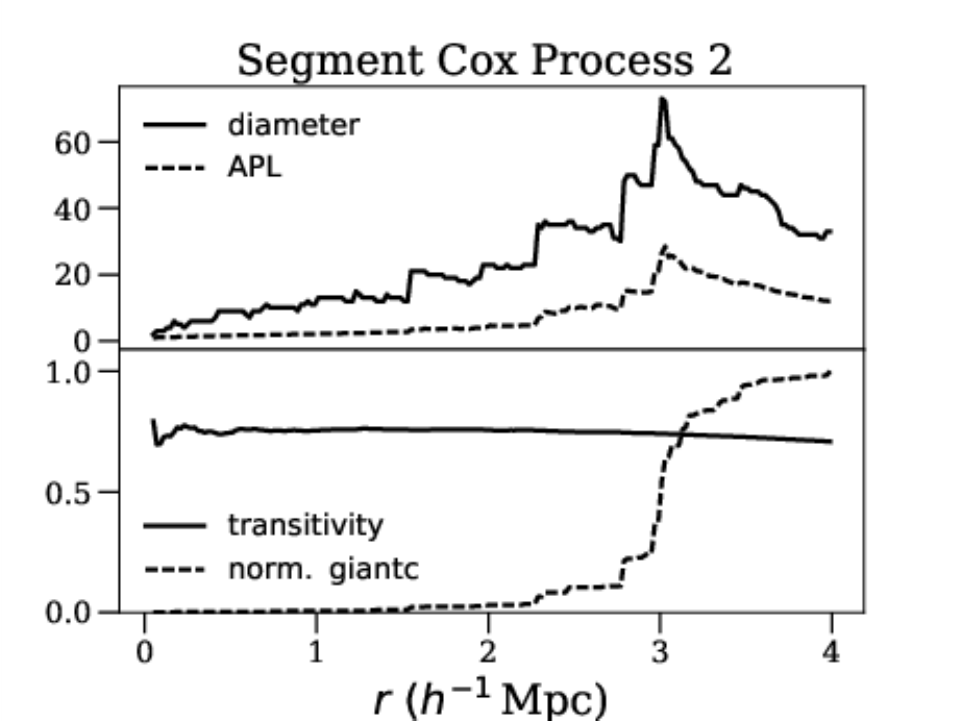}
  \caption{Network diameter, as well as APL, normalized giant component and
  transitivity, \textit{versus} $r$ for Segment Cox process 1 and 2, for a mean
  of ten realizations. The right  plot  shows the behavior of Segment
  Cox process for one realization. }
  %
 %
  \label{fgr:diam_cox}
\end{figure*}

The Figure \ref{fgr:diam_cox} shows the network diameter, the
APL, the giant component and the transitivity as a function of $r$ for the
Segment Cox process 1 and 2 on top.. The left and middle top  plots in Figure
\ref{fgr:diam_cox} are the average value of ten realizations for each
configuration, the right top plot is an example of one realization of the
Segment Cox Process 2.  We analyze the centrality measures with this example. We
observe that the maximum network diameter depends on the mean density of
segments: a lower density of segments produces maximum diameter at a higher $r$.
However, a higher density of point per segment produces a higher probability of
an extra peak on the network diameter before the maximum diameter. 

For RGG samples, we observe one peak at the $r$ that corresponds to the maximum
diameter and the emergence of the giant component, corresponding to the phase
transition between a disconnected and a fully connected network \citep{penrose2003random}. After this peak, the diameter decreases indicating a
diameter saturation and the network tends to a complete network \citep{hong2016discriminating}. This behavior is observed for RGG samples like
those of \citet{penrose2003random} and the Illustris simulated samples of
\citet{hong2016discriminating}. However, the galaxy data samples present a
different behavior: we distinguish other peaks before the maximum network
diameter.  The analysis of continuum percolation for Cox point
processes shows that the percolation probability presents a non-trivial
behavior. While for the Poisson point process the percolation probability tends
to 1 exponentially with the high-density or large-radius limit, for the Cox
process the regime of large-radius is not equivalent to that of a high-density \citep{hirsch2019continuum}. The transition phase is a function of both the
density and the radius.


\section{Clustering measures and multiscale behavior }
\label{sec:Meandeg}

The two-point correlation function $\xi(r) $ has been used
to analyze the large-scale distribution of galaxies with the aim to
determine the crossover to homogeneity of the distribution \citep{peebles1980large}. However, it assumes the homogeneity of the data \citep{coleman1992fractal}. To overcome this problem
the counts-in-spheres measurement $N( < r)$ is commonly used. The $N( < r)$ is
the average number of neighbors around a given object at a distance $r$. For a
homogeneous distribution $N( < r) \propto r^3$ in the three-dimensional case.
For a general distribution $N( < r) \propto r^{D_2}$, where $D_2$ is the
correlation dimension defined as
\begin{equation}
	D_2(r) \equiv \frac{d \ln N( < r)}{d \ln r} .
	\label{corr_dim}
\end{equation}
Since $N( < r)$ estimator is affected by edges effects and
completeness of the samples, \citet{scrimgeour2012wigglez} proposed the scaled
counts-in-spheres estimator $ \mathcal{N}(<r)$ to minimize these effects. The
scaled counts-in-spheres estimator is the ratio between $N( < r)$ and the same
measure for the random distribution $N_{\rm ran}( < r)$ using as center the data
positions. In terms of the $\mathcal{N}(<r) \propto r^{\gamma}$ with $ \gamma
= D_2 - 3$; the correlation dimension is then
\begin{equation}
	D_2(r) \equiv \frac{d \ln \mathcal{N}( < r)}{d \ln r} +3 .
	\label{corr_dim2}
\end{equation}

\citet{scrimgeour2012wigglez} analyzed the scaled counts-in-spheres estimator and the respective correlation dimension. Their  study showed that the correlation dimension is a robust indicator for the  homogeneity scale because this measure is not dependent on the correlation between widely separated $\mathcal{N}(<r)$ bins \citep{scrimgeour2012wigglez}.
 They concluded that the results exclude a fractal distribution with fractal
 dimension below  $D_2 = 2.97$ on scales up $\sim 80 h^{-1}$Mpc in the WiggleZ
 galaxy distribution. \citet{laurent201614} develop a scaled counts-in-spheres
 estimator based on Landy-Szalay (LS) estimator \citep{landy1993bias} of the
 two-point correlation function and applied it  to a large
 study using  BOSS DR12 quasar sample; this study also confirm the transition to
 homogeneity at large scales.
 \citet{gonccalves2018cosmic} compare four scaled counts-in-spheres estimators:
 the two previous estimators, the Peebles-Hauser (PH) scaled counts-in-spheres
 estimator (based on PH estimator of 2-point correlation function) and a
 variation of the original $\mathcal{N}(<r)$ developed by
 \citet{scrimgeour2012wigglez}. They used the BOSS data to make measurements of
 the angular homogeneity scale, confirming homogeneity scale above $\theta_h
 \sim 10^{\circ}$. \citet{avila2018scale} show that the derivation of the
 original  $\mathcal{N}(<r)$, Equation \ref{corr_dim}, is able to distinguish
 the presence of under- and over-densities in the sample and the estimator
 presents consistent results as the sample enlarges.
 
\begin{figure}
\centering
  \includegraphics[scale=.35]{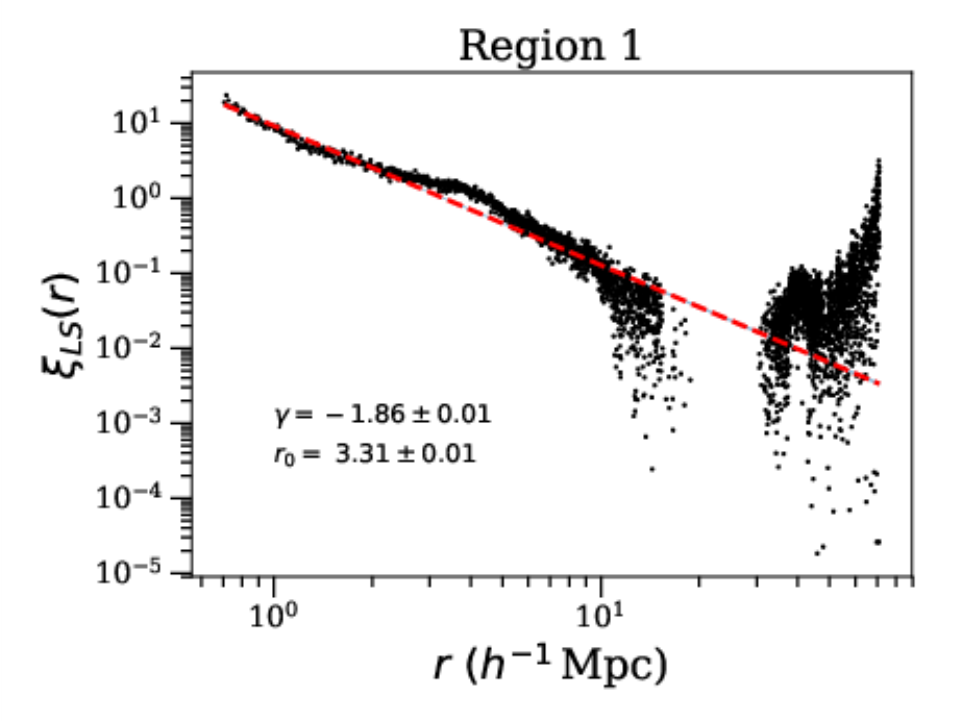}
  \includegraphics[scale=.35]{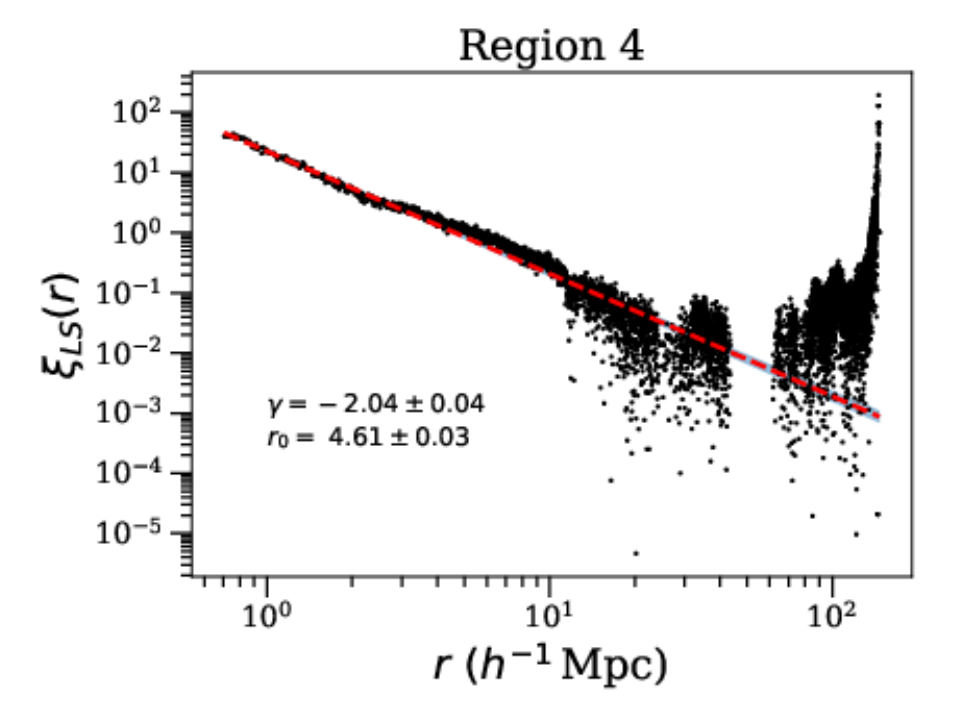}
  \caption{The LS two-point correlation $\xi_{LS}(r)$ for Region 1 and 4
  calculated with eight times more random points than the data sample. The
  dashed red line is the power-law fit to the LS correlation function
  $\xi_{LS}(r) = \left( r/r_0 \right)^{\gamma}$ calculated via non-linear least
  squares.}
  \label{fgr:2pcf_R1R4}
\end{figure}

\begin{figure*}
    \centering
    \includegraphics[width=.3\textwidth]{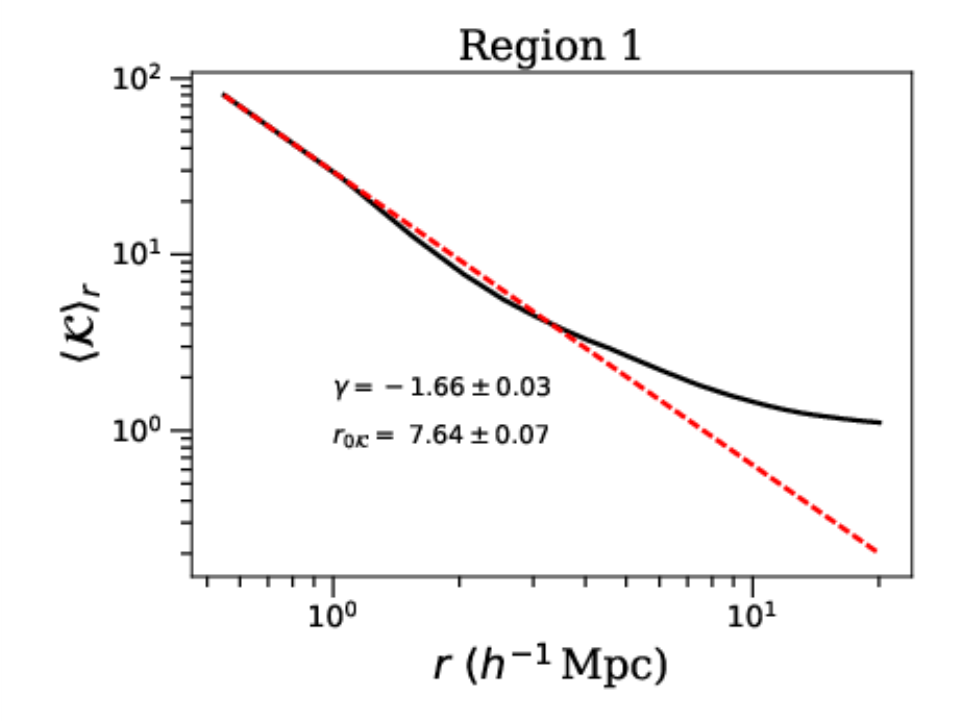}
    \includegraphics[width=.3\textwidth]{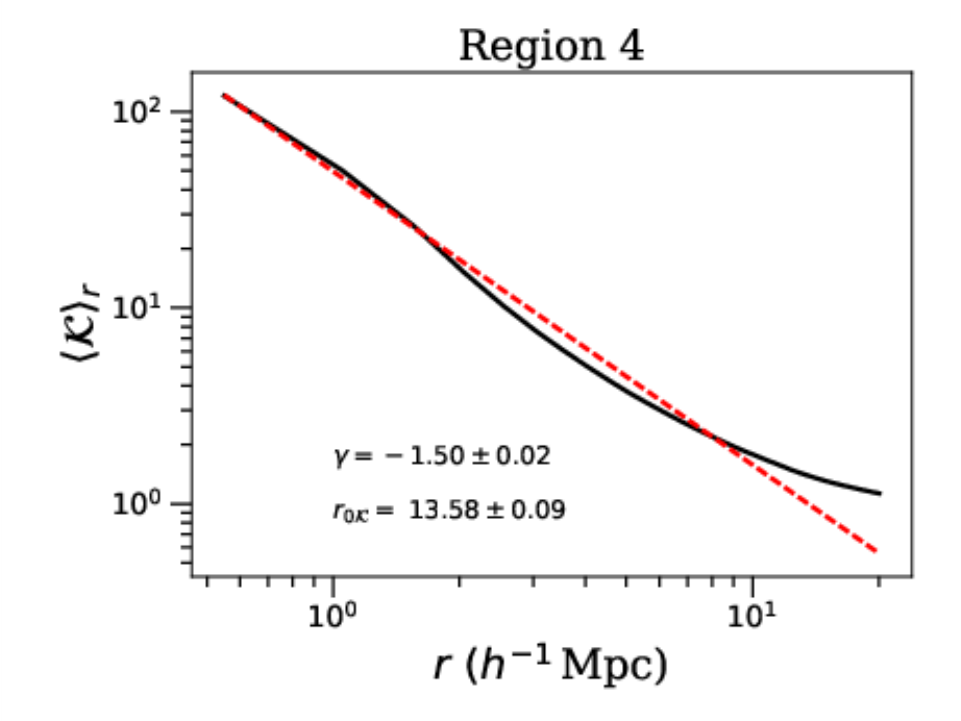} 
    \includegraphics[width=.3\textwidth]{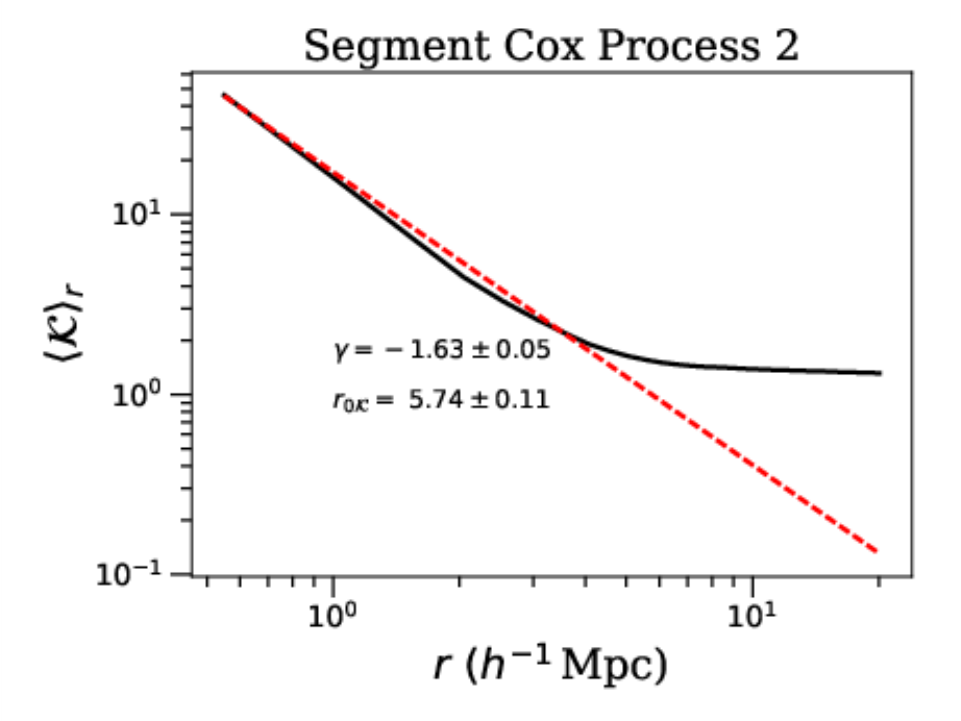}
  \caption{The scaled counts-in-spheres estimator $\mathcal{N}_k(<r) = \langle
  \mathcal{K} \rangle_r$ from Equation \ref{eq:scaled_meandeg} for Region 1,
  Region 4 and Cox Process 2 calculated until $r= 20 h^{-1}$Mpc with $\Delta r =
  0.5 h^{-1}$Mpc. The dashed red line is the power-law fit to the scaled
  counts-of-spheres $\langle \mathcal{K} \rangle_r = \left( r/r_{0\mathcal{K}}
  \right)^{\gamma}$ calculated via non-linear least squares.}
  \label{fgr:2pcf_mean}
\end{figure*}

 In this work, we propose the mean degree $\langle k \rangle$ as estimator for $N( < r)$ to analyze the fractal correlation dimension as a function of $r$. In fact, $N( < r)$ and $\langle k \rangle$ are both equivalents in terms of definition in a spatial network and represent the average number of neighbors around an object at a given radius $r$. The definition of mean degree is
\begin{align}
\label{eq:meandeg}
\langle k \rangle= \frac{1}{n}\sum_{i=1}^{n} k_i= \frac{1}{n}\sum_{i=1}^{n}\sum_{j\neq i} A_{ij},
\end{align}
where $n$ is the total number of nodes $i$ at the network,
$A_{ij}$ and $ k_i$ are defined in Equations \ref{eq:adj} and \ref{eq:deg}
respectively. The scaled counts-in-spheres estimator is
\begin{align}
\label{eq:scaled_meandeg}
\mathcal{N}_{k}(<r)= \langle \mathcal{K} \rangle_r = \frac{\langle k \rangle^{\rm data}_{r}}{\langle k \rangle^{\rm random}_{r}},
\end{align}
where $\langle k \rangle^{\rm random}_{r}$ is the mean degree as
a function of $r$ for the RGG sample and $\langle k \rangle^{\rm data}_{r}$ is
the mean degree as a function of $r$ for the data sample. The correlation
dimension follow the same definition of Equation \ref{corr_dim}. We compare the
power-law fit of the scaled counts-in-spheres estimator in Equation
\ref{eq:scaled_meandeg} with the power-law fit for the LS estimator of the
two-point correlation function. The results of the LS two-point correlation
function for Region 1 and 4 are showed in Figure \ref{fgr:2pcf_R1R4} with the
respective power-law fit, $\xi_{LS}(r) = \left( r/r_0 \right)^{\gamma}$. The
results of $\langle \mathcal{K} \rangle_r$ for Region 1, 4 and Segment Cox
Process 2 are showed in Figure \ref{fgr:2pcf_mean} with the respective power-law
fit $\langle \mathcal{K} \rangle_r = \left( r/r_{0\mathcal{K}}
\right)^{\gamma}$. 

\begin{figure}
\centering
  \includegraphics[scale=.35]{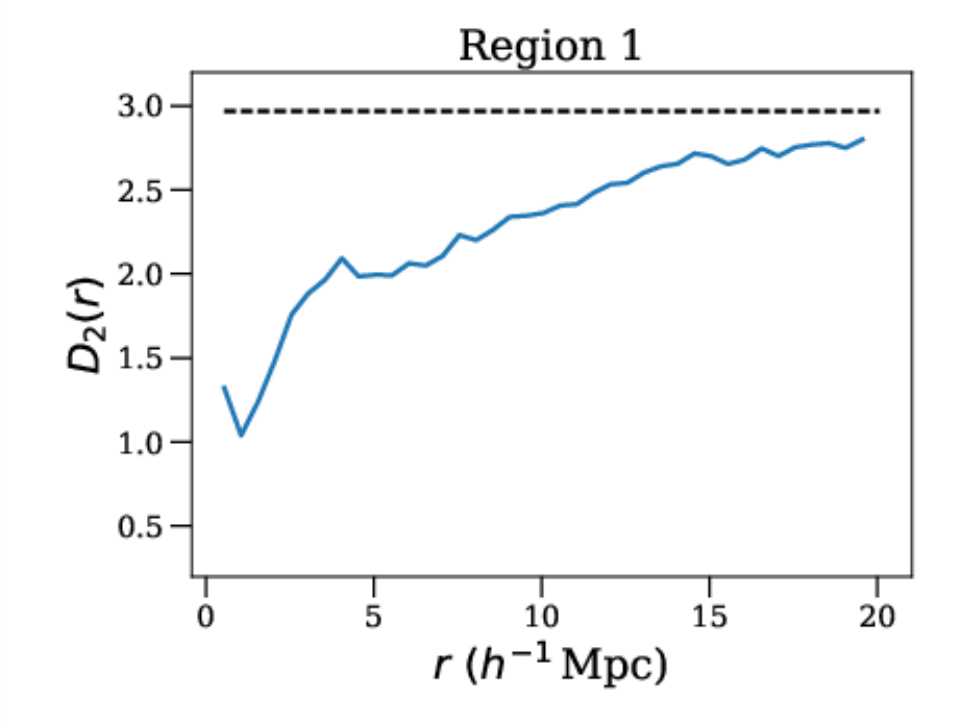}
  \includegraphics[scale=.35]{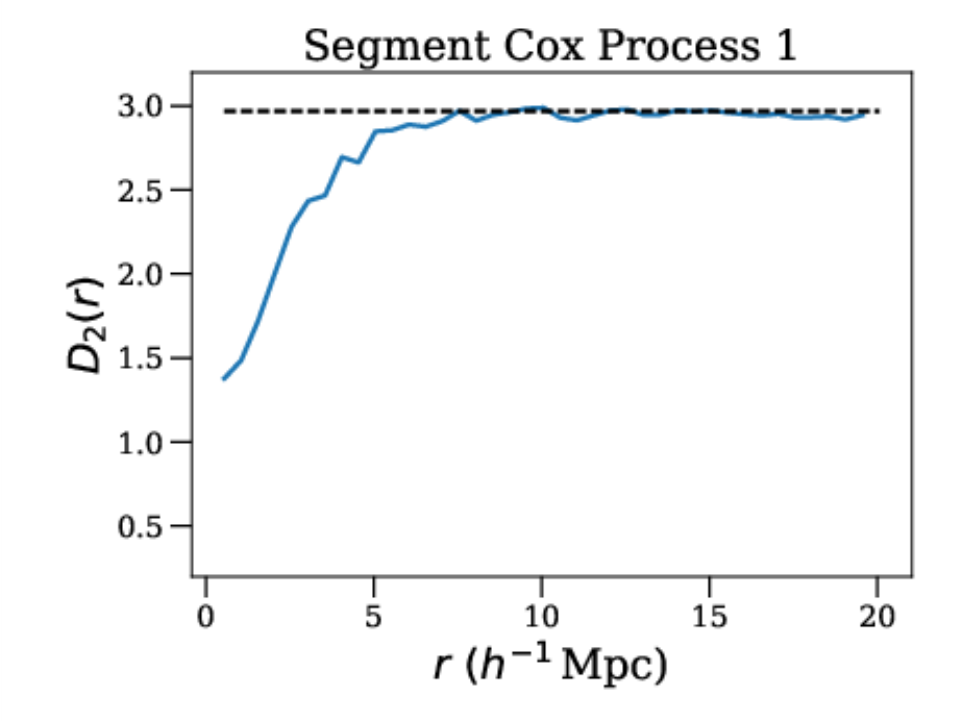}
  \includegraphics[scale=.35]{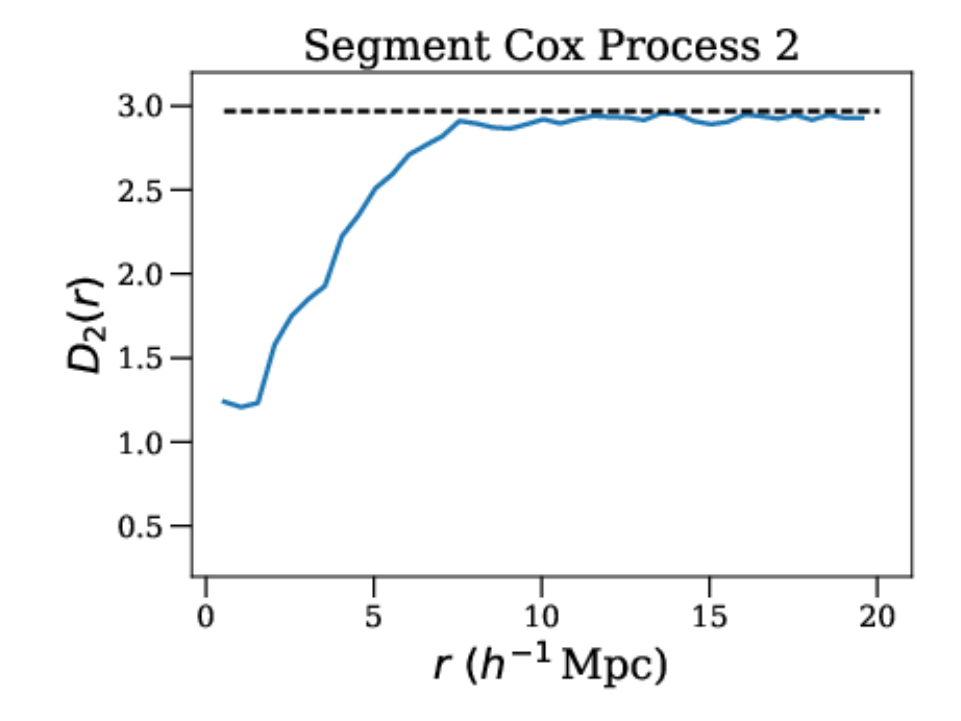}
  \includegraphics[scale=.35]{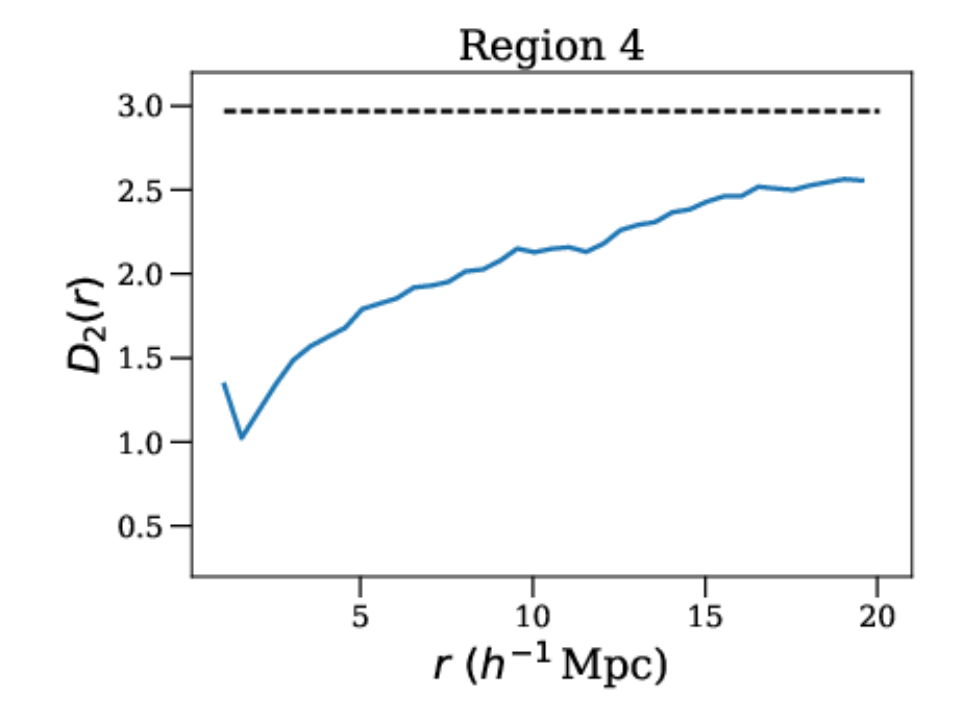}
  \caption{The blue line represents the correlation dimension as a function of
  $r$ obtained from Equation\ref{corr_dim} for the Equation
  \ref{eq:scaled_meandeg} until $r= 20 h^{-1}$Mpc with $\Delta r = 0.5
  h^{-1}$Mpc. The dashed black line represents $D_2(r)=99\% D_2^h$, where $D_2^h
  =3$, this is the criterion adopted in \citet{scrimgeour2012wigglez} to consider
  the transition to homogeneity of the distribution. In this plot we show
  $D_2(r)$ for Region 1, Region 4, Cox Process 1 and 2.}
  \label{fgr:D2mean}
\end{figure}

To analyze the behavior of the $D_2(r)$ for Region 1, Segment
Cox process 1 and 2, and Region 4, we measured the mean degree as a function of
$r$ until $r= 20 h^{-1}$Mpc with $\Delta r = 0.5 h^{-1}$Mpc. For the Segment Cox
process 1 and 2, we determine $D_2(r)$ for one realization of each configuration
with a RGG sample of same dimension and mean number of point. Identical
procedure is used for Region 1 and 4. The results are showed in Figure
\ref{fgr:D2mean}. 
 
The dashed line in Figure \ref{fgr:D2mean} represents
 $D_2(r)=99\% D_2^h$, where $D_2^h =3$; this is the same criterion adopted in \citet{scrimgeour2012wigglez} to consider the transition to homogeneity of the
 distribution. Our goal here is to obtain the correlation dimension via network
 theory and not to obtain the homogeneity scale of our samples. In fact, our
 samples are too small for this proposition. However, we observe that the
 homogeneity scale is achieved to the Segment Cox process 1 and 2 as expected,
 but it does not occur to Region 1 and Region 4 although both show a trend to
 arrive at the homogeneous scale for higher values of $r$. Comparing the $r_0$
 and $r_{0\mathcal{K}}$, we observe that $r_{0\mathcal{K}}$ is closer to the
 homogeneity scale obtained by $D_2(r)$ than $r_0$ for the Segment Cox Process.
 The $r_0$ is closed to the connectivity regime of the data sample network,
 i.e., all galaxies are connected in one component.
 
We are interested in small values of $r$ at the percolation
 phase, indicated by the maximums of the network diameter. These values are not
 detected with a power-law approximation to the two-point correlation function
 or the scaled counts-in-spheres estimator. However, the correlation dimension
 at this scale indicates a high clustering. The function $D_2(r)$ shows a
 multiscale behavior, i.e., the function gradually increases tending towards
 weaker clustering scales \citep{martinez1995multiscaling}. The values of the
 correlation function at the maximum network diameter of our samples are:
 $D_{2SC}(r3)=1.39$, $D_{2R1}(r3)=1.41$, $D_{2R2}(r3)=1.26$, $D_{2R3}(r4)=1.44$,
 $D_{2R4}(r4)=1.51$, $D_{2Cox2}(r2)=1.76$;  for the $r$ associated to
 superclusters are $D_{2SC}(r2)=1.73$, $D_{2R}1(r2)=1.33$, $D_{2R2}(r2)=1.21$,
 $D_{2R3}(r3)=1.30$, $D_{2R4}(r3)=1.31$, $D_{2Cox2}(r1)=1.58$.  We observe that
 the galaxy distribution presents a smaller correlation dimension than the
 Segment Cox Process 2 at the radii of maximums of the network diameter except
 for the SuperComa region. The correlation dimension of the Segment Cox Process
 2 presents a smaller variation between the connection radii than the galaxy
 distribution. 

\begin{figure}
\centering
  \includegraphics[scale=.35]{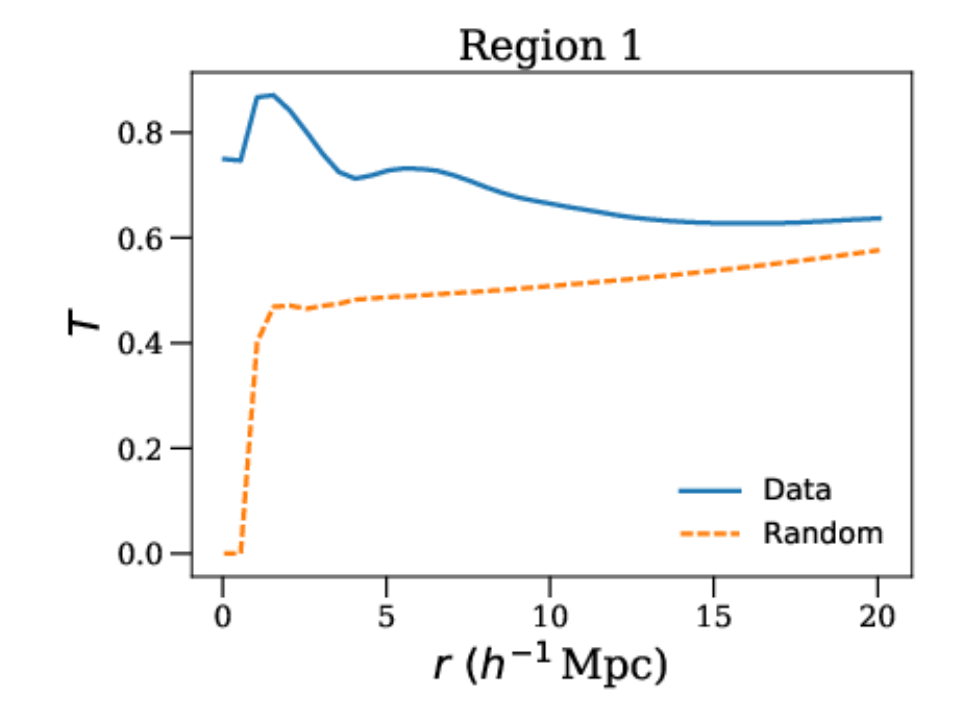}
  \includegraphics[scale=.35]{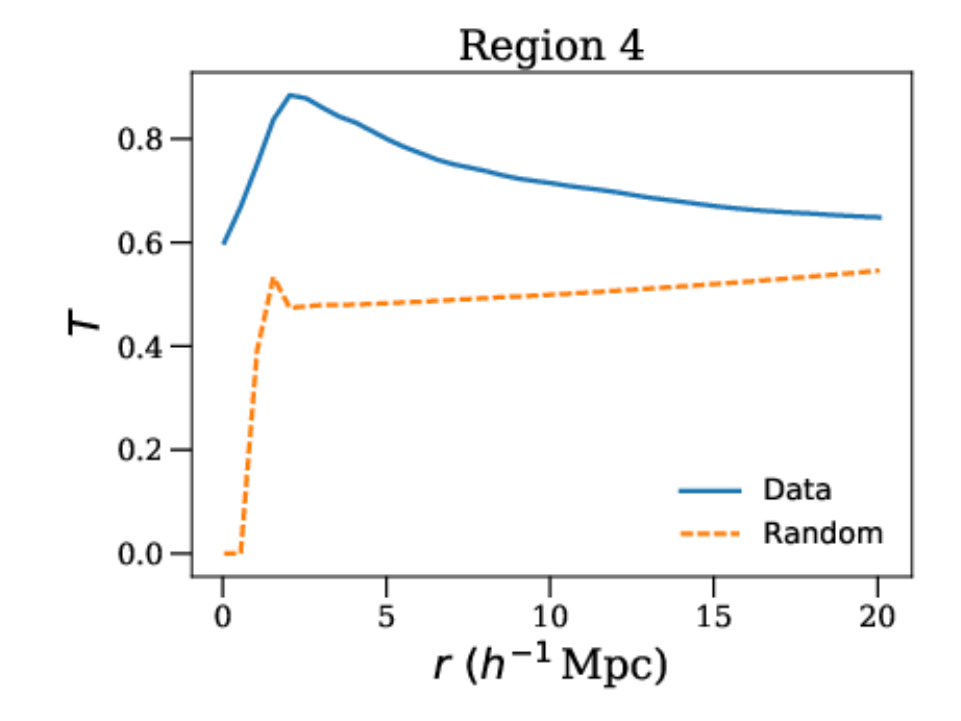}
  \includegraphics[scale=.35]{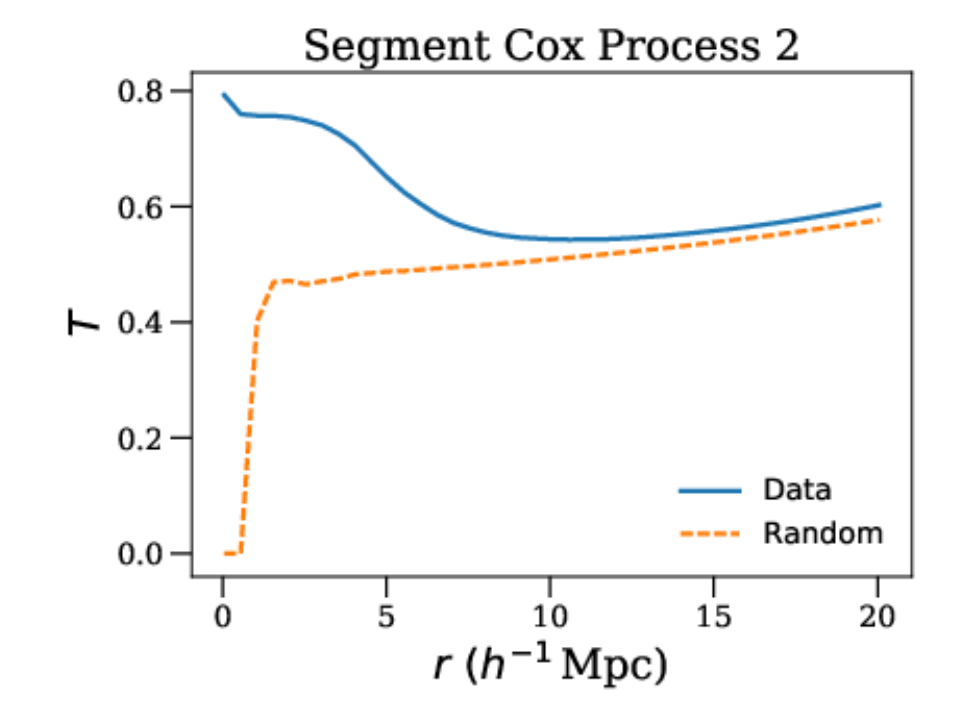}
  \caption{The transitivity as a function of $r$ until $r= 20 h^{-1}$Mpc with
  $\Delta r = 0.5 h^{-1}$Mpc. The blue line is the $T(r)$ for the data sample
  and the yellow dashed line is the $T(r)$ for the corresponding RGG sample.}
  \label{fgr:transiti}
\end{figure}

In network theory, the transitivity is a clustering indicator
 measuring the number of triangles formed in the network compared with all
 possible triangles. The Figure \ref{fgr:transiti} compare the transitivity as a
 function of $r$ for Region 1, 4 and the Segment Cox Process 2 with the
 respective measure for the RGG sample. This measure agrees with the correlation
 dimension, indicating higher clustering for Region 1, 4 and the Segment Cox
 Process 2 than for the RGG sample. The Segment Cox Process 2 presents smaller
 and more stable transitivity than Region 1 and 4 for small $r$, tending to the
 same behavior of the RGG sample for high $r$. The transitivity is used in
 \citet{hong2016discriminating} and \citet{hong2020constraining} to distinguish
 topologically different distributions.

\section{Results and Discussion}
\label{sec:results}

We measured the variation of the network diameter, the APL, the transitivity,
and the magnitude of the giant component as a function of the connection radius
$r$ for the galaxy network, the Segment Cox network  and
 the RGG sample with the same mean density of points
as the galaxy data sample (see Section \ref{sec:diamAPL}.)

We show the corresponding $r$ for the maximums of the network diameter as a
function of $r$ in Table \ref{tbl1} (for data samples and Segment Cox Process 2)
and in Table \ref{tbl4} (for RGG samples) with the respective values for the
network diameter, APL, number of galaxies in the giant component, mean local
clustering coefficient, transitivity, mean degree centrality and mean
betweeness centrality.

For RGG samples, we observed that the maximum network diameter is reached in a
critical $r$ and is related to the inflection point of the growth curve of the
giant component, in agreement with \citet{hong2016discriminating} (see Section
\ref{sec:diamAPL}). This point is reached  suddenly and is related
to the mean density of galaxies on the sample. After this peak the diameter
decreases, the number of nodes at the giant component remains approximately
constant, increasing the connections between the nodes at the same component,
indicating a diameter saturation: the network tends to a complete network \citep{hong2016discriminating}.

The development of the network diameter as a function of $r$ for the data
samples present more than one pronounced peak,
suggesting that this measure can distinguish a set of characteristics distances
of the spatial distribution of galaxies. The multi-peak behavior of the network
diameter function infers a distance scale, represented by $r$, for the
structures in the node distribution. The APL follows the same behavior. We
verify that the magnitude of the giant component grows in steps, indicating the
formation of the component with a representative fraction of the galaxies, each
step corresponding to the maximums on the network diameter function. The
emergence of the giant component in steps represents the inhomogeneities of the
galaxies distribution, in agreement with the finding of \citet{einasto1984structure} using FoF algorithms. The difference between the
methods is that FoF algorithms need an auxiliary method to find the best linking
length, which can be the largest linking length or the mean linking length.
\citet{einasto1984structure} used the magnitude of the giant component to find
the best linking length for each structure. Due to the saturation, the network
diameter function presents maximums at the connection radii that best represents
the structures on the sample. This behavior is also present in
the Segment Cox network, mainly in the segment Cox Process 2, where the mean
density of points per segment is higher than in Segment Cox Process 1.
Continuous percolation studies for Cox Process \citep{hirsch2019continuum} shows
a non trivial percolation evolution. The behavior of the galaxy samples we used
in this work have a similar percolation evolution.

Comparing the plots in Figure \ref{fig:diam} for the Coma Region and Super Coma
 Region, we observe that the peak around $r = 0.37 \ h^{-1}$Mpc represents the
 Coma cluster. The spatial network constructed with this $r$ 
 emphasizes the  cluster structure of galaxy distribution. For the Coma
 Region, the peak around $r = 1.47 \ h^{-1}$Mpc represents the percolation
 phase. For the Super Coma Region, the peak around  $r= 1.61\ h^{-1}$Mpc is
 related to the supercluster structures, and around  $r= 1.99\ h^{-1}$Mpc to the
 large-scale structure of the supercluster region, i.e., it gives the connection
 radius  at the percolation phase.
 Analyzing  the Super Coma Region and Regions 1 and 2, we observe that the
 maxima at the network diameter function have a similar $r$ even for samples
 with very different mean densities of nodes. For example, we can associate $r=
 0.37\ h^{-1}$Mpc as a connection radius for the clusters, $r= 1.61\ h^{-1}$Mpc
 to $r= 1.89\ h^{-1}$Mpc for superclusters, and $r= 2.0\ h^{-1}$Mpc for the
 percolation phase. 

The network diameter is very sensitive to the density of the nodes: a higher
density of nodes as the Super Coma Region has a characteristic connection radius
smaller than that of Regions 1 and 2, although both are the same order of around
$r= 2.0\ h^{-1}$Mpc. The network diameter function points out the region of the
sample with more statistical relevance with relation to the density of nodes
(see also Appendix \ref{sec:appendix2}). The APL is less sensitive to small
peaks in density than the network diameter but marks well the points of
emergence of the giant component. The correlation dimension and
transitivity for small connection radii indicates high clustering at these
scales (see Section \ref{sec:Meandeg}).

It is important to note that as the redshift increases (Region 3 and 4, Figure
\ref{fig:diam}), we observe an increase in the values of the connection radius
for the peaks on the network diameter. The main reason for this behavior is the
lack of completeness of the catalog. The increase in the connection radius
compensates the missing faint galaxies in the catalog.

In the next subsections, we discuss the statistical measurements for a network
constructed with the connection radii inferred from the network diameter function behaviour.

\begin{table*}
\centering

\caption{Values of the network diameter (diam), APL ($\langle d \rangle$), giant
component (Giant{C}), mean local clustering coefficient ($C$), transitivity
($T$),  mean degree $\langle k \rangle$, and  mean betweenness $\langle BC
\rangle$, for the connection radii $r$ that maximizes the network
diameter.}
\label{tbl1}

\begin{tabular}{lrrrrrrr}
\hline
   & diam & $\langle d \rangle$& Giant{C} & $ C $ & $T$ & $\langle k \rangle$ & $\langle BC \rangle$\\
\hline
 Coma &  &   &  &  & &  &  \\
$r_{1}=0.37 h^{-1} $Mpc & 34  & 7.26 & 685  & 0.37 & 0.51 & 6.65 & 1\,014  \\
$r_{2}=1.47 h^{-1} $Mpc & 35  & 7.29 & 1145 & 0.69 & 0.72 & 155.23 & 2\,800   \\[.2cm]
 Super Coma &  &   &  &  & &  &  \\
$r_{1}=0.37 h^{-1} $Mpc & 34  & 6.64 & 685 & 0.33 & 0.58 & 3.28 & 214  \\
$r_{2}=1.61 h^{-1} $Mpc & 69 & 19.91 & 3290 & 0.70 & 0.76 & 53.68 & 13\,420  \\
$r_{3}=1.99 h^{-1} $Mpc & 72  & 21.19 &5557  &0.72& 0.83 & 71.05 & 76\,025   \\[.2cm]
 R1 &  &   &  &  & &  &  \\
$r_{1}=0.37 h^{-1} $Mpc & 15  & 2.82 & 58 &0.29&0.69& 1.77 & 5  \\
$r_{2}=1.85 h^{-1} $Mpc &37  & 8.15 & 447 & 0.68 & 0.86 & 12.00 & 447   \\
$r_{3}=2.07 h^{-1} $Mpc & 100  & 34.15 & 2161 & 0.70 & 0.84 & 13.94 & 146\,760   \\[.2cm]
 R2 &  &   &  &  & &  &  \\
$r_{1}=0.25 h^{-1} $Mpc & 17  &3.60 & 61 &0.17&0.66& 1.01 & 4 \\
$r_{2}=1.89 h^{-1} $Mpc & 52  & 16.66 & 1689  & 0.70 & 0.89 & 17.67 & 5\,073   \\
$r_{3}=2.09 h^{-1} $Mpc & 92  & 30.49 & 3194 &0.70 &0.87 & 20.02 & 4.1\ $10^9$ \\[.2cm] 
 R3 &  &   &  &  & &  &  \\
$r_{1}=0.37 h^{-1} $Mpc & 28  & 6.04 & 187 & 0.23 & 0.66 & 1.38 & 20  \\
$r_{2}=2.21 h^{-1} $Mpc & 67 & 14.53 & 1592  &0.67 &0.85 & 17.59 & 2\,866   \\
$r_{3}=2.73 h^{-1} $Mpc & 89  & 25.29 & 3571 &0.70 &0.87 & 23.02 & 2.8\ $10^{12}$ \\ 
$r_{4}=3.15 h^{-1} $Mpc & 102  & 32.24 & 5833 &0.71 &0.86 & 27.94 & 3.3\ $10^{16}$ \\[.2cm] 
 R4 &  &   &  &  & &  &  \\
$r_{1}=0.45 h^{-1} $Mpc & 16 & 3.48 & 88 & 0.23 & 0.66 & 1.41 & 7  \\
$r_{2}=2.43 h^{-1} $Mpc &  73 & 17.85 & 987  & 0.65 & 0.88 & 13.81 & 1\,849    \\
$r_{3}=2.93 h^{-1} $Mpc & 110  & 31.99 & 2288 & 0.68 & 0.86 & 17.45  & 260\,678   \\
$r_{4}=3.41 h^{-1} $Mpc & 111 & 41.52 & 4851 &0.70 & 0.85 & 21.66 & 7.3\ $10^{15}$ \\ 
 Segment Cox Process 2 &  &   &  &  & &  &  \\
$r_{1}=2.31 h^{-1} $Mpc & 34 & 7.33 & 306 & 0.81 & 0.75 & 9.60 & 223  \\
$r_{2}=3.01 h^{-1} $Mpc &  73 & 26.39 & 2568  & 0.79 & 0.74 & 14.80 & 7\ $10^{8}$  \\ 
\hline
\end{tabular}

\end{table*}

\begin{table*}
\centering
\caption{Values of the network diameter (diam), APL ($\langle d \rangle$), giant
component (GC), mean local clustering coefficient ($C$),transitivity ($T$),  mean degree $\langle k \rangle$, and  mean betweenness $\langle BC
\rangle$, for the connection radius $r$ that maximizes the network diameter in
RGG samples.}
\label{tbl4}

\begin{tabular}{lrrrrrrrr} 
\hline
   & diam & $\langle d \rangle$& G{C} & $ C $ & $T$ & $\langle k \rangle$ & $\langle BC \rangle$\\
\hline \\
 Random Coma &  &   &  &  & &  &  \\
$r_{C_U}=1.30 h^{-1} $Mpc & 66 & 25.80   &  906  & 0.40 & 0.48 & 3.08  & 7\,002 \\[.2cm]
Random Super Coma &  &   &  &  & &  &  \\
$r_{SC_U}=2.02 h^{-1} $Mpc & 209  & 75.47 & 3270 & 0.37 & 0.47 &  2.84  & 52\,097 \\[.2cm]
 Random R1 &  &   &  &  & &  &  \\
$r_{R1_U}=2.39 h^{-1} $Mpc & 127  & 39.42 & 2556 & 0.36 & 0.47 & 2.83 & 26\,285 \\[.2cm]
Random R2 &  &   &  &  & &  &  \\
$r_{R2_U}=2.33 h^{-1} $Mpc & 191 & 69.70 & 2706 & 0.37 & 0.48 & 2.90 & 48\,383 \\[.2cm]
Random R3 &  &   &  &  & &  &  \\
$r_{R3_U}=2.97 h^{-1} $Mpc & 157  & 57.12 &5247 &0.38& 0.48 & 2.99 & 97\,557   \\[.2cm]
Random R4 &  &   &  &  & &  &  \\
$r_{R4_U}=3.27 h^{-1} $Mpc & 194  & 75.46 & 3480 &0.38 &0.48& 2.99 & 75\,744\\
\hline

\end{tabular}
\end{table*}

\subsection{Degree Centrality}
\label{subsec:deg}

\begin{figure}
\centering
\includegraphics[width=.49\columnwidth]{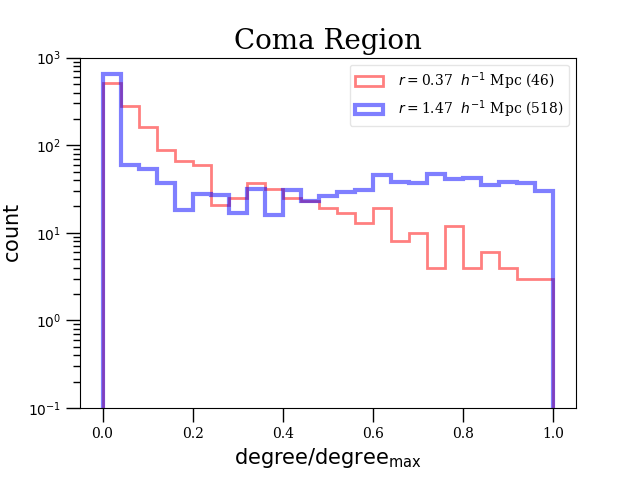}
\includegraphics[width=.49\columnwidth]{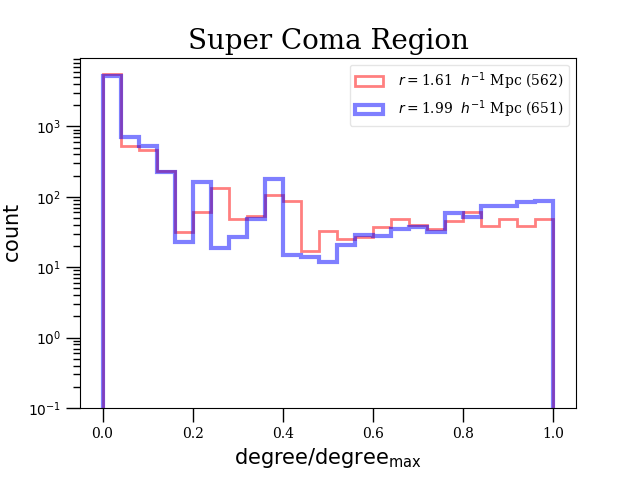}
\includegraphics[width=.49\columnwidth]{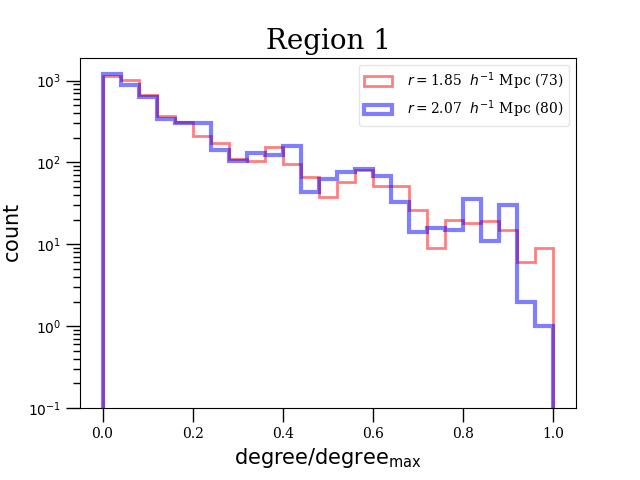}
\includegraphics[width=.49\columnwidth]{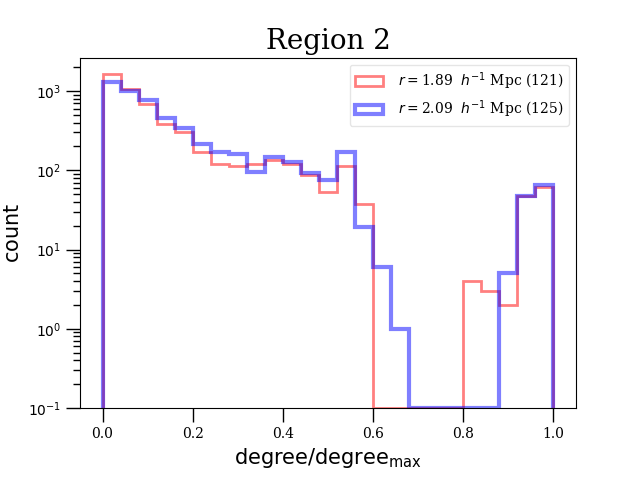}
\includegraphics[width=.49\columnwidth]{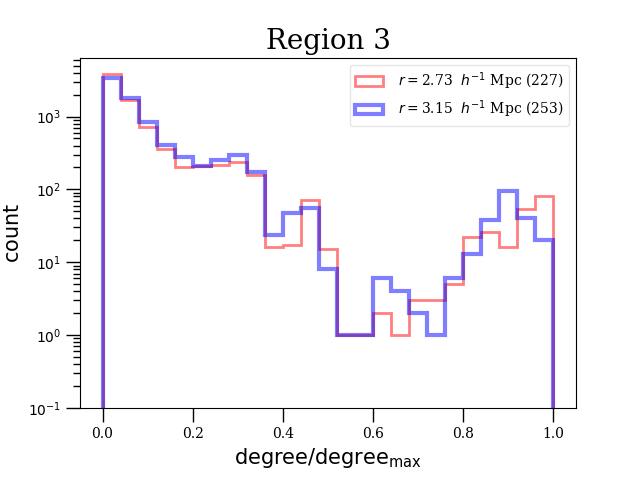}
\includegraphics[width=.49\columnwidth]{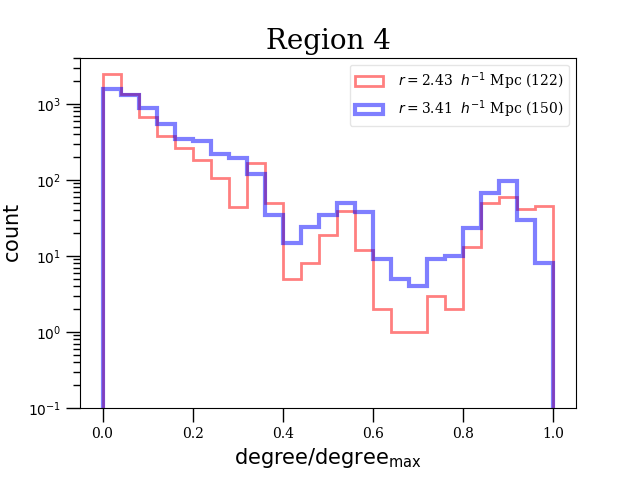}
\caption{ Degree centrality distribution of the networks constructed with the values of $r$ shown in the legend for Coma and Super Coma regions, and Regions 1, 2, 3 and 4. The number in parentheses is the maximum degree for the corresponding network used to normalize the degree.}
\label{fig:degree2}
\end{figure}

Figure \ref{fig:degree2} shows the degree distribution for the regions in the
study. For comparison, we exhibit the degree centrality for two values of $r$
corresponding to an intermediary peak on network diameter (red lines) and the
$r$ for the maximum network diameter (blue lines). We normalized the results by
the maximum degree of each distribution (the value is indicated in the legend,
in parentheses). The degree distribution presents an excess of nodes with a high
degree at the samples with dense structures, as we see in the Super Coma Region
and Regions 2, 3, and 4. Region 1 presents structures with a more uniform
density of galaxies. 

In general, the degree distribution is not too sensitive to the variation of $r$
for small $r$. For the data sample, the degree probability
distribution can be fitted by a power-law distribution $P(k) \propto r^{\gamma}$
with $ \gamma \simeq D_2 - 3$ for degree $k > 9$ until $r \sim 3.5 h^{-1}$Mpc.
The degree probability distribution for the RGG sample and the
Segment Cox Process follows a binomial distribution. The Segment Cox Process
presents a deviation of this distribution for large degrees. The increase of
$r$ produces an increase in the density of links between galaxies and,
consequently, an increase in the mean degree and in the high degrees.
We showed that the evolution of the mean degree as a function of
$r$ is equivalent to the count-of-spheres estimator presenting different
evolution between the data sample, the Cox sample and the RGG sample. The
correlation dimension is obtained of this estimator by the Equation
\ref{corr_dim}. 

\begin{figure*}
\centering
\includegraphics[width=.49\textwidth]{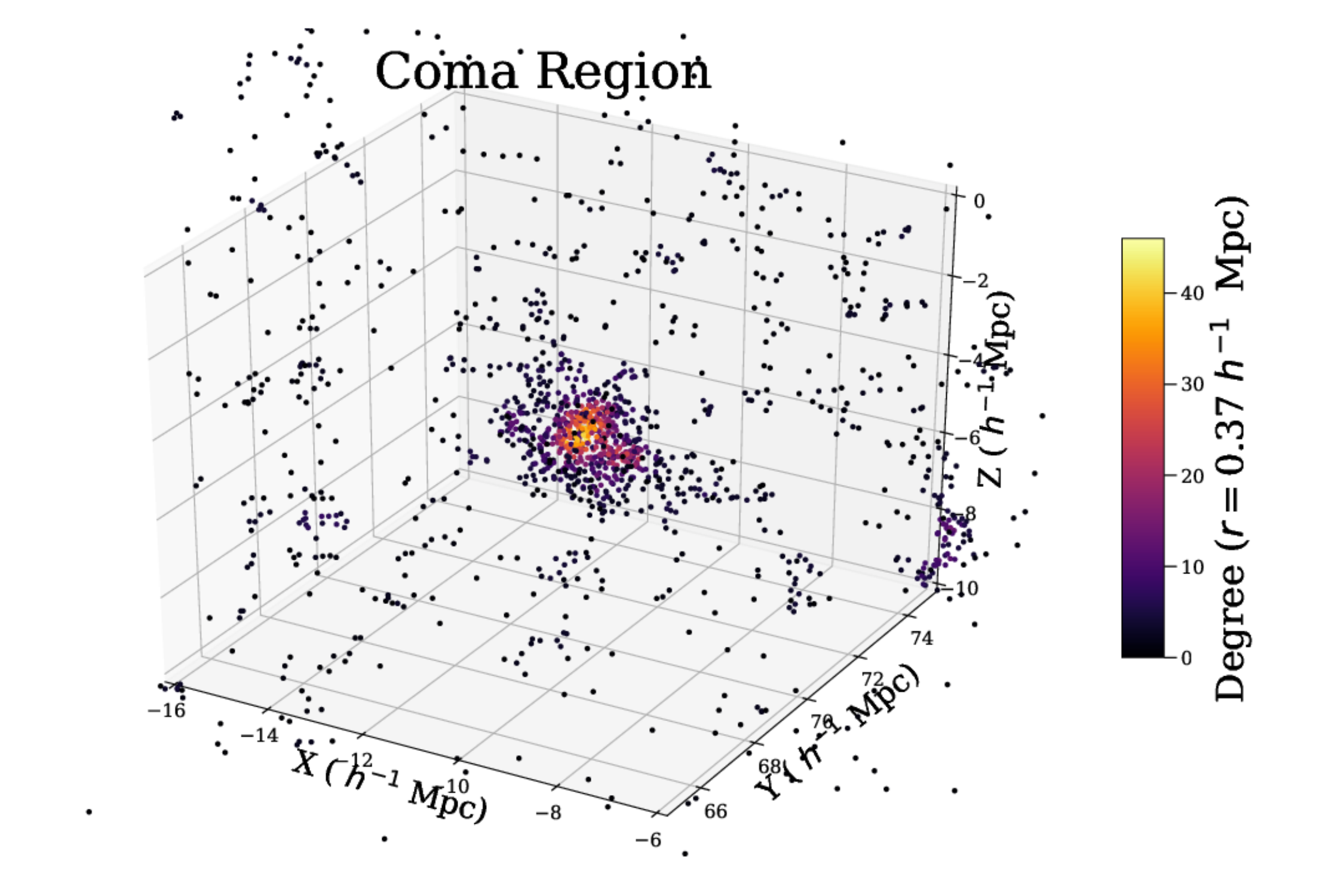}
\includegraphics[width=.49\textwidth]{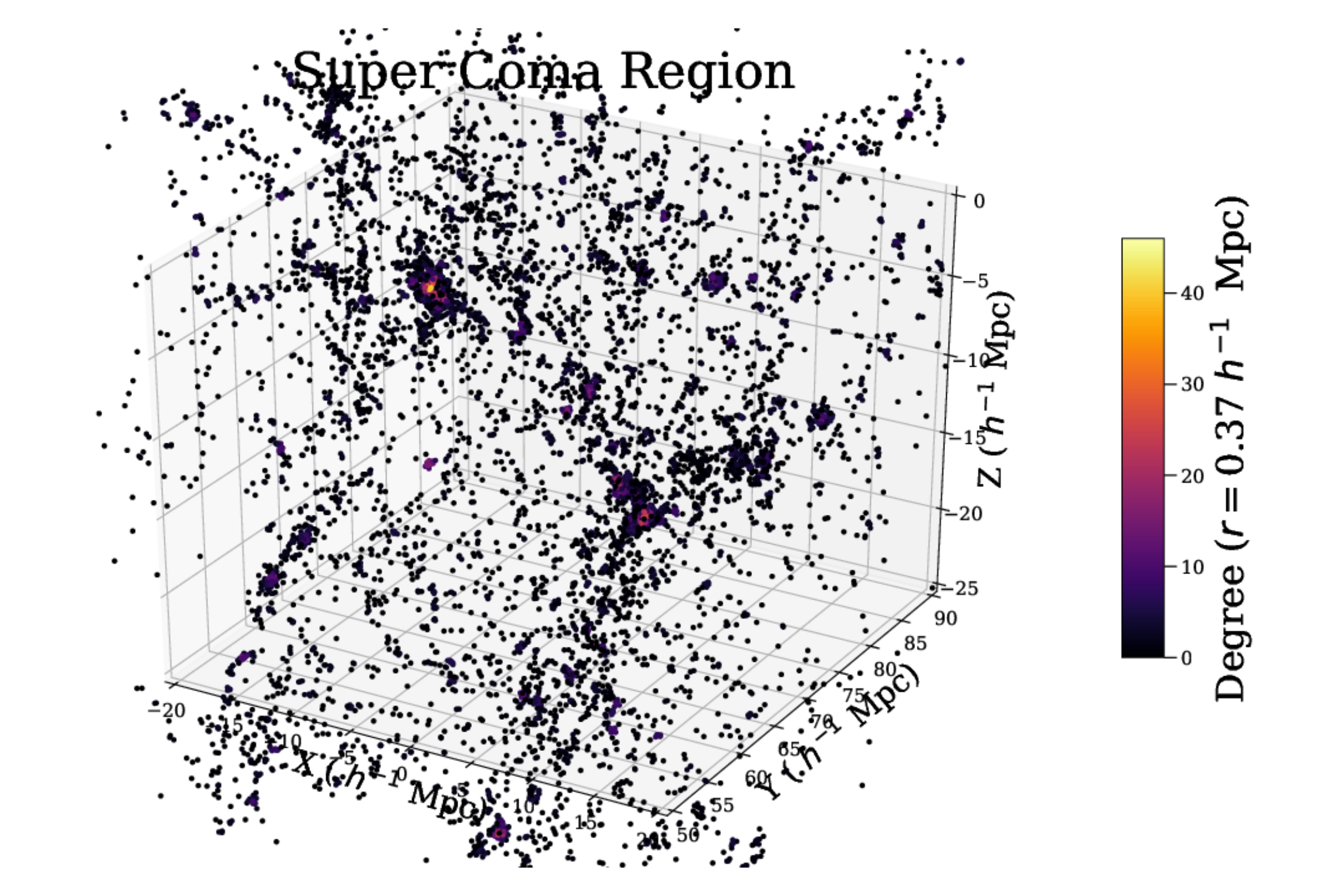}
\includegraphics[width=.49\textwidth]{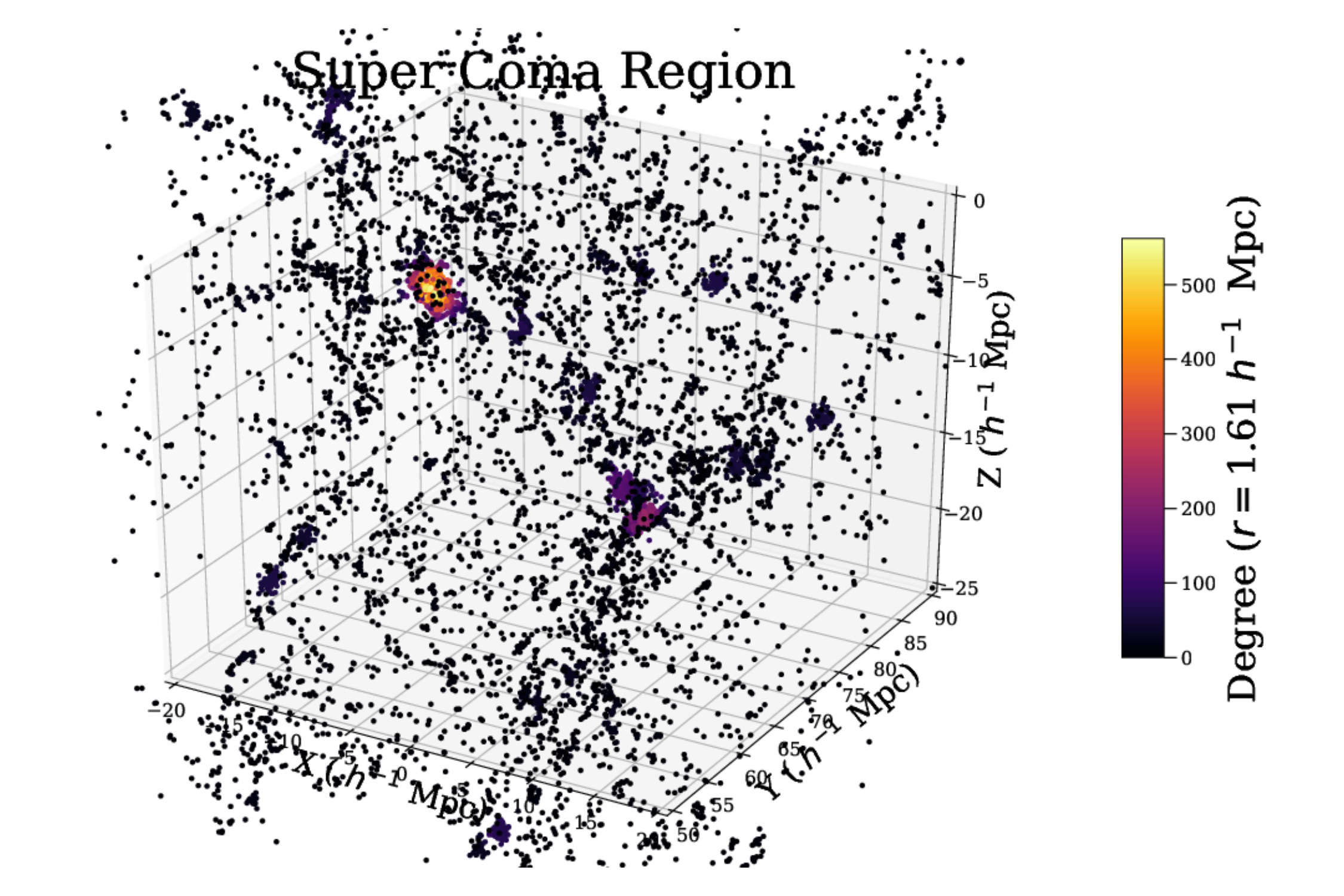}
\includegraphics[width=.49\textwidth]{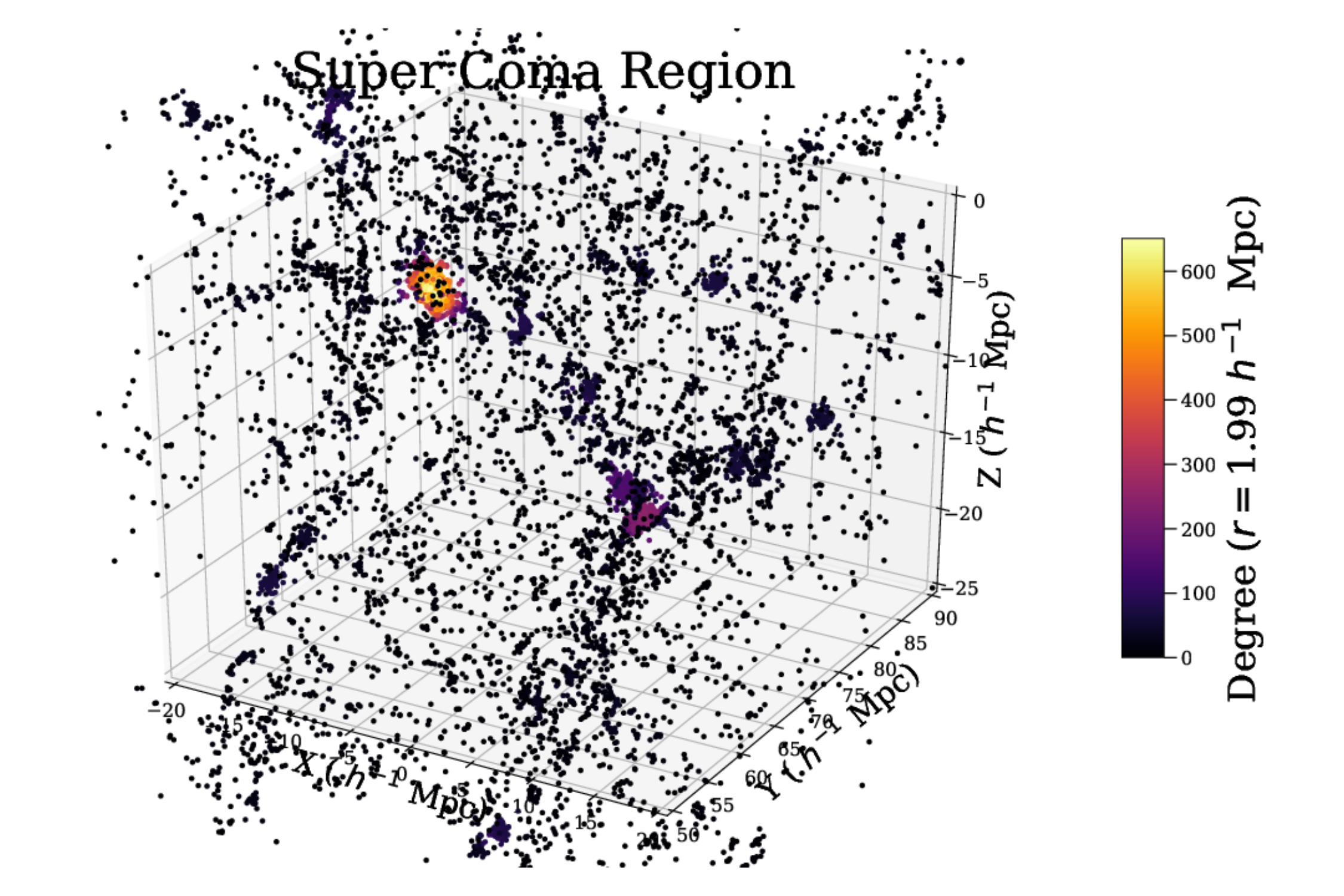}
\caption{Coma and Super Coma regions in color scale as a function of the degree centrality for the values of $r$ shown in the legend; yellow color represents the galaxies with the highest degree.}
\label{fig:supercoma3Ddegree}
\end{figure*}

Figure \ref{fig:supercoma3Ddegree} shows the differences between high degree
centrality regions for each $r$ corresponding to the maximum network diameter.
The core of the Coma Region presents a high degree; we observe the same behavior
in the Super Coma Region, where we also have a second region with a high degree
corresponding to the Leo cluster.  In Figures \ref{fig:centralRanregion} and
\ref{fig:Cox2distrib} (top), we show the degree distribution histogram and the
respective three-dimensional visualization of degree centrality for the RGG
samples and Segment Cox Process 2 corresponding to Region 1 dimensions.

\begin{figure*}
\centering
\includegraphics[width=0.3\textwidth]{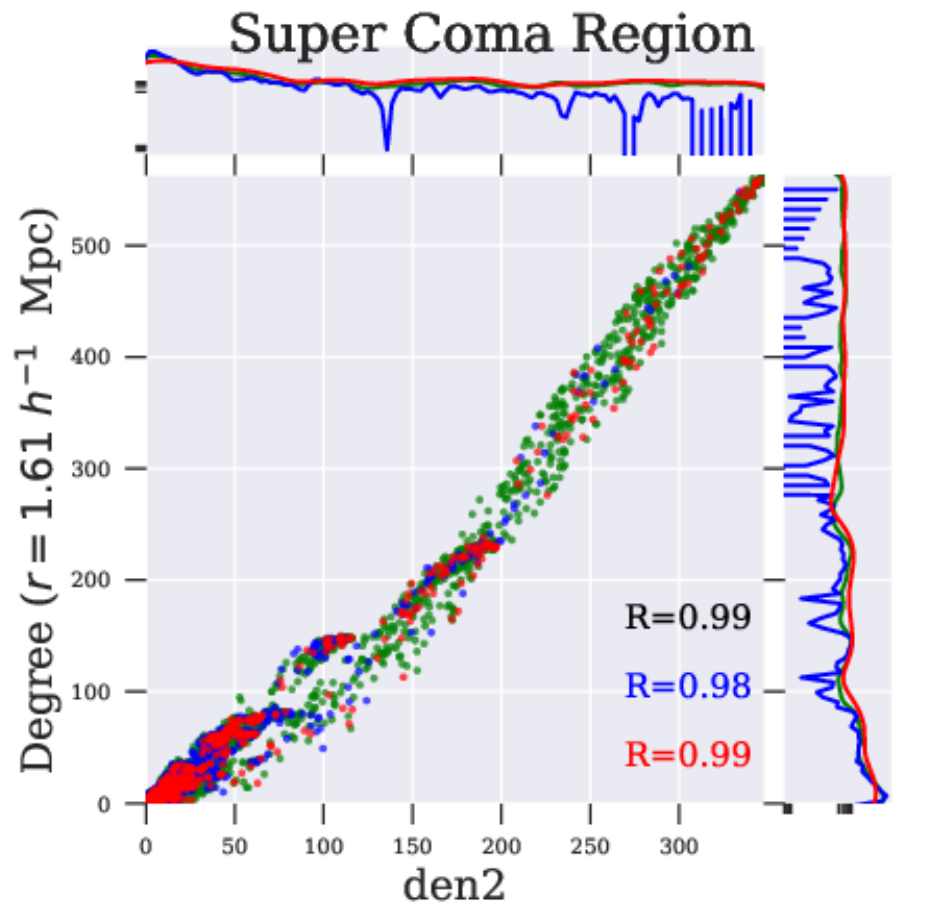}
\includegraphics[width=0.3\textwidth]{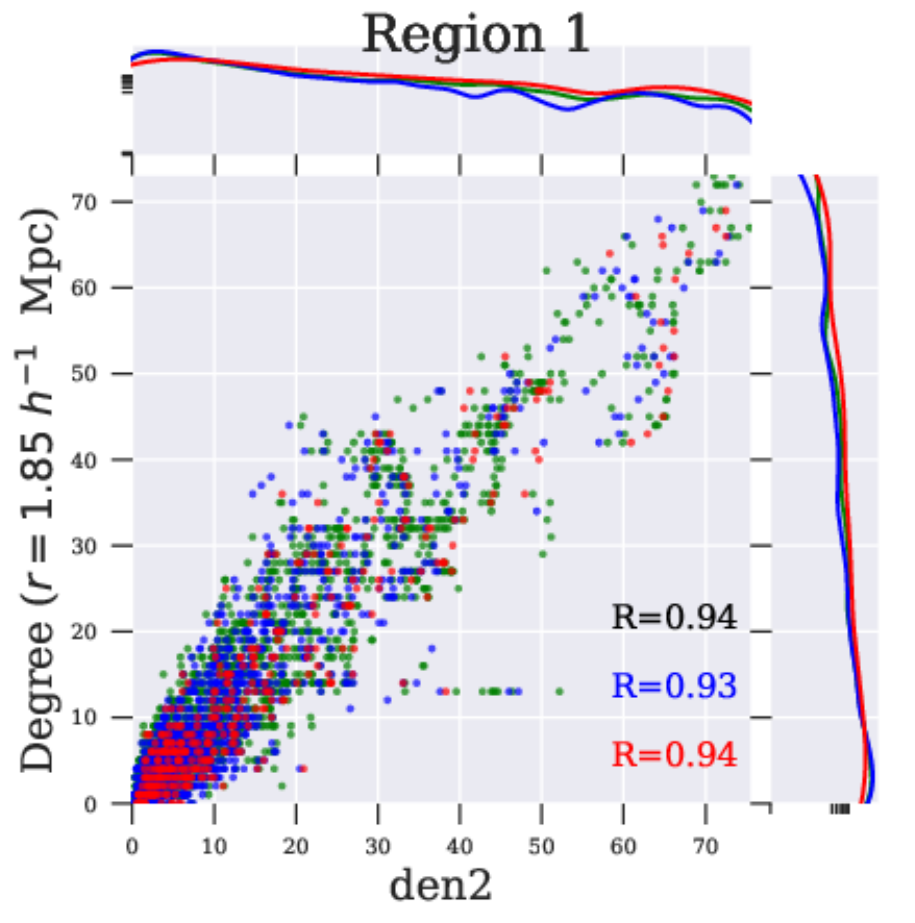}
\includegraphics[width=0.3\textwidth]{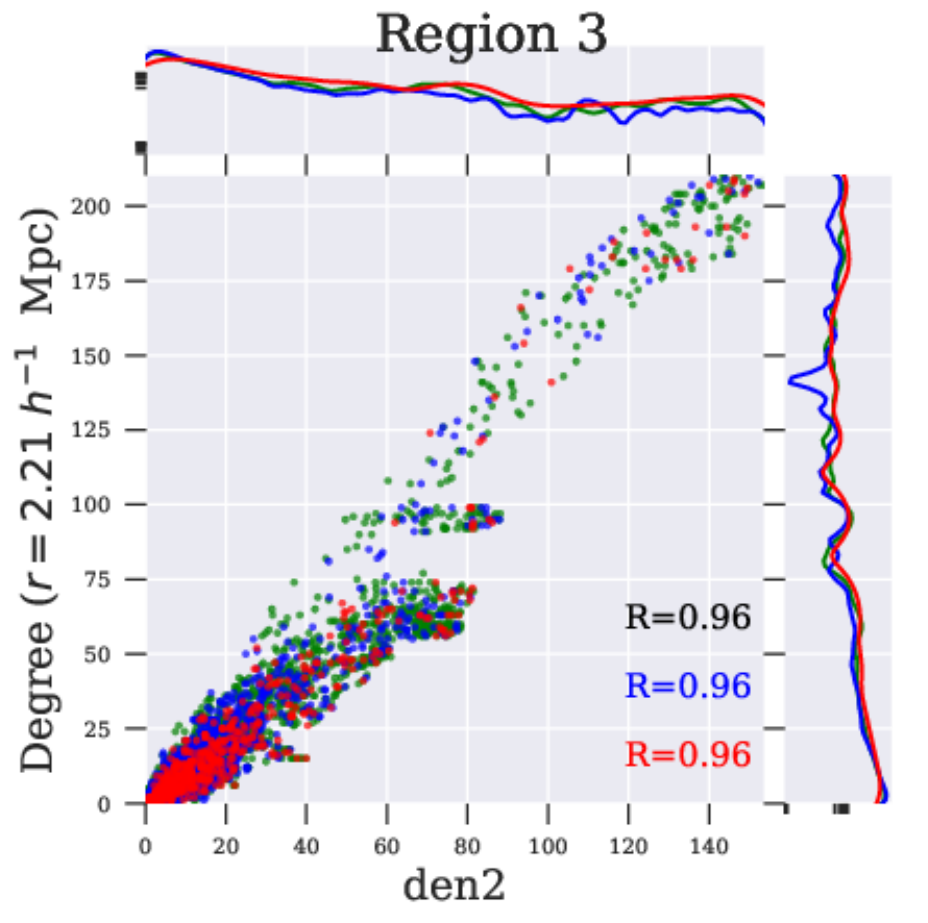}
\includegraphics[width=0.3\textwidth]{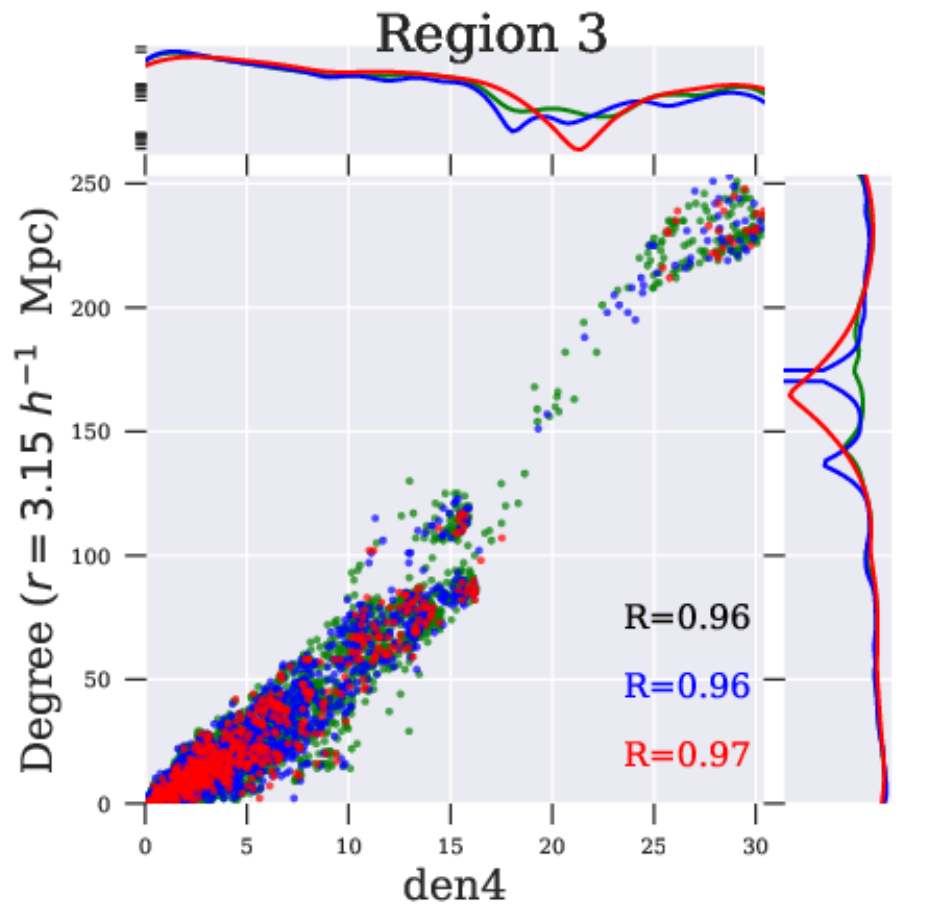}
\includegraphics[width=0.3\textwidth]{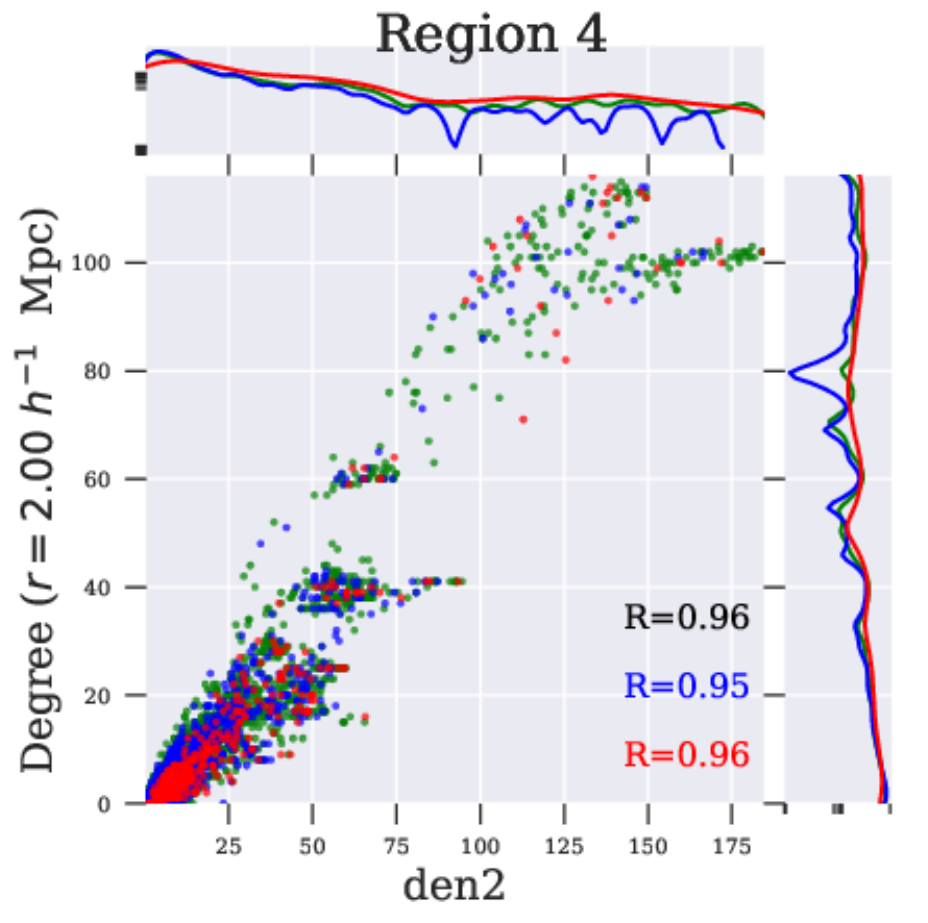}
\includegraphics[width=0.3\textwidth]{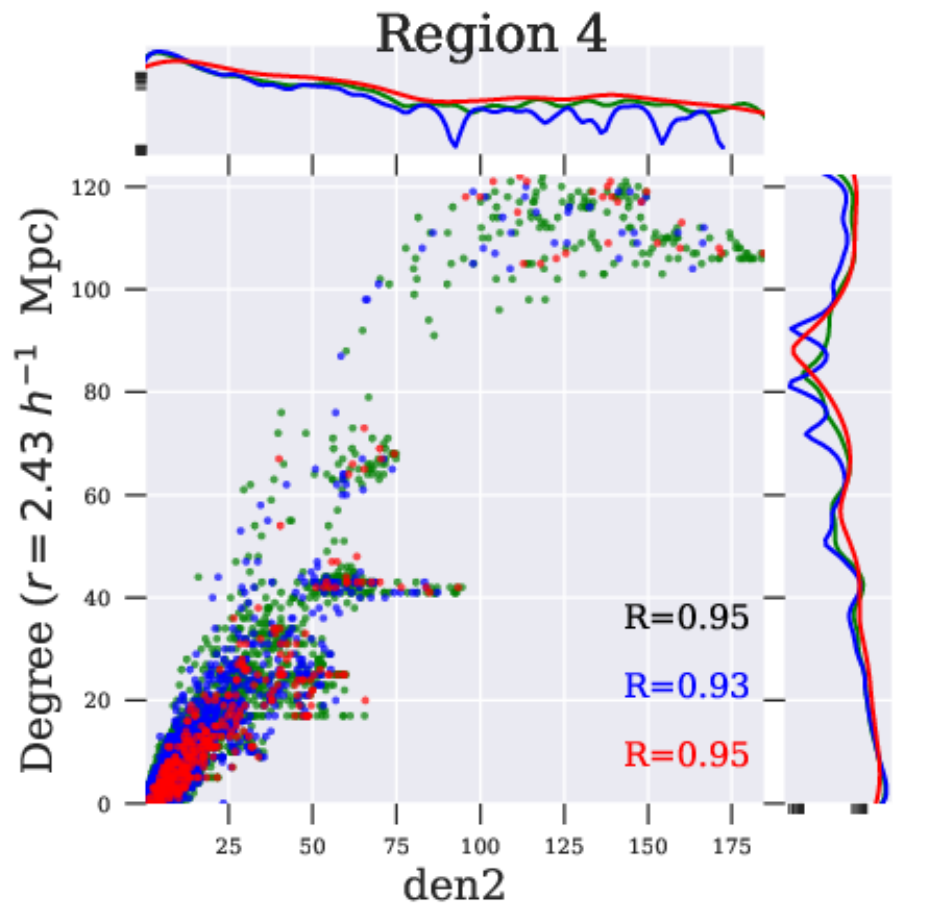}
\caption{Relationship between normalized environmental density den$a$ (for a kernel with smoothing scale $a$) of Ref.~\citenum{tempel2014flux} and degree centrality (for a $r$ corresponding to the maximums at the network diameter, except for Region 4 with $r=2.00 h^{-1}$Mpc ) for each galaxy on Super Coma regions and Regions 1, 3 and 4, respectively. The values for the Pearson correlation $R$ between the degree centrality and the normalized environmental density are shown in the plots. The Pearson correlation is in dark color for all galaxies, blue for the spirals, and red for the elliptical galaxies.  The smoothing scales are $a=2\ h^{-1}$ Mpc for den$2$ and $a=4\ h^{-1}$ Mpc  for den$4$; galaxies are identified by blue points (spiral), red (elliptical) and green (uncertain) according to Ref.~\citenum{lintott2008galaxy}. The side figures shows the cumulative distribution along the perpendicular axis.}
\label{fig:dendeg}
\end{figure*}

We compare the degree centrality with the normalized environmental density of
the galaxy developed by \citep{liivamagi2012sdss}. They converted the spatial
position of galaxies into a luminosity density field using a kernel sum with a
smoothing scale $a$. To estimate the luminosity field they multiplied the kernel
amplitudes by the weighted galaxy luminosity, which estimates the amount of unobserved luminosity by multiplying the observed
luminosity of a galaxy and a distance-dependent weight factor. The ratio between
the luminosity density and the average luminosity density of galaxies over a
region is the normalized environmental density. Figure \ref{fig:dendeg}  shows
the plots of the galaxy degree for the network constructed with a given $r$
versus the normalized environmental density of \citet{tempel2014flux} for each
galaxy. The data were jittered to reveal overlapping data. We have chosen the
kernel with smoothing scale similar to the $r$ in question. The agreement
between them depends on the choice of $r$ and of the kernel smoothing scale used
for measuring the luminosity density \citep{tempel2014flux}. We also calculate
the Pearson correlation coefficient shown inside the plots (in dark color for
all galaxies, blue for the spirals, and red for the elliptical). The Pearson
correlation coefficient indicates a strong correlation between degree centrality
and the normalized environmental density.  For example, the better agreement
between the galaxy degree and normalized environmental density in the Super Coma
region is for $r = 1.61 \ h^{-1} $Mpc and $r = 1.99 \ h^{-1} $Mpc with $a=2 \
h^{-1}$ Mpc. The Regions 3 and 4, with connection radius $r = 3.15 \ h^{-1} $Mpc
and $r = 3.41 \ h^{-1} $Mpc, have a better match with $a=4 \ h^{-1}$ Mpc for the
kernel smoothing scale of luminosity density. We verify the relationship between
the connection radius $r$ and the kernel scale $a$ by constructing a spatial
network using $r=2 \ h^{-1}$ Mpc and $r=2.43 \ h^{-1}$ Mpc in Region 4 and
compare both with $a=2 \ h^{-1}$ Mpc. We observe the better agreement for $r$
and $a$ with the same value. It is important to mention that the catalog does
not provide the normalized environmental density for intermediary kernel
smoothing scale. 

\citet{liivamagi2012sdss} observed that the selection of better luminosity
density thresholds is related to the spatial properties of the region in the
study. For small structures, a small kernel smoothing scale is necessary to
detect the structures. We observe that the same is valid for $r$, where the
chosen $r$ reflects the size of the largest components of the network. A peak in
small $r$ in the network diameter function identifies small structures and the
$r$ for the maximum network diameter characterizes the overall structure of the
sample. The degree distribution of the network of this $r$ correlates to the
normalized environmental density with a kernel smoothing scale of similar size.

We also tested the distribution of morphological types of galaxies classified by
galaxy zoo listed in \citet{tempel2014flux}. In the marginal plots of Figure
\ref{fig:dendeg} we have the cumulative distribution for each variable and each
morphological class, as given by \citet{lintott2008galaxy}. In general, at
higher degrees, we have more  elliptical than spiral galaxies, which confirms
the density-morphology relation \citep{dressler1980galaxy}.

Thus, the network degree is then a good proxy for local density at a given scale, more
simple to calculate than the normalized environmental density, and could be used
instead of it in environmental studies. Establishing the correlation between the
well-studied normalized environmental density of the galaxy distribution and the
parameters of the network theory allows us to develop alternative tools to the
study of the large-scale structures.

\subsection{Betweeness Centrality}
\label{subsec:bet}

\begin{figure*}
\centering
 \includegraphics[width=0.4\textwidth]{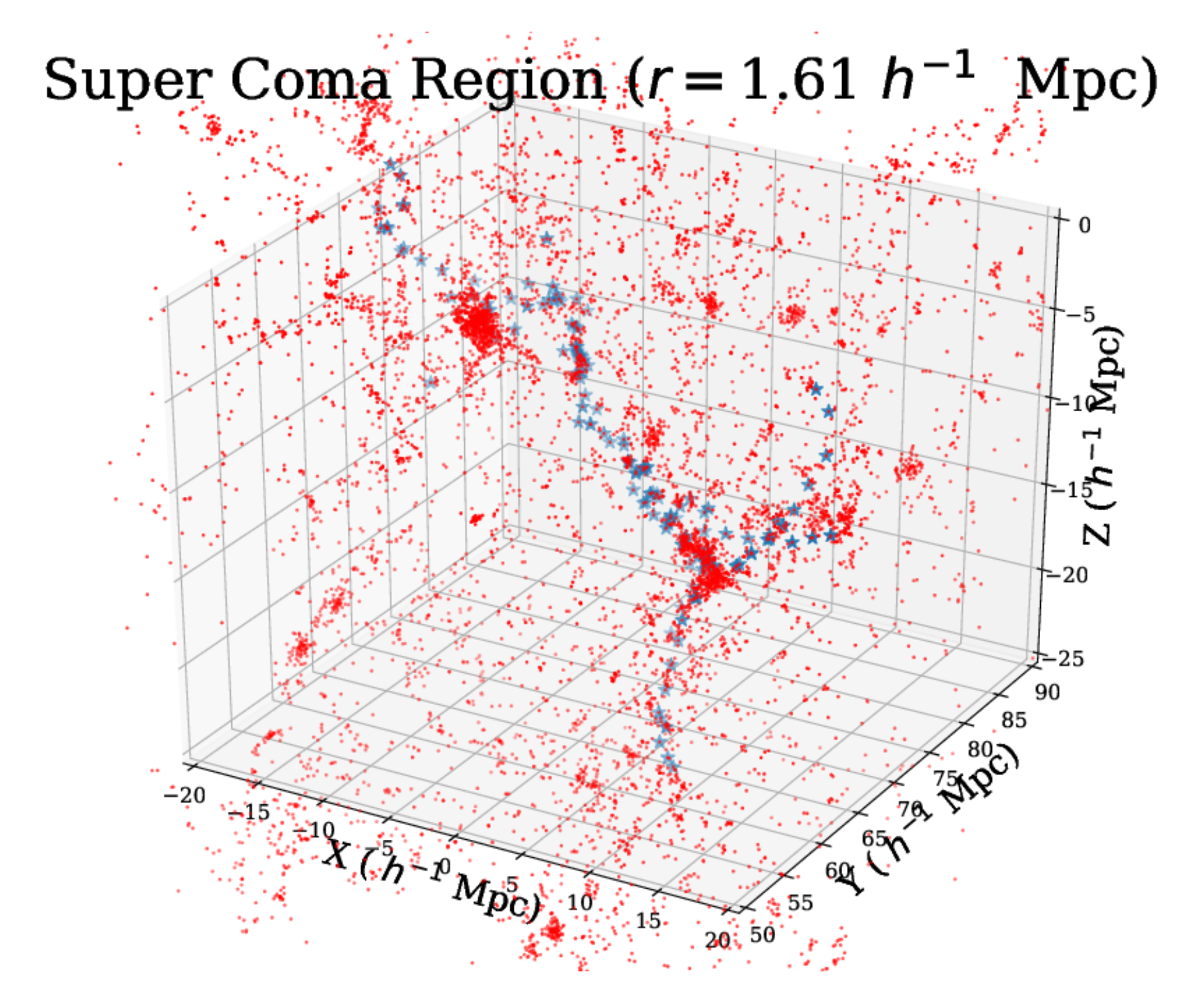}
 \includegraphics[width=0.4\textwidth]{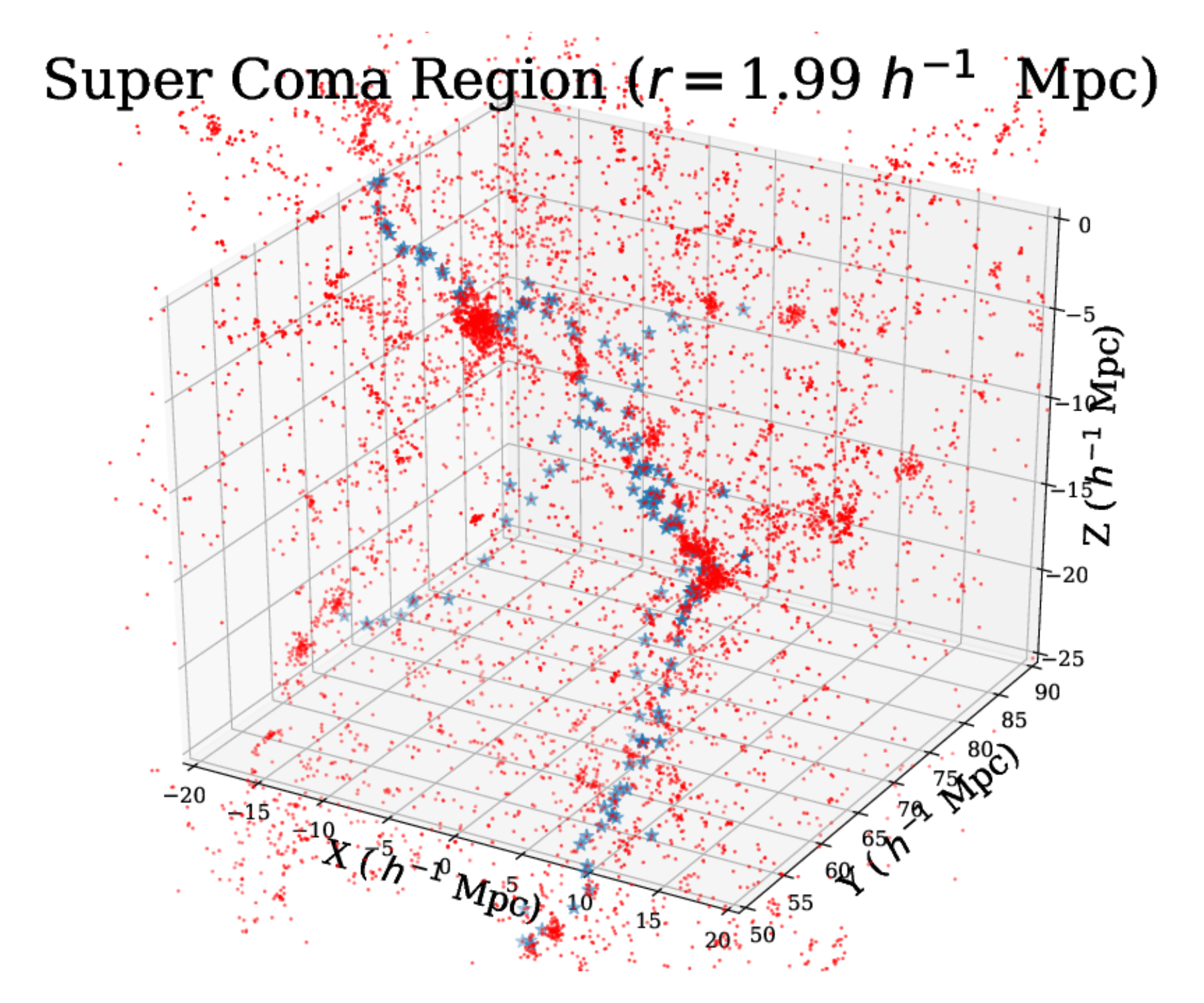}
  \includegraphics[width=0.45\textwidth]{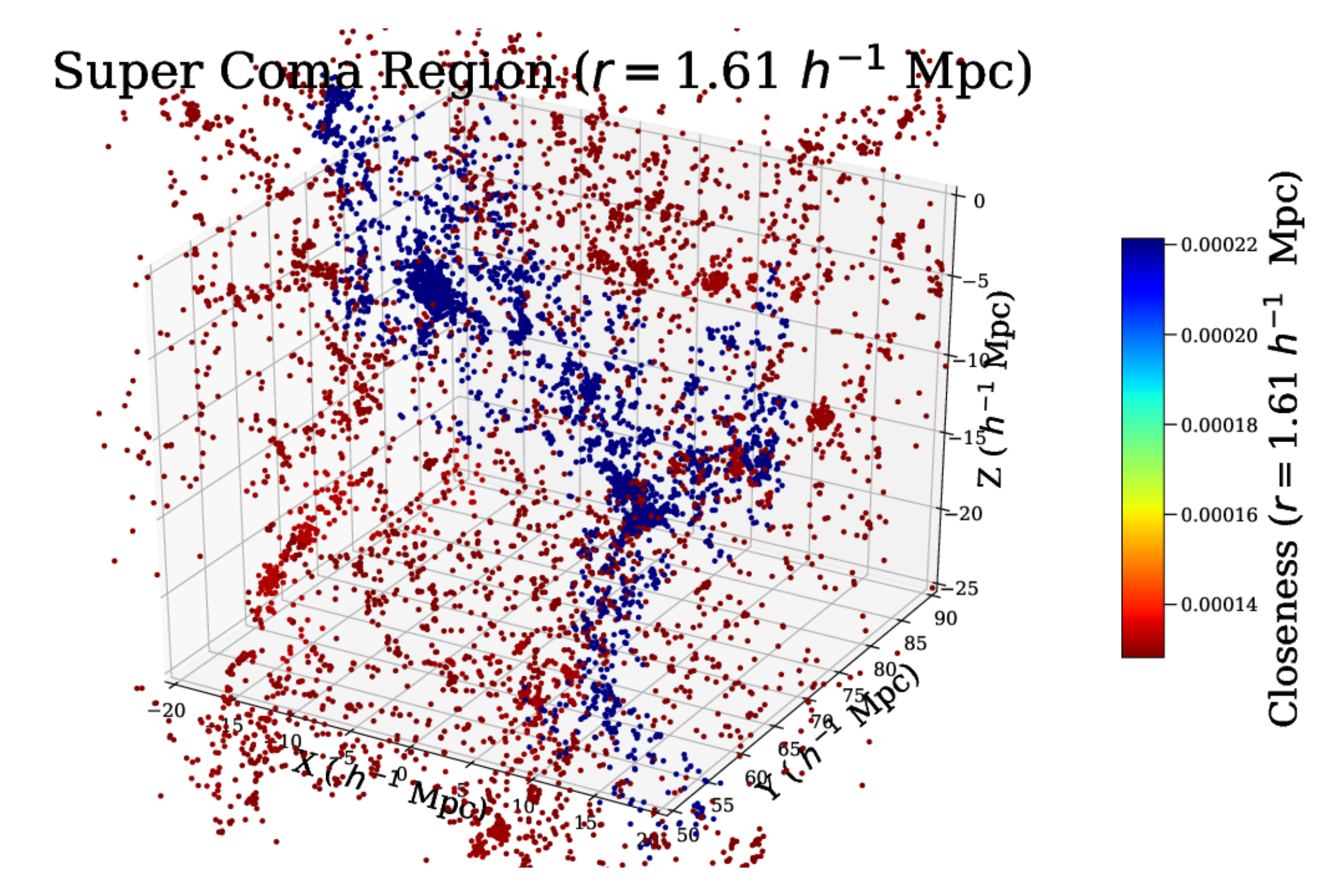}
 \includegraphics[width=0.45\textwidth]{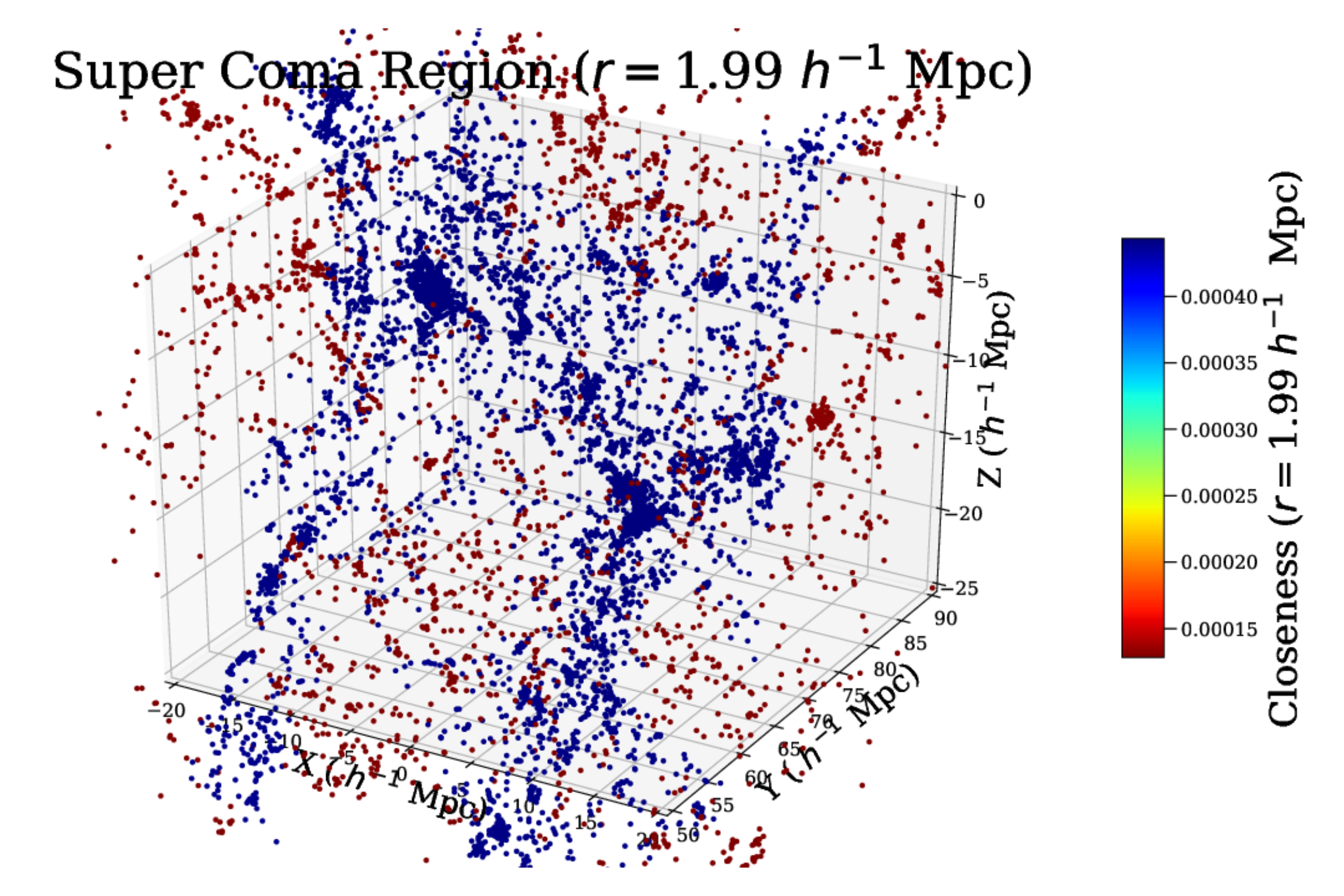}
 \caption{ The networks are constructed for the Super Coma Region with two
 values of the connection radius. Top figures: The red points are all galaxies
 in the sample and, the blue stars are the galaxies whose  BC is greater than $\langle BC \rangle \times 10.0$ for the region.
 Bottom figures: The sample with color scale corresponding to the values of the
 closeness centrality. }
 \label{fig:supercomaBCCC}
 \end{figure*}

\begin{figure*}
 \centering
 \includegraphics[width=0.4\textwidth]{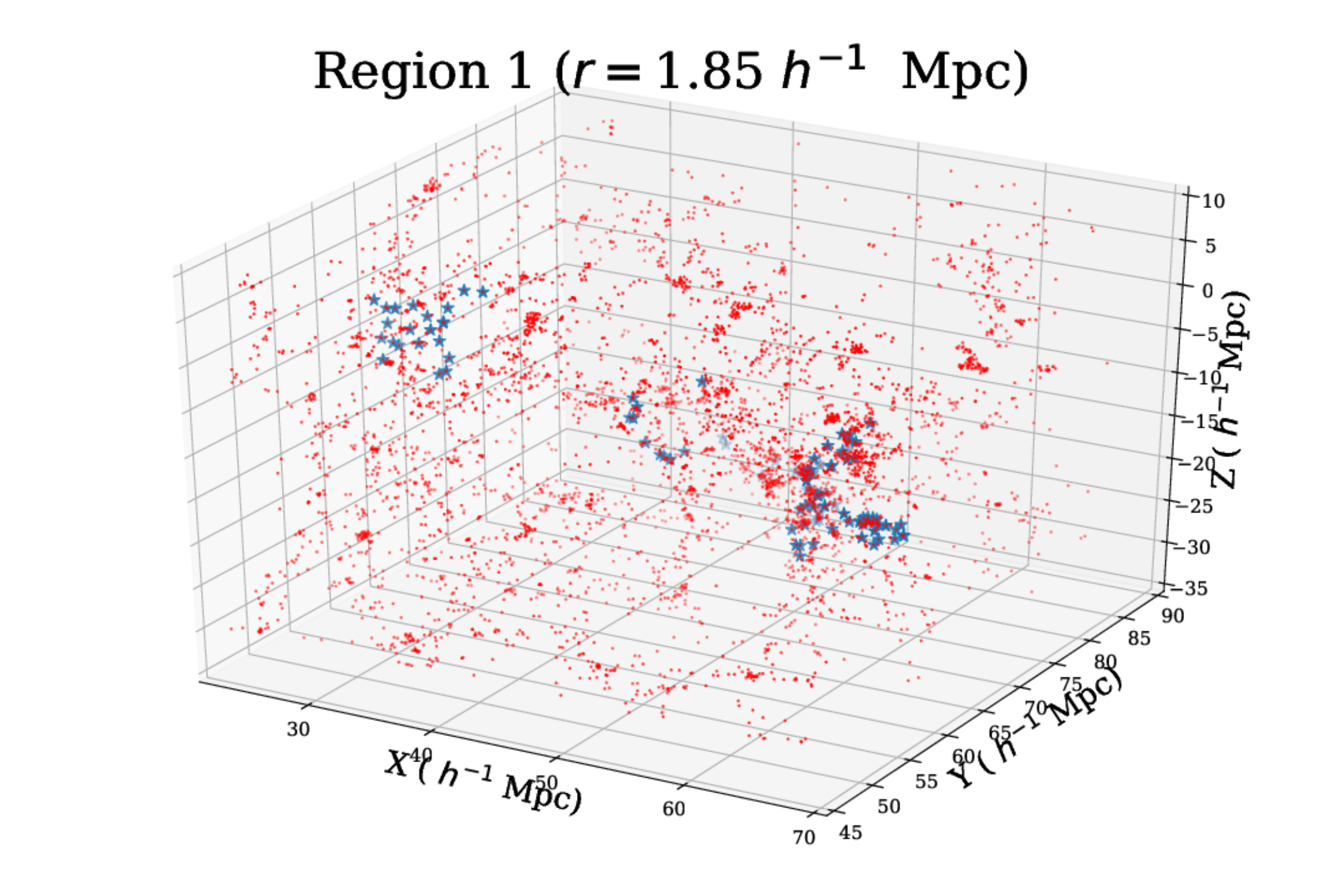}
 \includegraphics[width=0.4\textwidth]{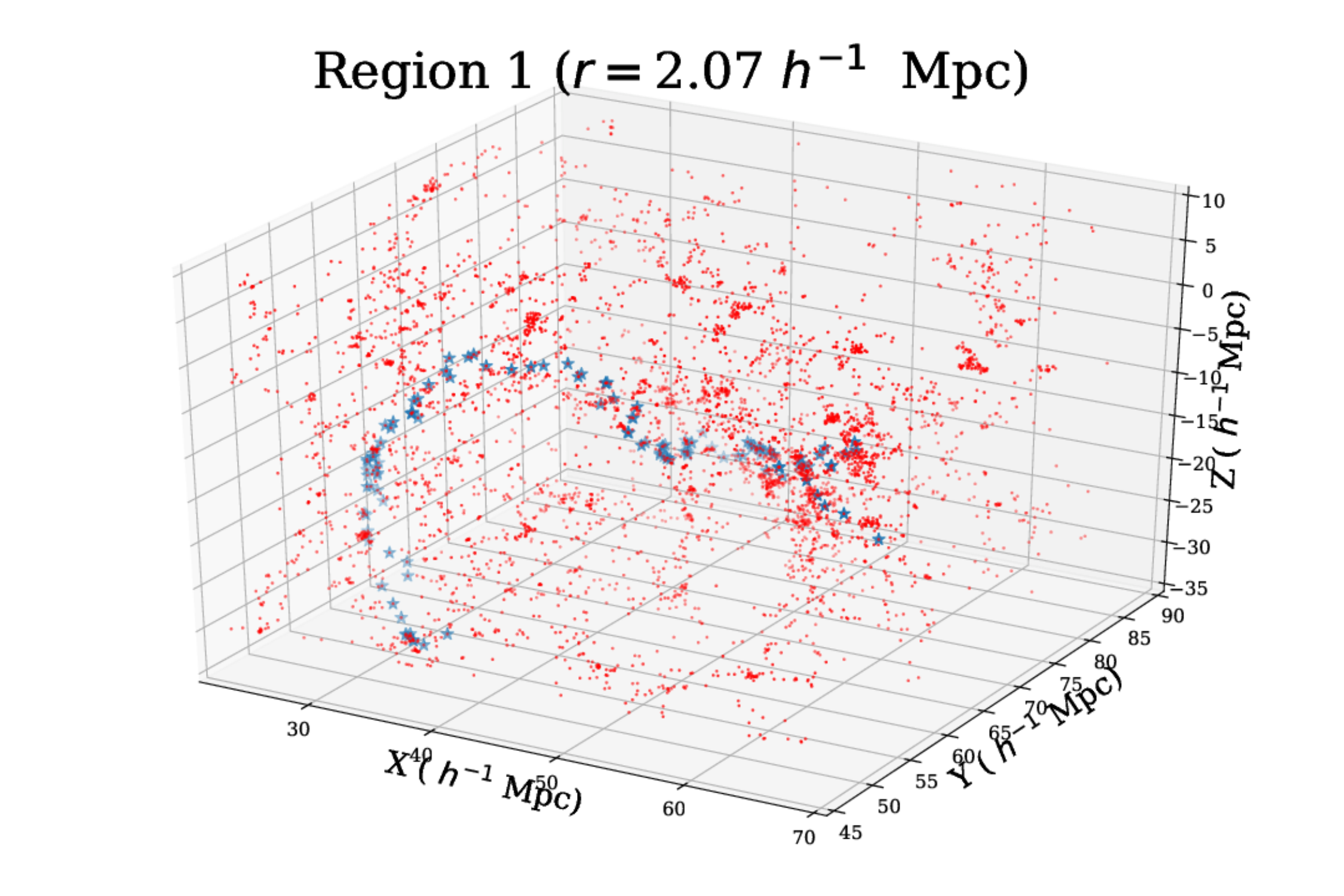}
  \includegraphics[width=0.45\textwidth]{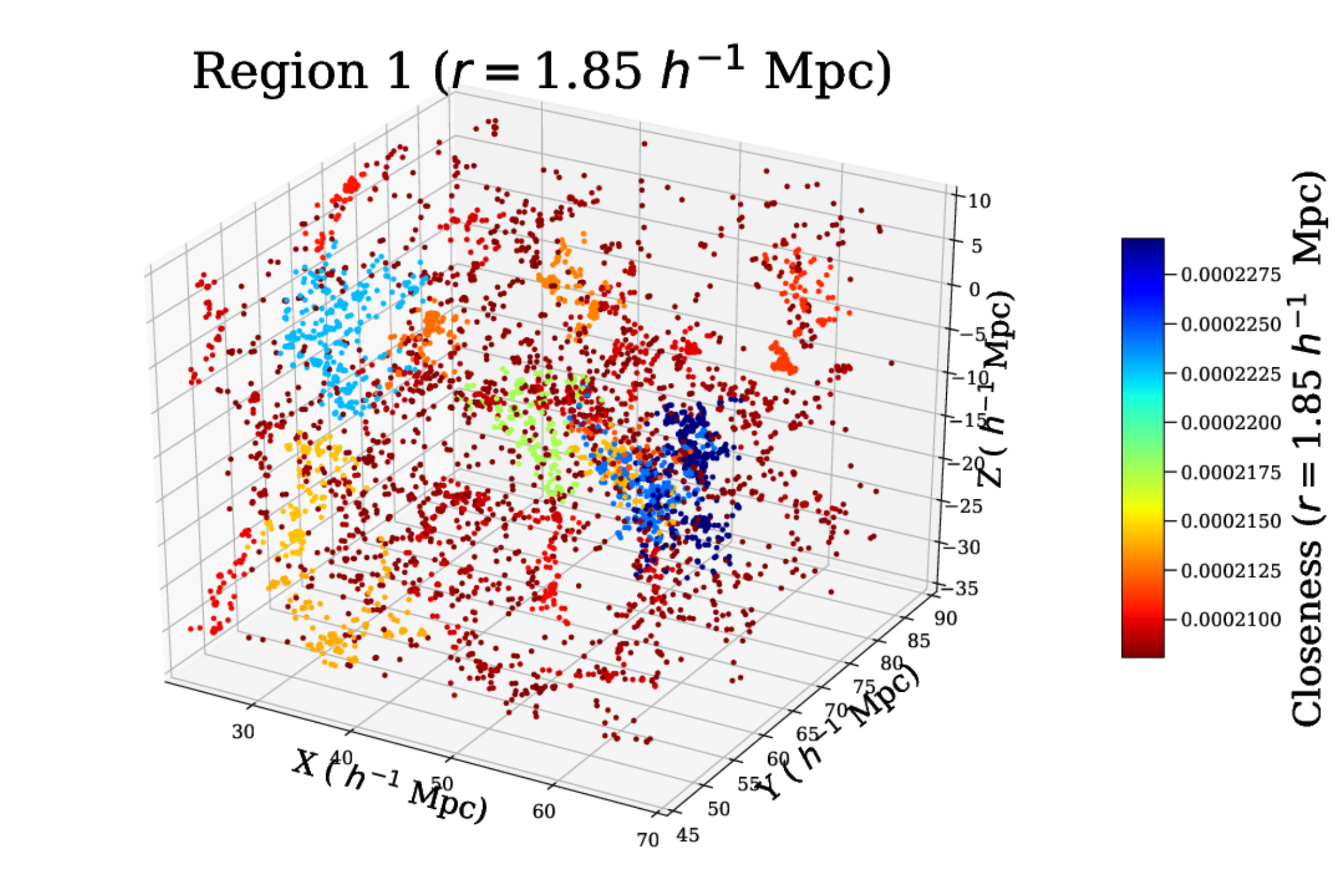}
 \includegraphics[width=0.45\textwidth]{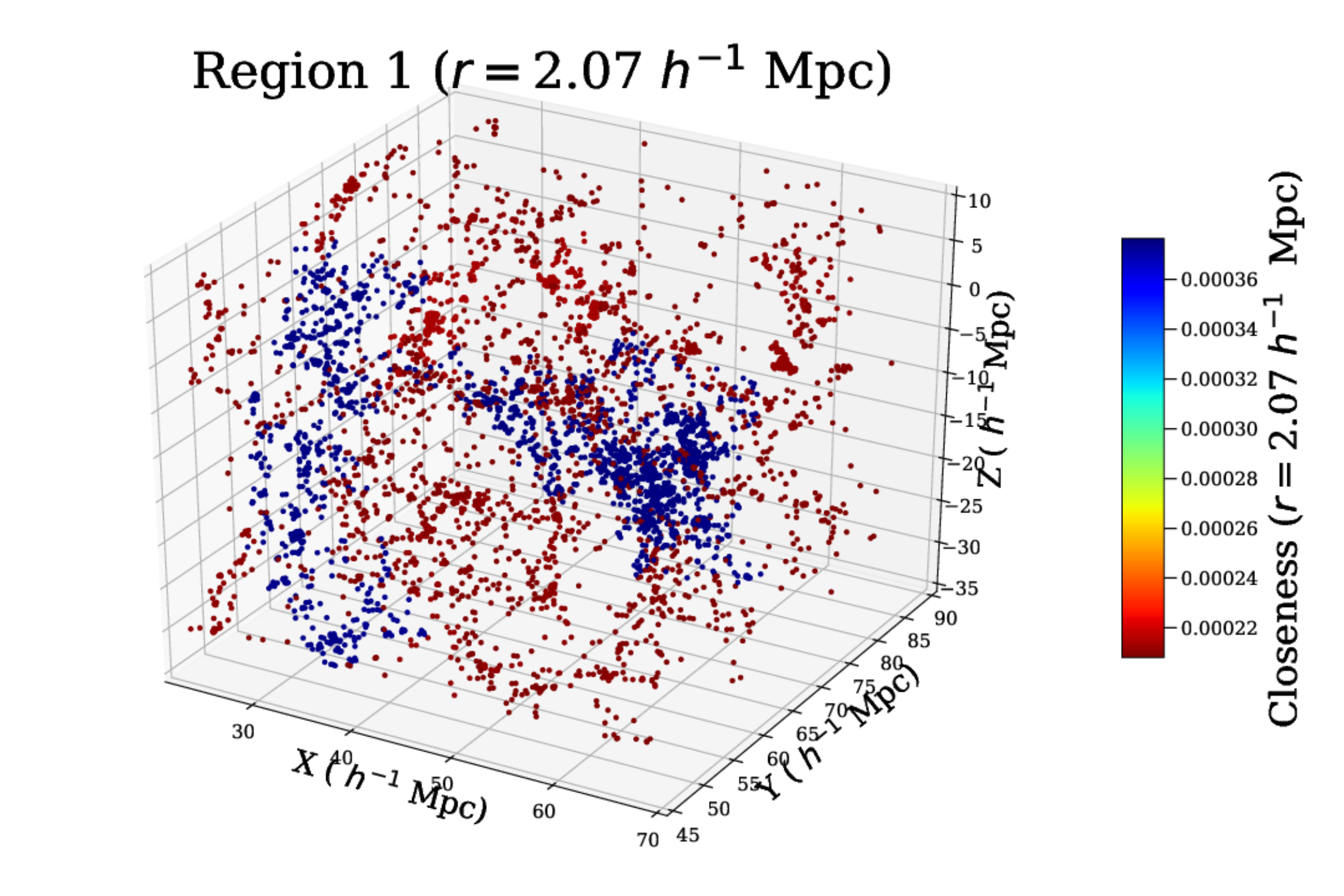}
 \caption{ The networks are constructed for the  Region 1 with two values of the
 connection radius. Top figures: The red points are all galaxies in the sample
 and, the blue stars are the galaxies whose  BC is
 greater than $\langle BC \rangle \times 10.0$ for the region. Bottom figures:
 The sample with color scale corresponding to the values of the closeness
 centrality.}
 \label{fig:R1BCCC}
 \end{figure*}

\begin{figure*}
 \centering
 \includegraphics[width=0.4\textwidth]{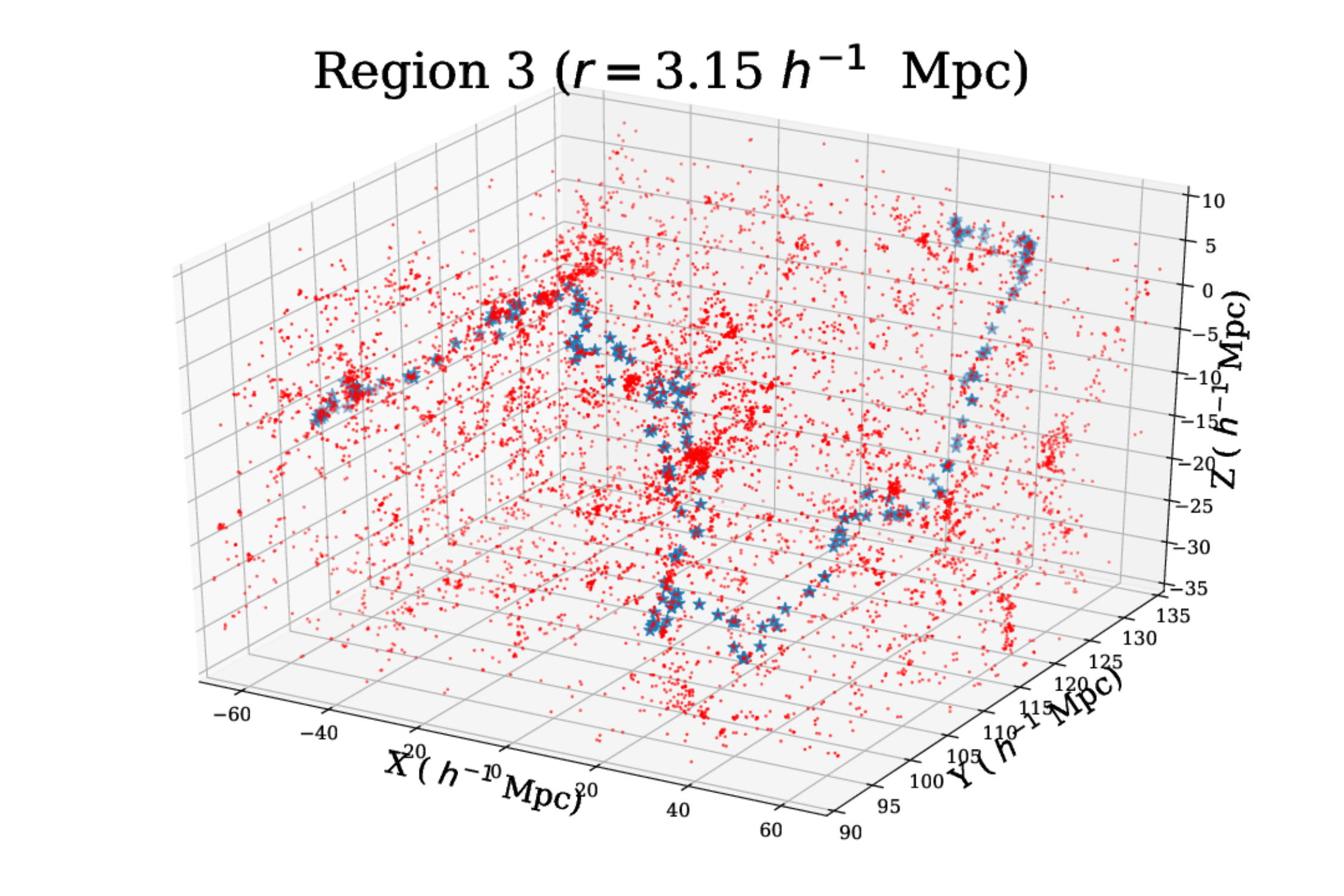}
 \includegraphics[width=0.4\textwidth]{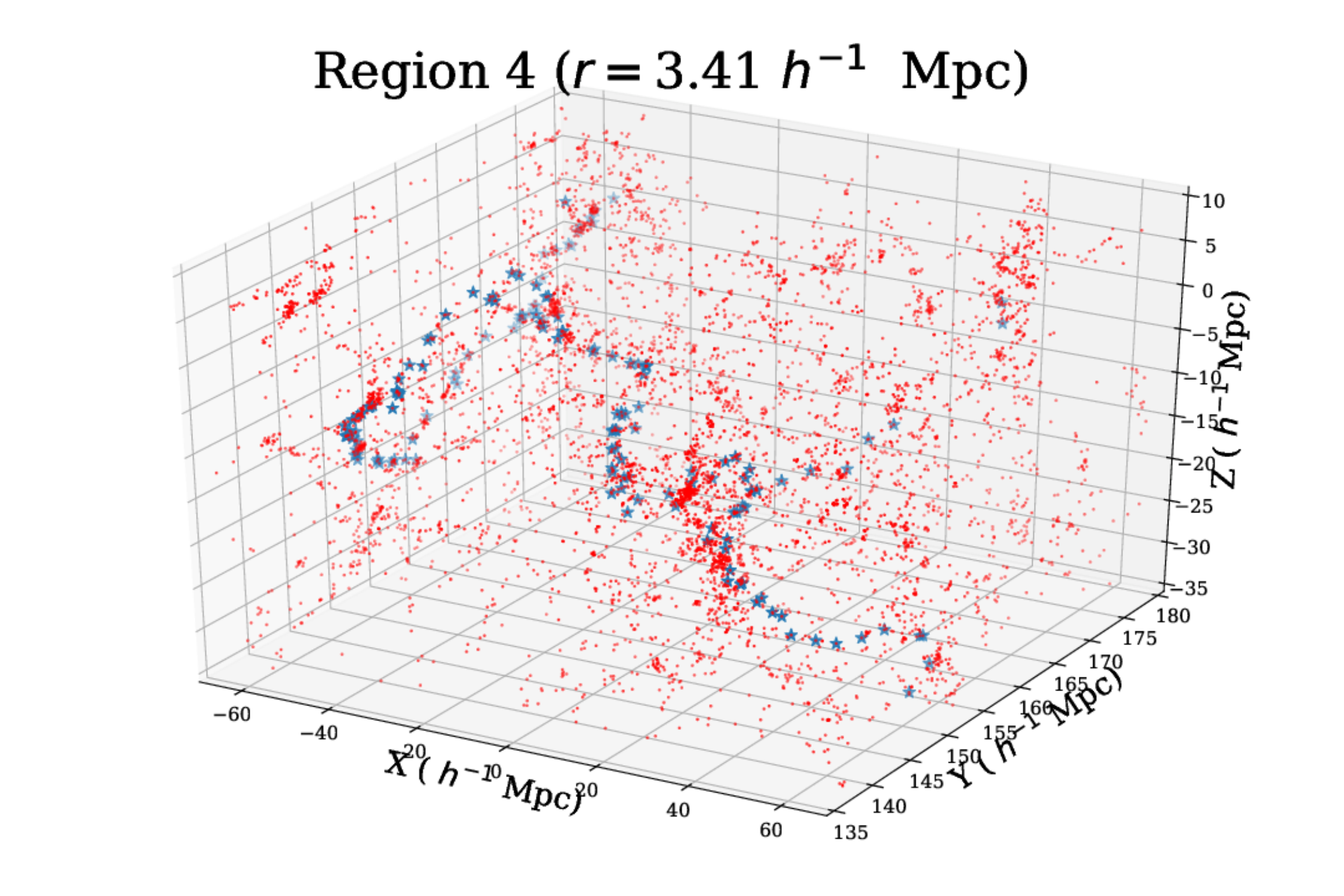}
 \caption{ Region 3 and 4 respectively. The red points are all galaxies in the
 sample and, the blue stars are the galaxies whose 
 BC is greater than $\langle BC \rangle \times 10.0$ for the region. }
 \label{fig:R3R4BC}
 \end{figure*}

The high values of betweeness centrality (BC) characterize galaxy bridges
connecting dense regions in the sample \citep{hong2015network}, tracing very
well the filamentary structures for network produced with $r$ of the maximum
diameter. The galaxies with high BC are, in most part, inside  the giant
component. The betweeness and degree are complementary centrality measures. The
galaxies with high BC connect galaxies of high
degree centrality. The structure presented in Figures  \ref{fig:supercomaBCCC},
\ref{fig:R1BCCC} (top), and \ref{fig:R3R4BC}  emphasizes the galaxies path that
connects the network. Thus we show the galaxies whose BC  measure is greater than $\langle BC \rangle \times 10.0$ for the
region. The Figure \ref{fig:centralRanregion} (middle) shows the behavior of the
BC  for a Random Region 1. 

It is important to emphasize that the $r$ that maximizes the network diameter
for the data sample can differ from the $r$ for the RGG samples with the same
mean density of galaxies (compare Table \ref{tbl1} with Table \ref{tbl4} for
Region 1 and Random Region 1).  The choice of the $r$ based on the mean density
of a random distribution distorts the global measures such as the BC. We compare the network constructed for Region 1
with $r=2.07 \ h^{-1} $Mpc and with $r=2.39 \ h^{-1} $Mpc that corresponds to
the maximum diameter for the RGG sample with the same dimensions and mean
density of Region 1. The BC  presents higher values
for the galaxy network constructed with the $r$ at a peak in the network
diameter function. In this case for $r=2.07 \ h^{-1} $Mpc we find $\langle BC
\rangle=146760.53$ and for $r=2.39 \ h^{-1} $Mpc the mean BC is $\langle BC
\rangle=22665.08$. By definition, the BC  is a
function of the number of paths that pass through a node. As a consequence, the
network constructed with $r$ that maximizes the network diameter produces just
the necessary connections between galaxies, optimizing the BC  of the nodes.  For $r=2.39 \ h^{-1} $Mpc, the network produces
unnecessary connections, growing the routes between galaxies and weakening the
BC. \citet{kirkley2018betweenness} showed that the
BC  distribution of random planar graphs presents a
bimodal regime consisting of an underlying tree structure for high BC nodes and
a low BC regime arising from the presence of loops providing local path
alternatives. We believe that similar behavior occurs in our samples. The
network constructed with the connection radii that maximize the network diameter
produces high values for BC  that correspond to an
MST structure on the largest components of the network. We intend to explore
more this centrality measure and its relation with the MST method, as developed
by \citet{alpaslan2014galaxy}, and the Bisous model, developed by
\citet{tempel2014detecting}, in a next study.

The appropriate choice of the $r$ to represent the large-scale structure of the
galaxy distribution allows us to use a simple undirected and unweighted network
to trace the filamentary structures. The choice of the $r$ based on the mean
density of a random distribution requires to adjust the construction of the
network with the Euclidean distance as a weight of the edges connecting the
nodes, as  in \citet{hong2015network}. 

 In networks constructed with $r$ from  the intermediary peaks on network diameter
 function, we observe that  in some cases, the galaxies with high BC
 belong to the border of regions with high degree
 in the largest connected components of the network, as we see in Region 1 with
 $r=1.85 \ h^{-1} $Mpc. In this case, the high values of BC  only connect regions of high density and do not trace filamentary
 regions as it does in the Super Coma region with $r=1.61 \ h^{-1} $Mpc.

\begin{figure*}
\centering
\includegraphics[width=0.35\textwidth]{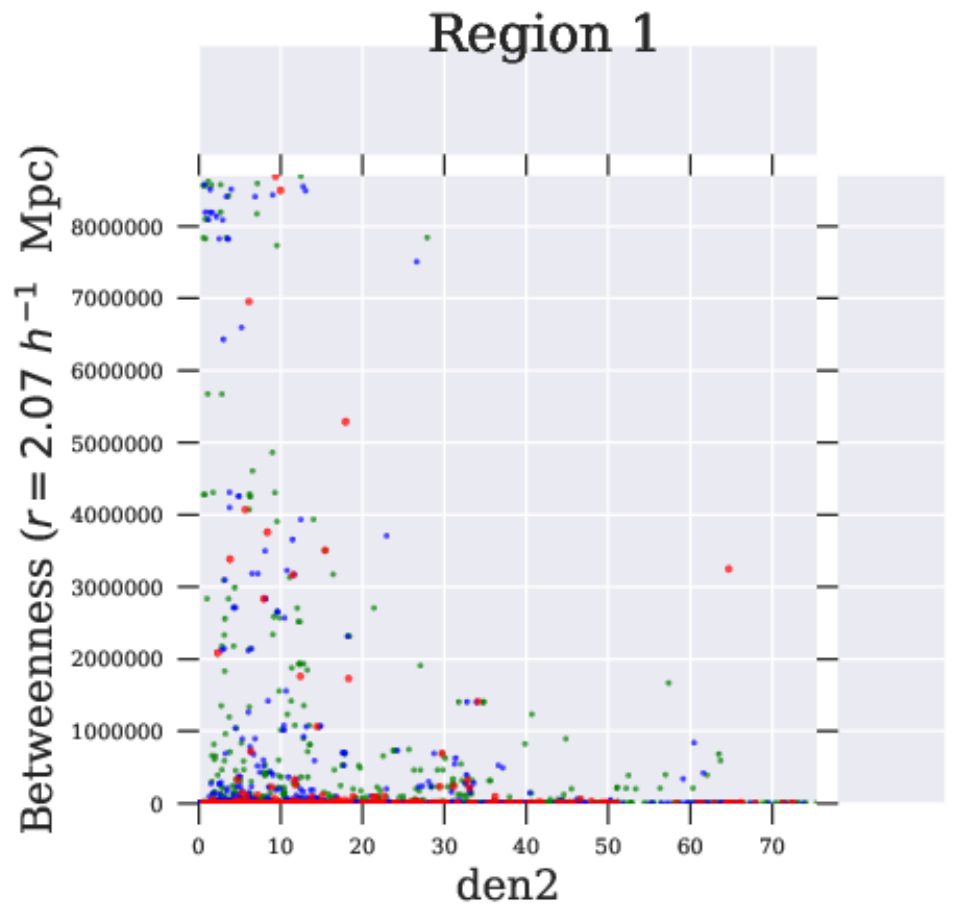}
\includegraphics[width=0.32\textwidth]{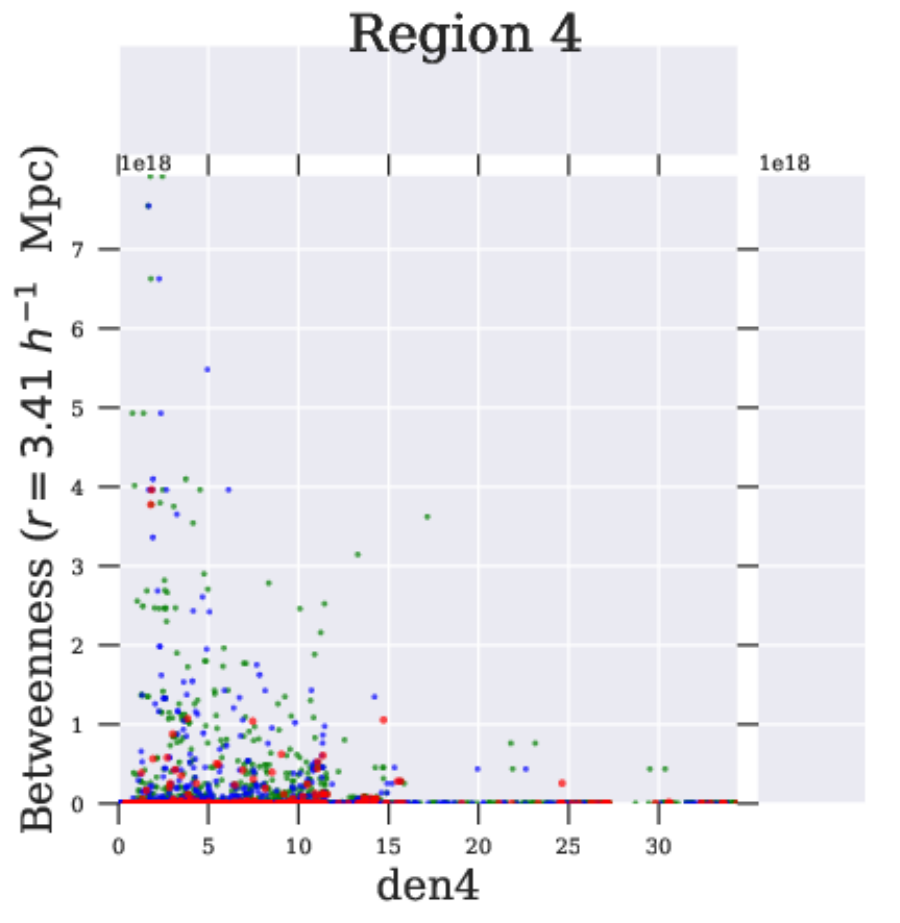}
\caption{ Relationship between luminosity density field (for a smoothing scale
$a$) of Ref.~\citenum{tempel2014flux} and betweenness centrality (for a $r$
corresponding to the maximum network diameter) for each galaxy on Region 1 and 4
respectively. The  kernel scale $a=2 \ h^{-1}$ Mpc  for den$2$; galaxies are
identified by blue points (spiral), red (elliptical) and green (uncertain)
according to  \citet{lintott2008galaxy}.}
\label{fig:denbet}
\end{figure*}

Figure \ref{fig:denbet}, similar to Figure \ref{fig:dendeg},  shows the plots of
BC  constructed with a given $r$ versus the
normalized environmental density: the galaxies with high BC  belong to regions with low luminosity density.

\subsection{Closeness Centrality}
\label{subsec:clos}

Figures  \ref{fig:supercomaBCCC} and \ref{fig:R1BCCC} show the closeness
centrality in a range of colors for two values of $r$. The distribution of the
closeness centrality is highly bimodal for the network constructed with the $r$
corresponding to the maximum diameter, in agreement with \citet{hong2015network}
results. This measure divides the sample according to the galaxies proximity:
galaxies with the highest closeness belongs to the giant component of the
network (a large connected component represented by the blue points structure on
Figure \ref{fig:supercomaBCCC} (bottom) associated to the Coma supercluster).
The Coma supercluster extracted by \citet{liivamagi2012sdss} is similar to ours.
However, the closeness centrality for all RGG samples also presents the bimodal
behavior at the maximum network diameter, suggesting that it is an expected
characteristic  in spatial distributions at the emergence of the giant component
(see Figure \ref{fig:centralRanregion} (bottom) for closeness distribution).
This construction composes a network with one large connected component (the
giant component) formed with a significant number of the galaxies of the sample
and with small groups of galaxies (small connected components) or isolated
galaxies. This connection radius is the most adequate to study the large-scale
structure of the sample.

\begin{figure*}
\centering
\includegraphics[width=0.4\textwidth]{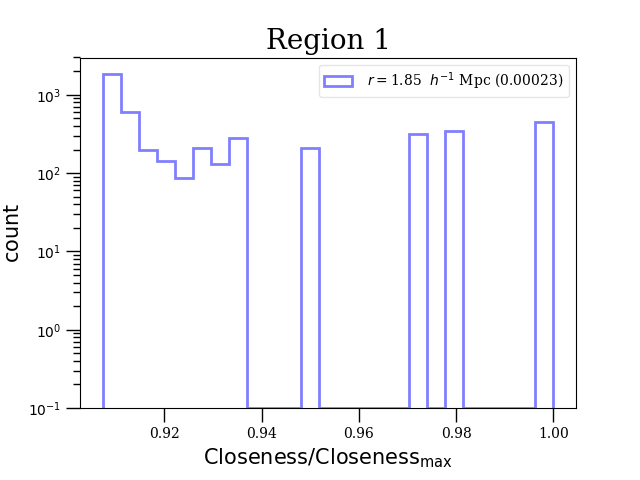}
\includegraphics[width=0.4\textwidth]{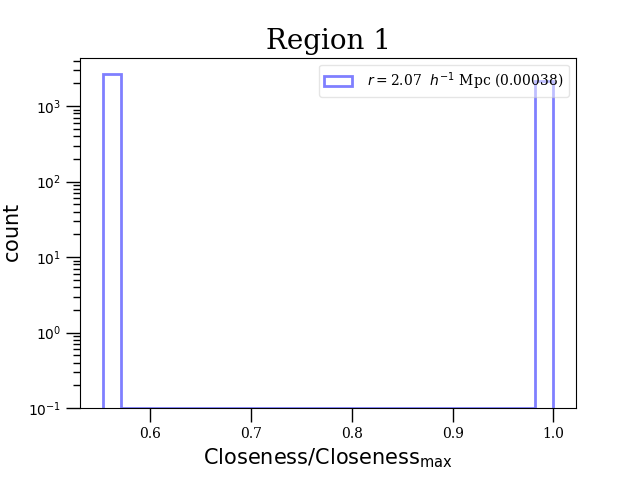}
\caption{Closeness centrality distribution for Region 1 for two connection radius indicated in the legend.}
\label{fig:R1histCC}
\end{figure*}

\begin{figure*}
\centering
\includegraphics[width=0.6\textwidth]{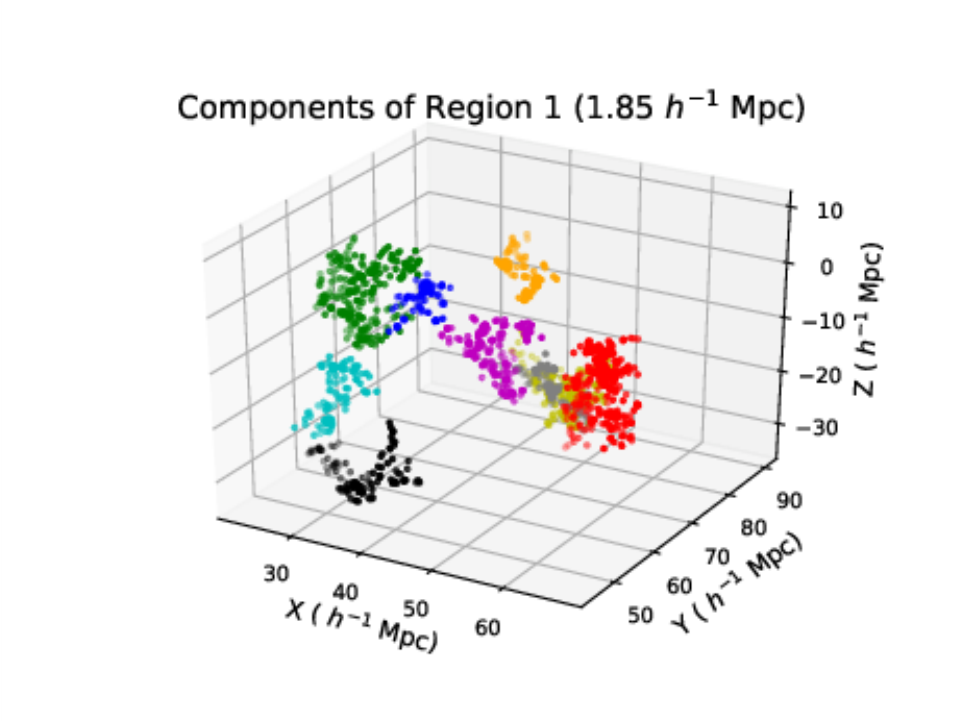}
\caption{The nine largest connected components formed in the Region 1 with
$r=1.85  \ h^{-1}$ Mpc and at least 100 galaxies by component. The giant
component (red points) has 447 galaxies. The yellow component has 350 galaxies,
the green and the magenta have 320 and 211 galaxies respectively.}
\label{fig:components}
\end{figure*}

At the small connection radii indicated by the peaks in the network diameter
function, the closeness centrality distribution does not
present bimodal behavior as a rule (see Figure \ref{fig:R1histCC}). The
closeness centrality distribution reflects the distinct connected components
formed in the network with the respective connection radius, and the galaxies
with the highest closeness centrality belong to the largest connected
components. The $r$ corresponding to the first peak on the network diameter are
associated with the largest connected components to clusters of galaxies and the
ones formed with the $r$ corresponding to the second peak to superclusters of
galaxies. For example, see Figures \ref{fig:R1BCCC} and \ref{fig:R1histCC} for
Region 1 with $r=1.85  \ h^{-1}$ Mpc. The color scale represents the values of
the closeness centrality, where the galaxies with higher closeness are in blue.
In Figure \ref{fig:components},  we show the nine largest connected components
compounded by at least 100 galaxies for Region 1 and $r=1.85  \ h^{-1}$ Mpc.
Between them, we have the giant component with 470 galaxies and three components
with 350, 320 and 211 galaxies, respectively.  These four components are exactly
the higher values of the closeness centrality distribution shown in Figure
\ref{fig:R1histCC}. The components with high closeness centrality, associated to
superclusters regions, also present the galaxies with the highest degree and
 BC. As a product of this association, we can
produce a catalog of superclusters only by extract the largest connected
components of the network. In Appendix \ref{sec:appendix2}, we show the largest
components for the Region 3 and 4 and the discussion about the size relevance. 

\subsection{Clustering Coefficient}
\label{subsec:clustcoeff}

In our analyses, the transitivity presents higher values for data samples than
RGG samples, indicating high clustering in these data samples
(compare Table \ref{tbl1} with Table \ref{tbl4} and Figure \ref{fgr:transiti}).
The transitivity at different connection radii also presents different behaviors
between the data, the Segment Cox Process and RGG samples. In
data samples, the transitivity grows with $r$ to a maximum and then stabilizes
at  $\sim 0.8$, after that slowly decreases (see Section \ref{sec:Meandeg} for discution).  For the RGG samples, the
transitivity is very stable, remaining at the value of $\sim 0.5$ as $r$ grows.

While transitivity is a global measure, the local clustering coefficient
measures the connectivity between neighbors of a node. Following the definition
in Eq. \ref{eq:clucoef}, the local clustering coefficient is not defined for an
isolated node ($k_{i} = 0 , 1$). Studies suggest that $C$ is higher on real
networks than for random networks with similar dimensions \citep{ravasz2003hierarchical}. We also observe this behavior in the comparison
between the data samples and RGG samples. The high clustering coefficient in
real networks is related to the modular organization in these networks.

\begin{figure*}
\centering
\includegraphics[width=0.4\textwidth]{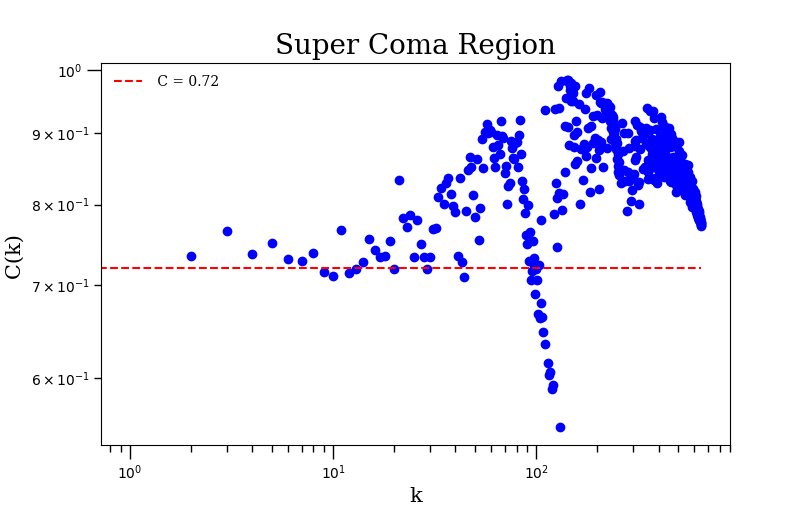}
\includegraphics[width=0.4\textwidth]{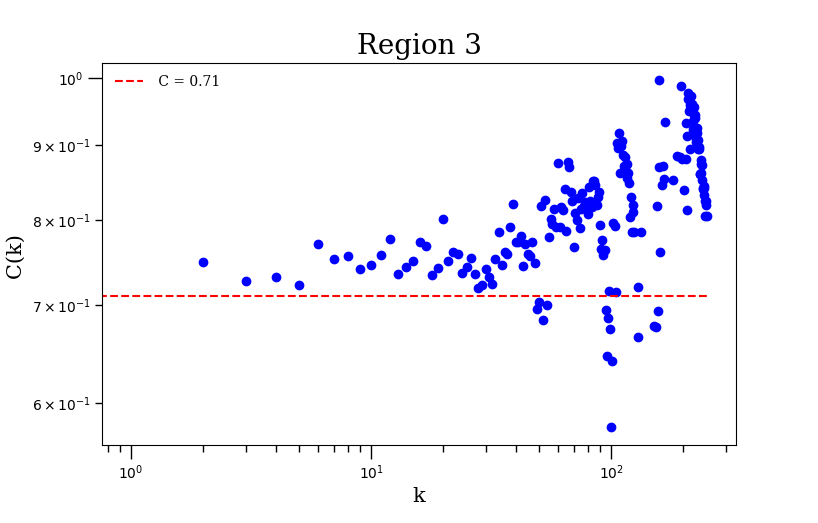}
\caption{The mean local clustering coefficient for each degree as a function of the degree in blue points and the mean local clustering coefficient of the sample in red line. }
\label{fig:clust}
\end{figure*}

\citet{ravasz2003hierarchical} simulated a hierarchical network and showed that
the mean local clustering coefficient of a node with $k$ links follows the
scaling law $C(k) \sim k^{-1}$. They argue that this scaling law quantifies the
coexistence of a hierarchy of nodes with different degrees of clustering.
However, this scaling law is absent from geographically organized networks. 

For our samples, the mean local clustering coefficient  as a
function of the degree, shown in Figure \ref{fig:clust},  is independent of the
degree of the node. In spatial networks constrained by distance, as is the
galaxy network, this is expected because nodes with a high degree belong to
densely interconnected clusters of galaxies. Our results are in agreement with
the real spatial networks presented by \citet{ravasz2003hierarchical}. However,
it is not clear if the evolution of the mean local clustering coefficient as a
function of the degree can represent the hierarchical organization in a spatial
network. This question needs more investigation, not only for the spatial galaxy
network but also for spatial networks in general. 

\section{Conclusions}
\label{sec:conclusions}

In this work, we have analyzed the distribution of galaxies in several regions
from the catalog of \citet{tempel2014flux} using the tools from network theory, with the aim of  improving our understanding of the matter distribution in clusters. We have attained the following conclusions: 

\begin{itemize}
    \item[--]The value of the connection radius $r$ at the maximum network
    diameter is an optimal parameter to capture the large-scale structure of the
    network, occurring  at the percolation phase and the
    emergence of the giant component. 
    The choice of the connection radius based
    on the mean density of random networks can produce unnecessary connections
    distorting some measures. The connection radius chosen by this method varies
    with the network in the study, thus minimizing selection effects caused by
    flux-limited inhomogeneities as in \citet{tago2008groups} and
    \citet{tago2010groups}.  
   Although the $r$ at the maximum network diameter shows better results to
   analyze the large-scale structure of the network, the network diameter
   function indicates relevant structures at smaller $r$, useful for cluster and
   supercluster analyses. As a product of the analyses of smaller $r$, we have a
   list of connected components of the network that can be associated with
   supercluster regions. The advantage of choosing $r$ by its relation with the
   network diameter is letting the sample free to show its relevant scales 
   (see, e.g., Figure \ref{fig:diam}), and consequently, gives a more accurate
   description of the relevant connection formed on network sample.

    \item[--] The degree centrality of each object is tightly related to the
   luminosity density obtained by \citet{tempel2014flux}. Our analysis indicates
   that the same order of magnitude of the connection radius and the kernel
   smoothing scale of the density estimation improves the correlation between
   degree centrality and normalized environmental density (see Figure
   \ref{fig:dendeg}). The degree centrality is simpler to calculate than the
   luminosity density, and it enables us to implement a range of complementary
   network measures providing more information on the galaxy distribution. We
   suggest the use of the degree centrality as a measure in substitution of
   normalized environmental density. In addition, it is possible
   to derive the correlation dimension with the evolution of the mean degree as
   a function of $r$, connecting network theory and  traditional clustering
   metrics.

    \item[--] The galaxies of high closeness centrality belong to the largest
    components of the network and, according to the $r$ pointed in the network
    diameter function, can be associated with supercluster regions (Figure
    \ref{fig:components}). 
     
    \item[--] The galaxies with high  BC connect
    galaxies of high degree centrality and thus can be used to map the spine of
    the spatial distribution (Figure \ref{fig:supercomaBCCC}). The
     BC is a measure more susceptible to the choice
    of the connection radius since its mean values is maximum at the
    maximum of the network diameter as a function of $r$. 
    
    \item[--]  The transitivity and the mean local clustering coefficient
    present higher values for the spatial network of galaxies than for RGG
    samples, indicating a clustering
    organization at the data samples. However, the spatial network of galaxies
    does not follow the scaling law for $C(k) \sim k^{-1}$  that is
    characteristic of a hierarchical organization. This behavior occurs in some
    real networks, but it is also absent in other spatial networks,
    therefore the analyze of $C(k) \sim k^{-1}$ is inconclusive
    for the galaxy distribution.

\end{itemize}

We establish a comparison between a real three-dimensional galaxy network (data
sample) and a RGG sample. The degree and betweeness distributions present
higher mean values for the galaxy network than for RGG samples using the same
criteria for the choice of the connection radius. The closeness distribution
does not show a significant difference. Our study specifies a method
 for the  choice of the connection radius for  galaxy
distribution analysis. We find a connection radius statistically relevant to
each structure of the samples in the study, and the centrality measures for
the spatial network constructed reflect this choice. We do not analyze all
possible uses of each centrality measure. However, we highlight  their
differences, how they are related, and main uses on each scale.  
\appendix{}

\section{Random and Segment Cox processes}
\begin{figure*}[h]
\centering
 \includegraphics[width=0.3\textwidth]{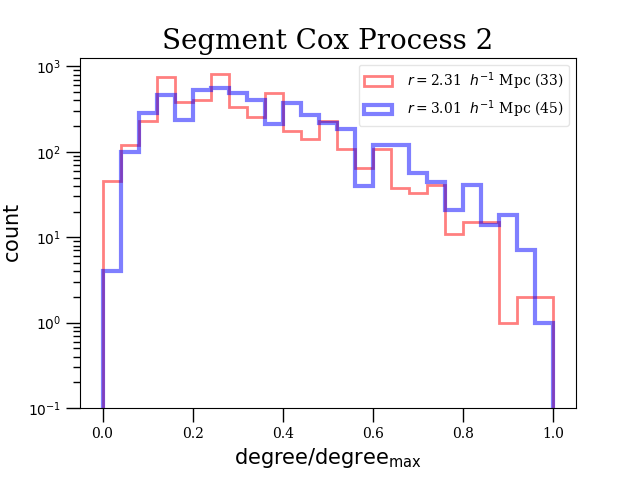}
 \includegraphics[width=0.3\textwidth]{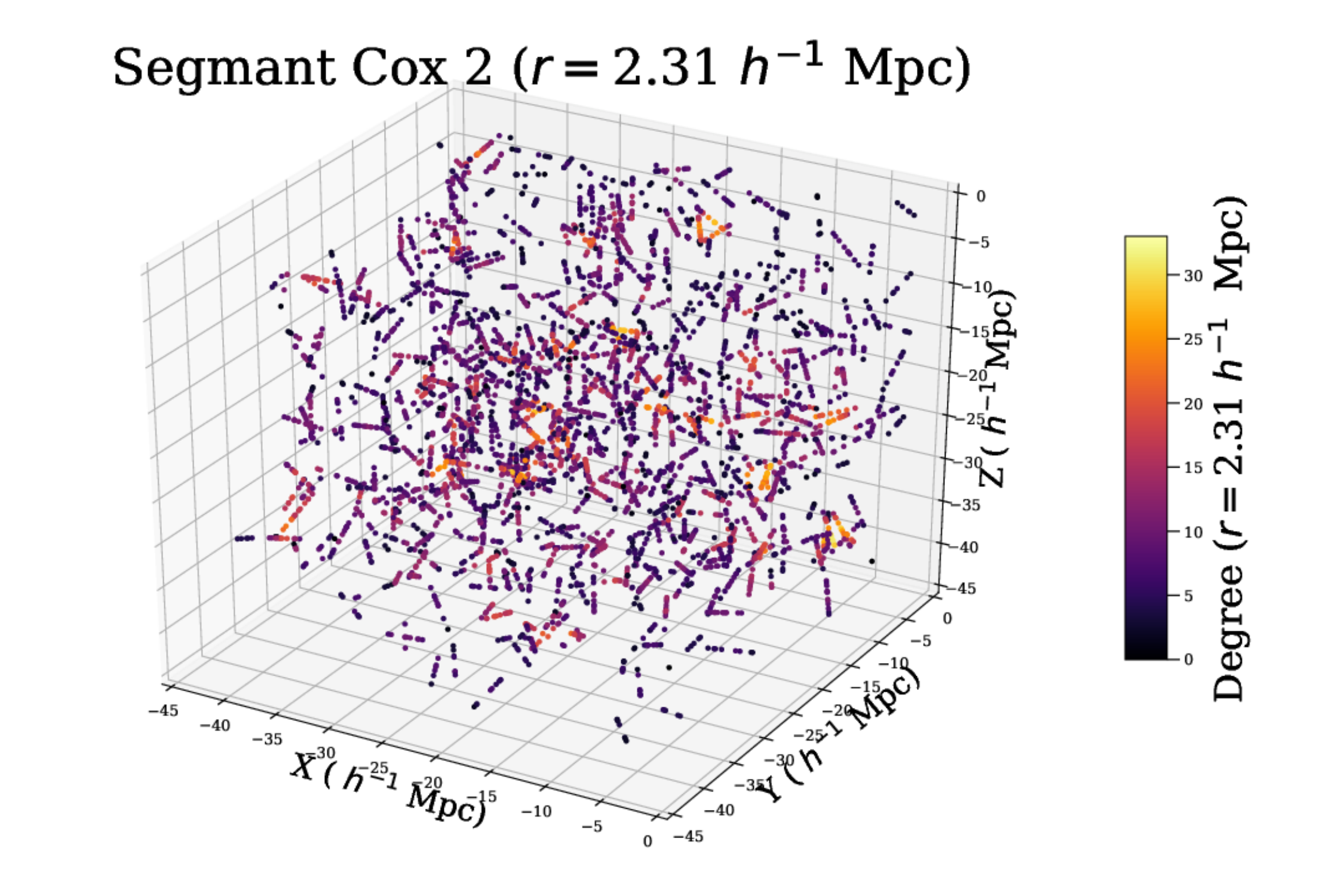}
 \includegraphics[width=0.3\textwidth]{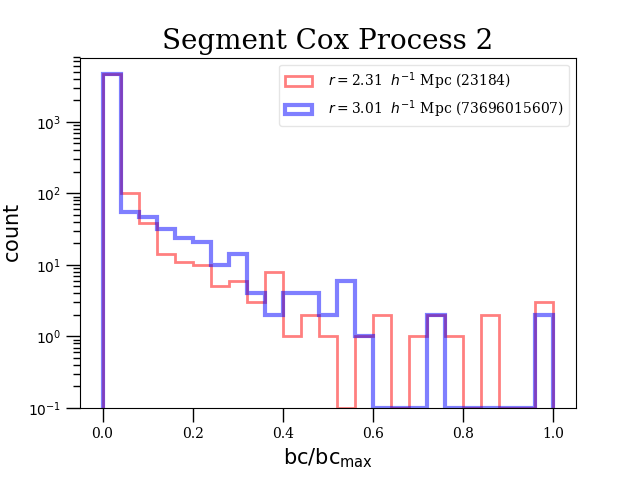}
 \includegraphics[width=0.3\textwidth]{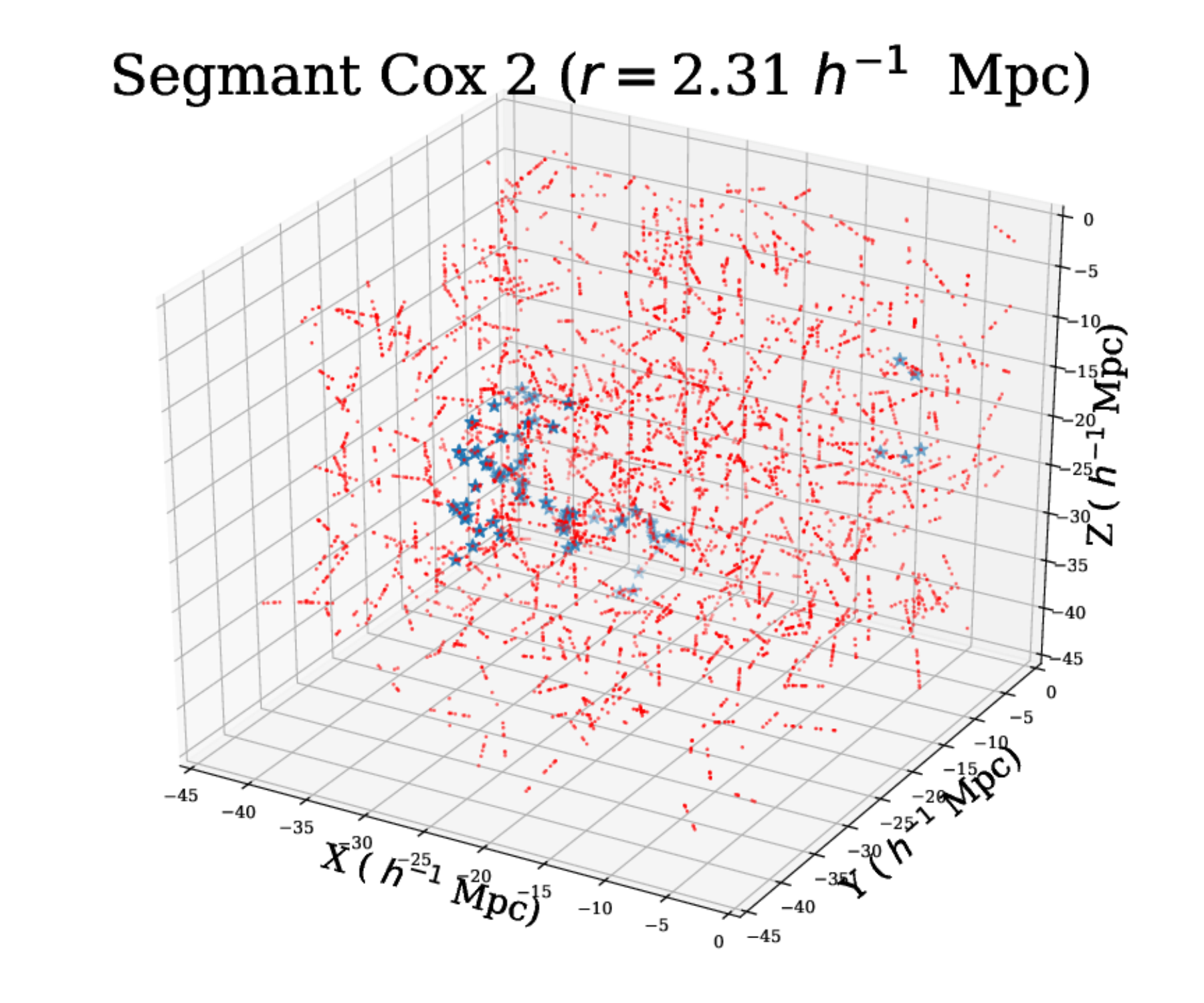}
 \includegraphics[width=0.3\textwidth]{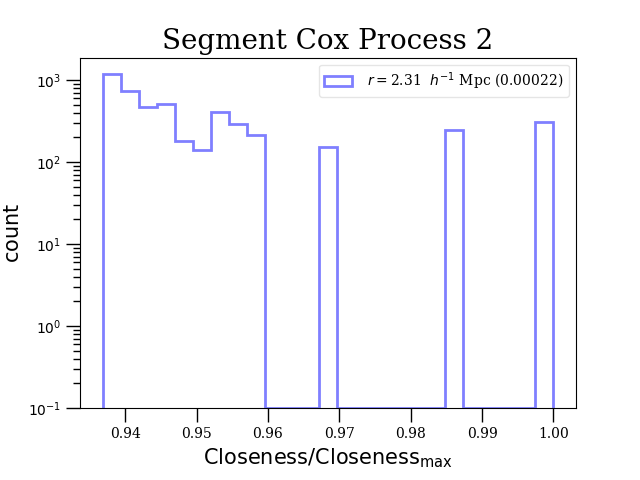}
 \includegraphics[width=0.3\textwidth]{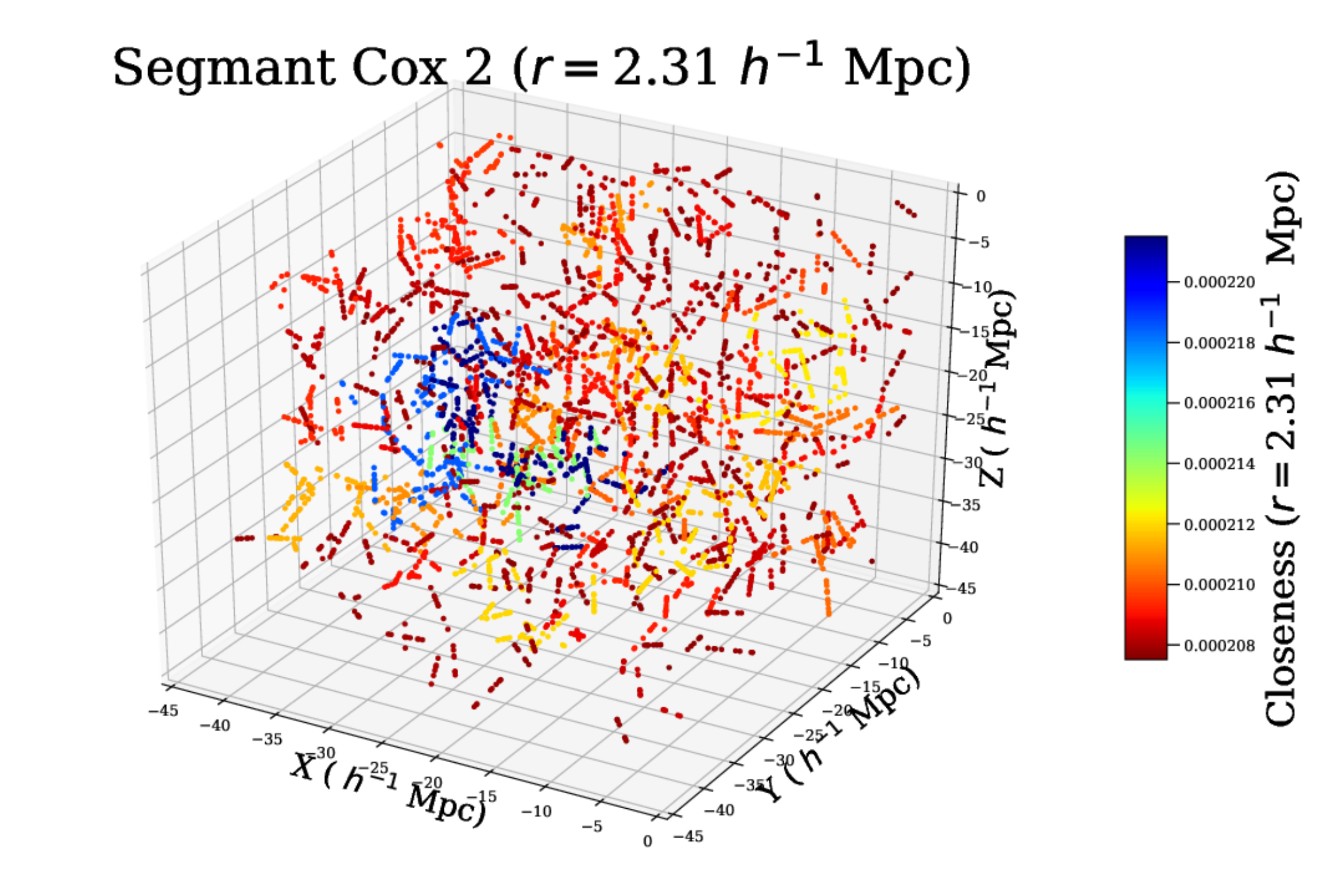}
\caption{Segment Cox process 2 distribution with the same dimensions and points (galaxies) of the Region 1. In the left panel: the centrality distributions of Degree, Betweenness and Closeness, respectively. In the right panel: the respective three-dimensional distribution of points with high values of the centrality measures in color scale for Degree and Closeness and blue star to Betweenness centrality greater than $\langle BC \rangle \times 10.0$ for the region. }
\label{fig:Cox2distrib}
\end{figure*}

\begin{figure}
\centering
 \includegraphics[width=0.35\textwidth]{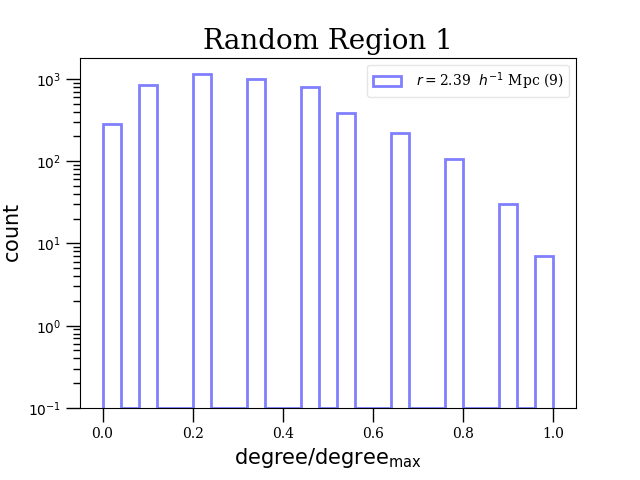}
 \includegraphics[width=0.35\textwidth]{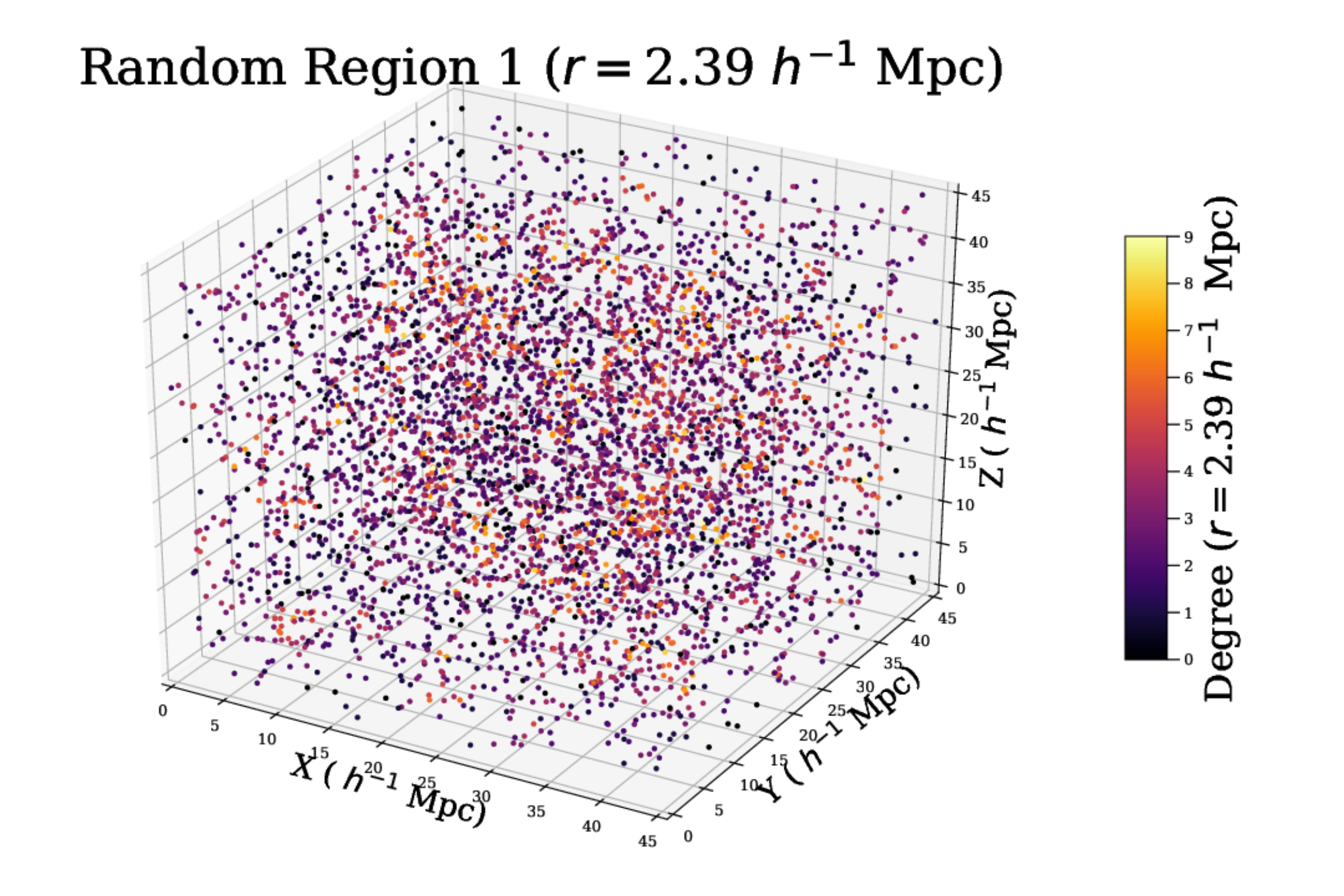}
 \includegraphics[width=0.35\textwidth]{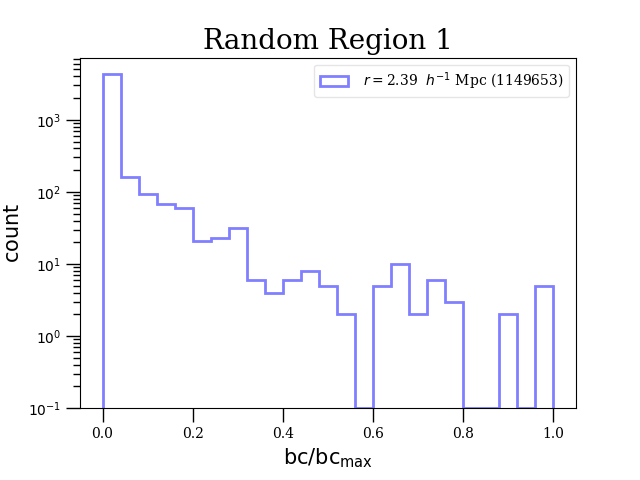}
 \includegraphics[width=0.3\textwidth]{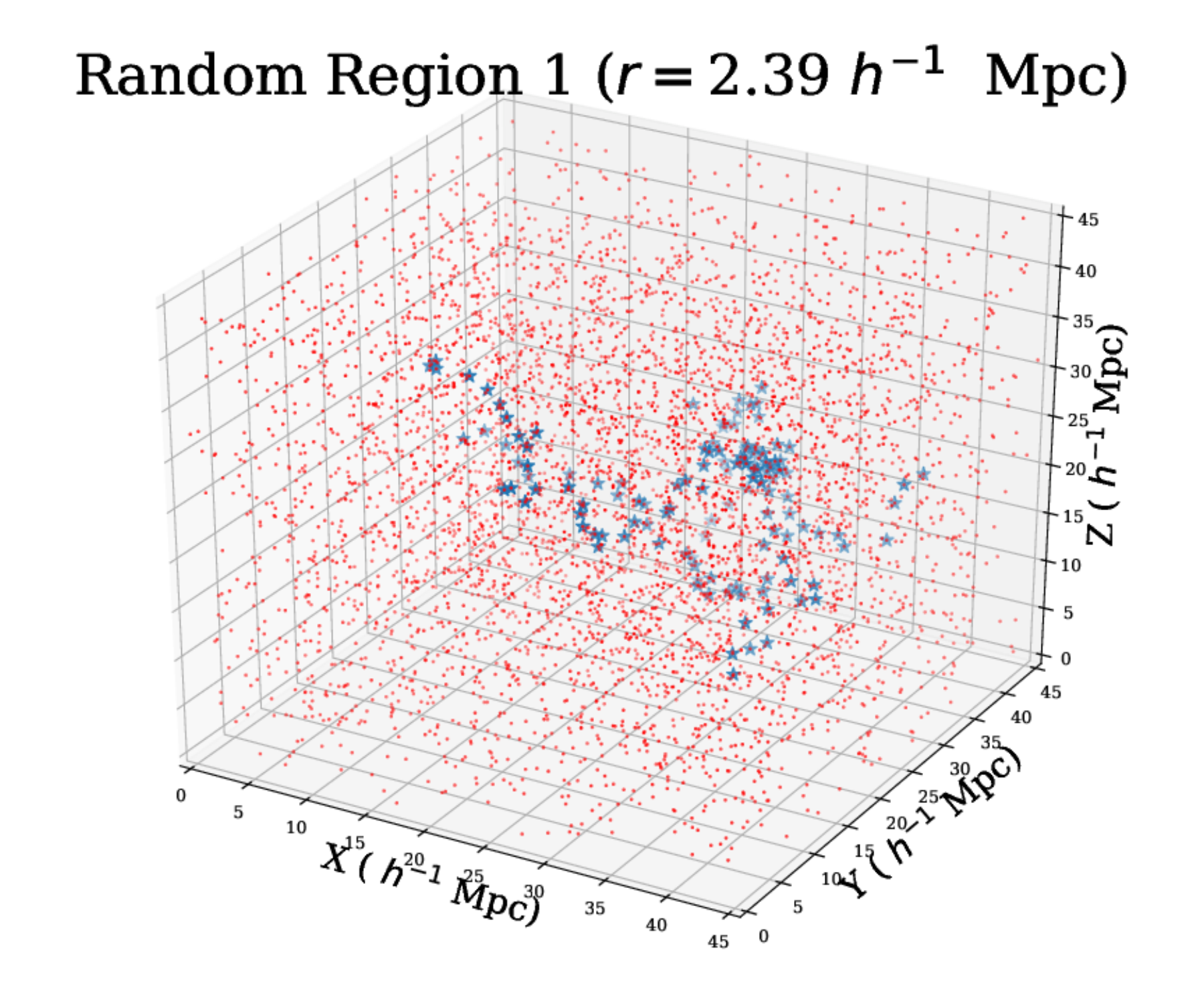}
 \includegraphics[width=0.35\textwidth]{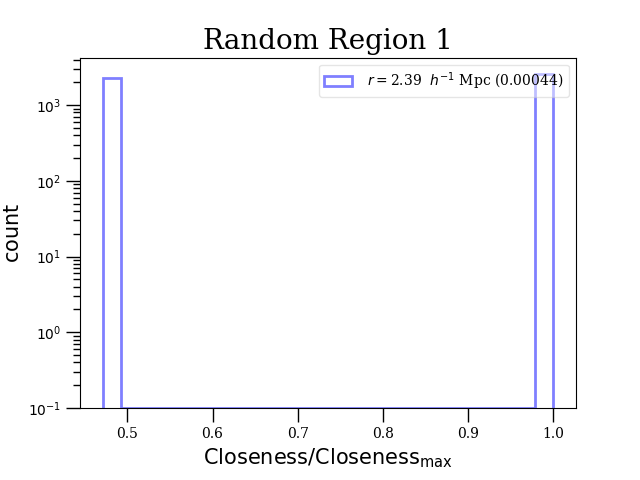}
 \includegraphics[width=0.35\textwidth]{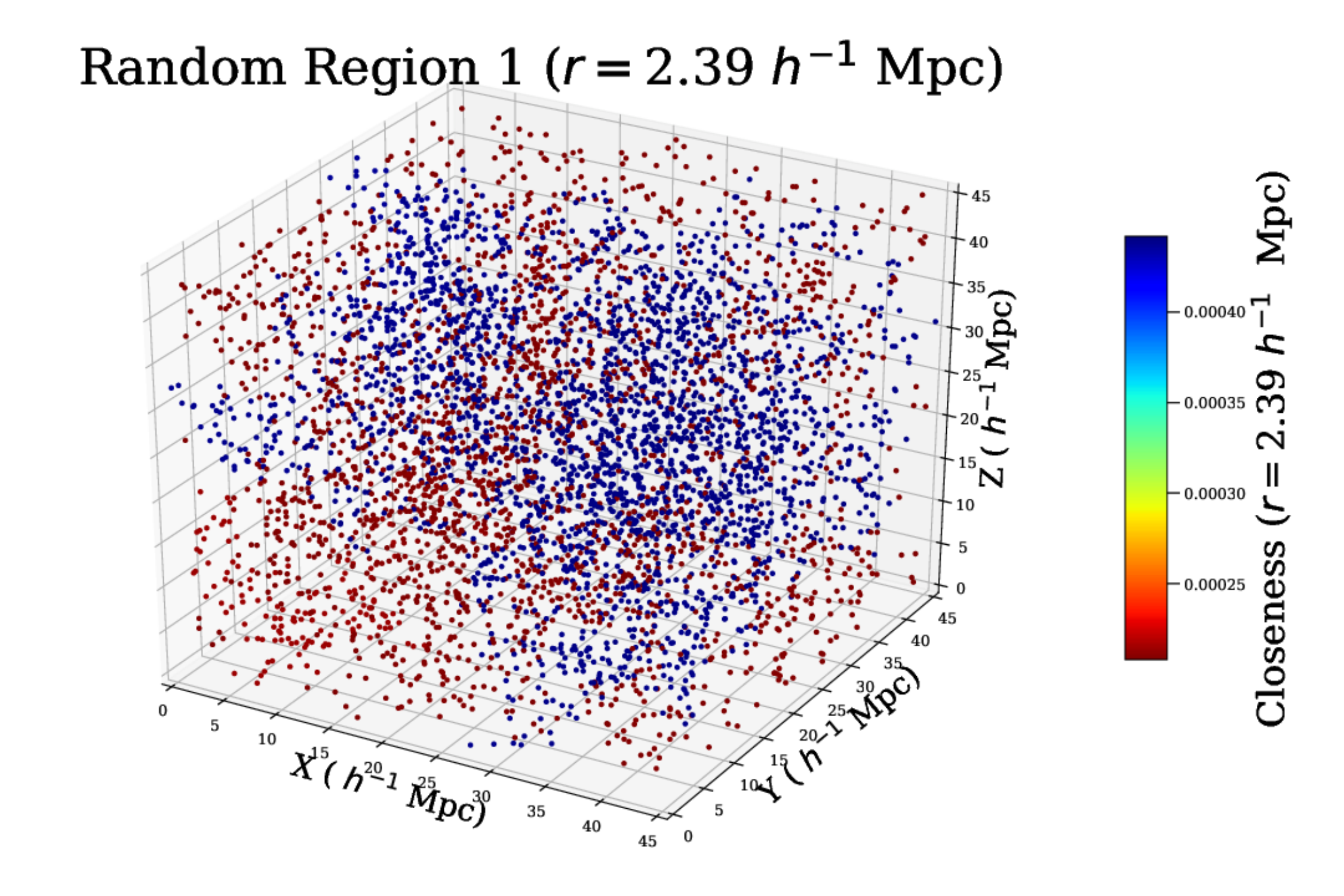}
\caption{Random uniform distribution with the same dimensions and points (galaxies) of the Region 1. In the left panel: the centrality distributions of Degree, Betweenness and Closeness, respectively. In the right panel: the respective three-dimensional distribution of points with high values of the centrality measures in color scale for Degree and Closeness and blue star to Betweenness centrality greater than $\langle BC \rangle \times 10.0$ for the region. }
\label{fig:centralRanregion}
\end{figure}

\section{Sample Sizes Relevance}
\label{sec:appendix2}
\begin{figure}
\centering
 \includegraphics[width=0.32\textwidth]{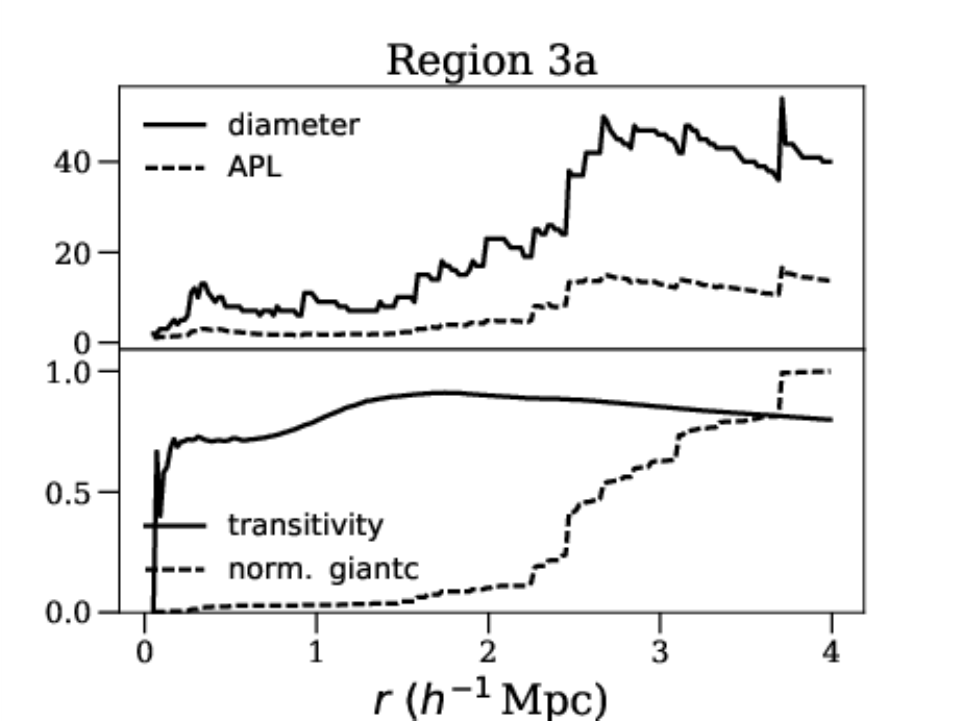}
 \includegraphics[width=0.32\textwidth]{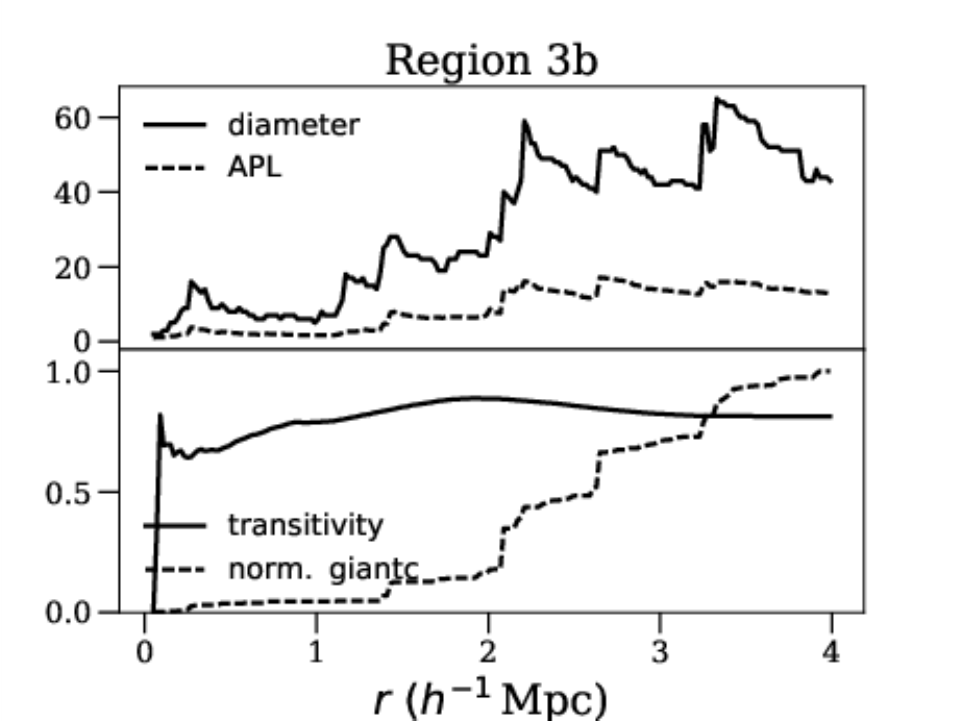}
 \includegraphics[width=0.32\textwidth]{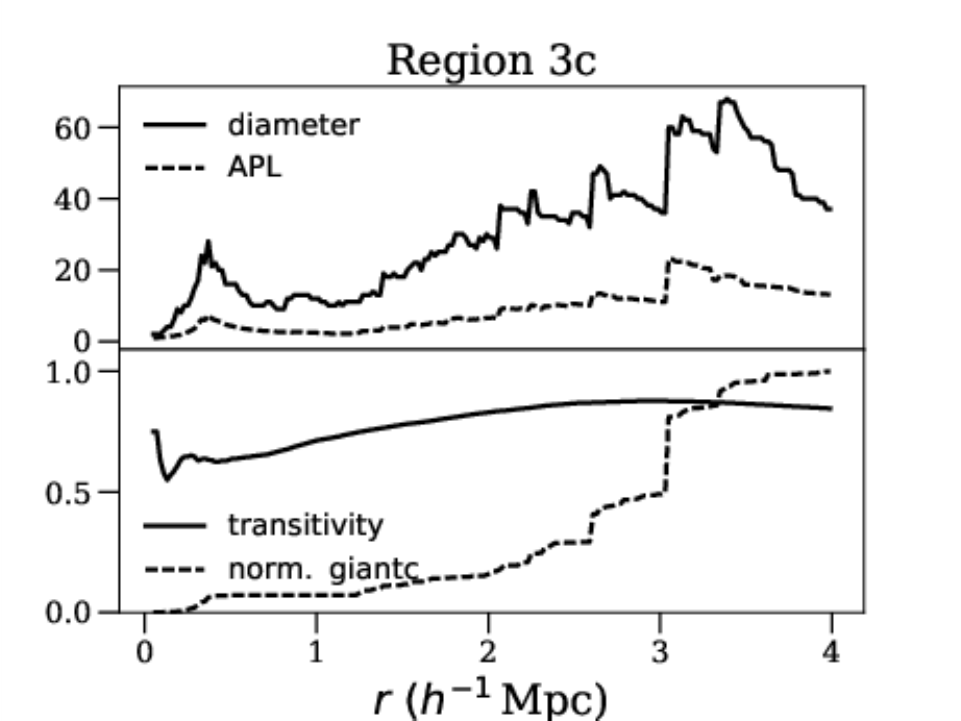}
\caption{Network diameter, average path length, normalized giant component  and transitivity as a function of the connection radius $r$ for the Regions 3a, 3b, 3c.}
\label{fig:diamregion3abc}
\end{figure}

\begin{figure}
\centering
 \includegraphics[width=0.32\textwidth]{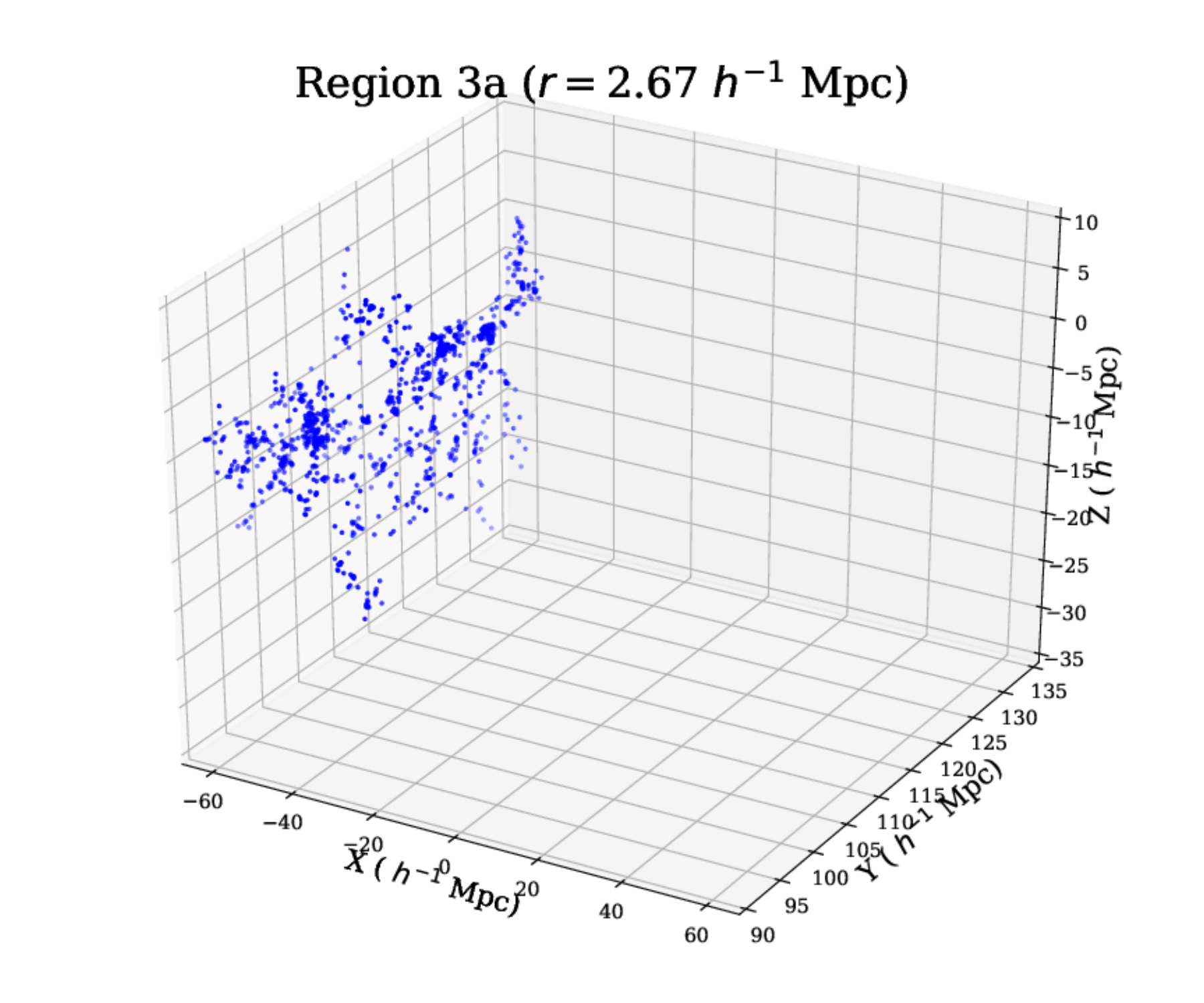}
 \includegraphics[width=0.32\textwidth]{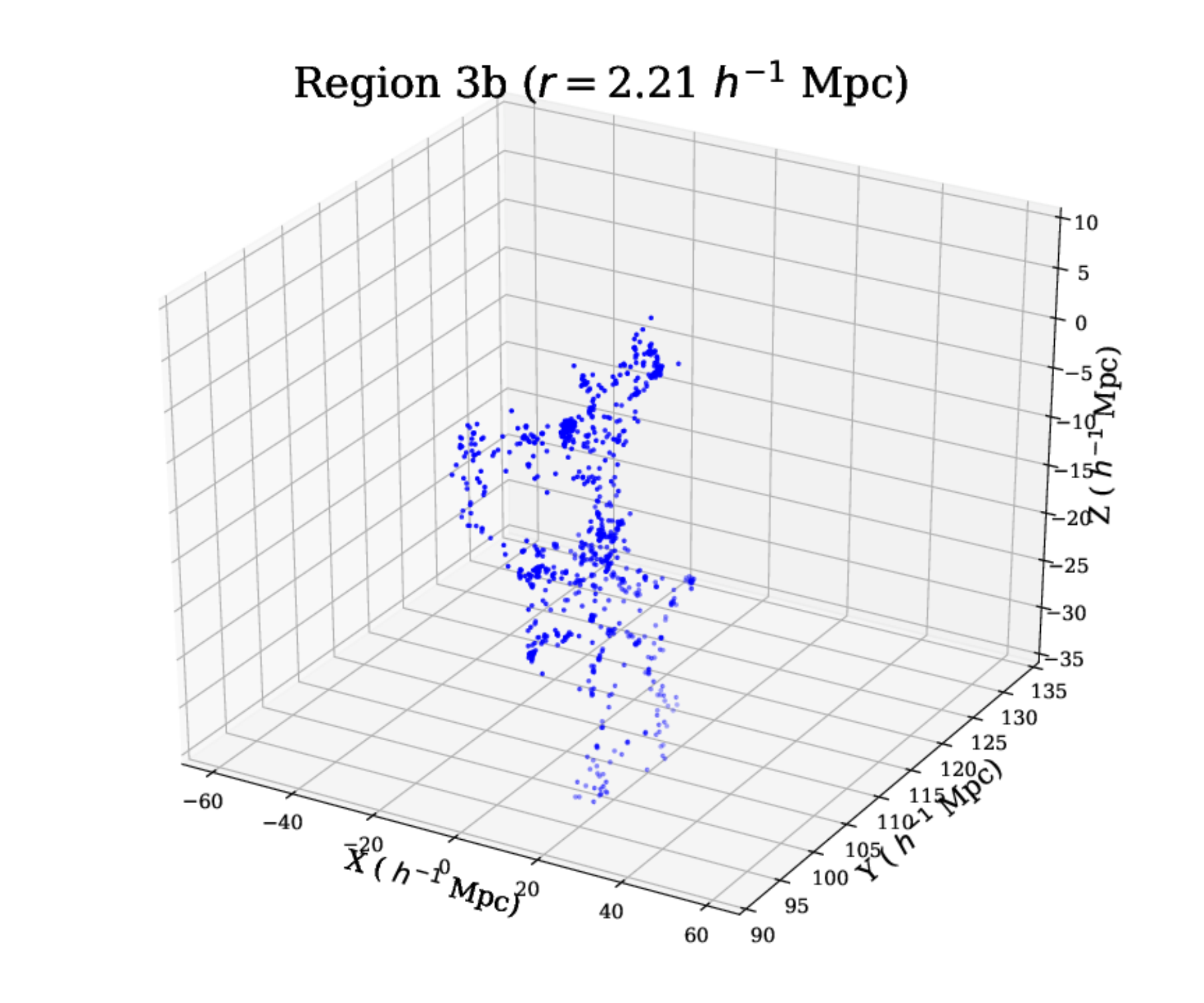}
 \includegraphics[width=0.32\textwidth]{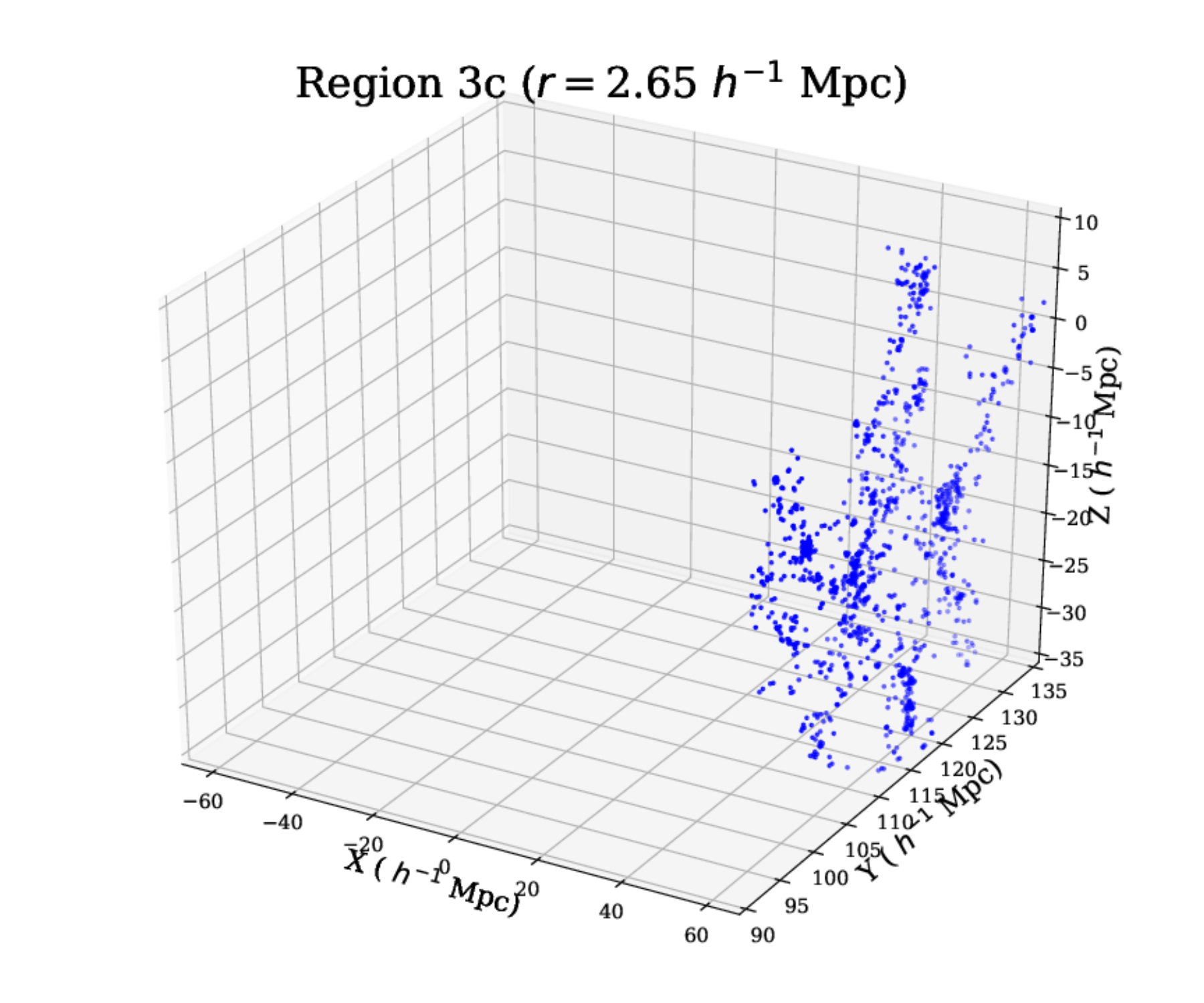}
\caption{Regions 3a, 3b, 3c, only the giant component of each sample.}
\label{fig:region3giant}
\end{figure}

\begin{figure}
\centering
 \includegraphics[width=0.32\textwidth]{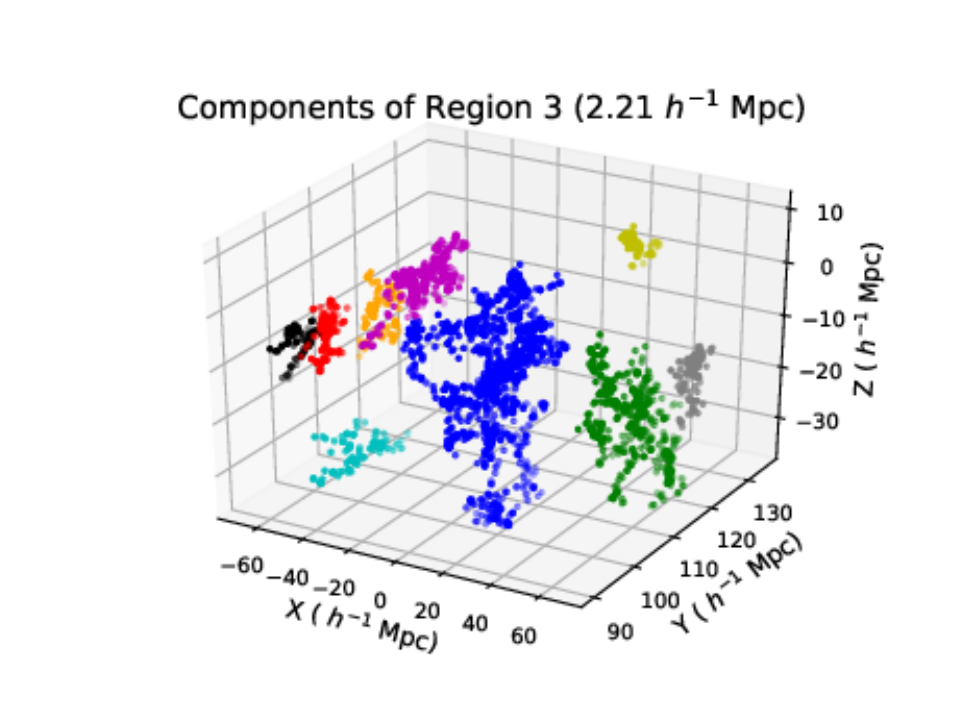}
 \includegraphics[width=0.32\textwidth]{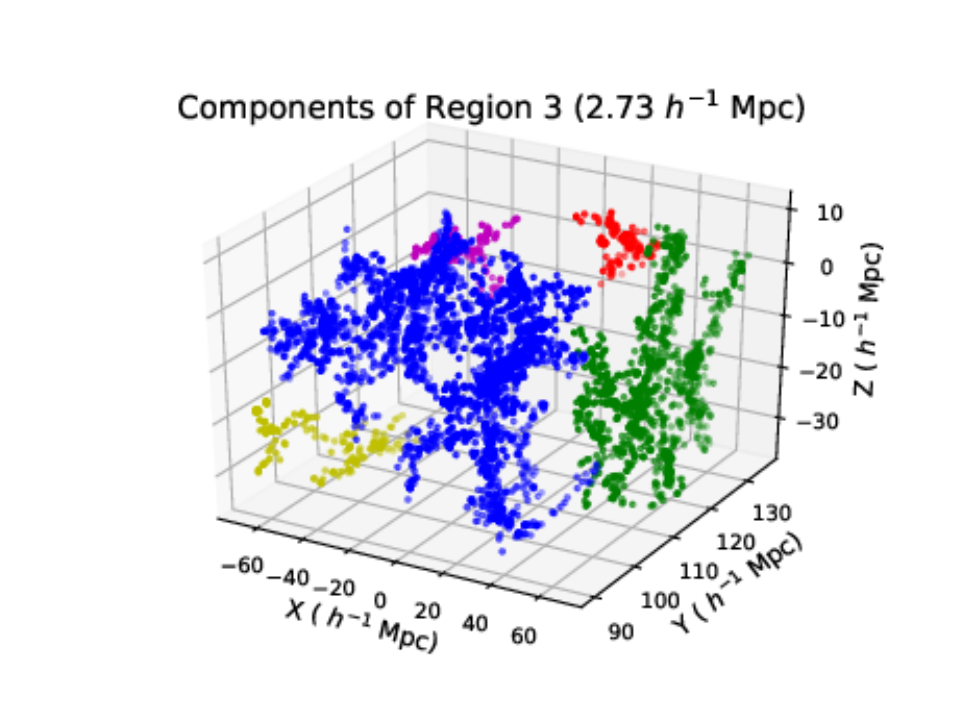}
 \includegraphics[width=0.32\textwidth]{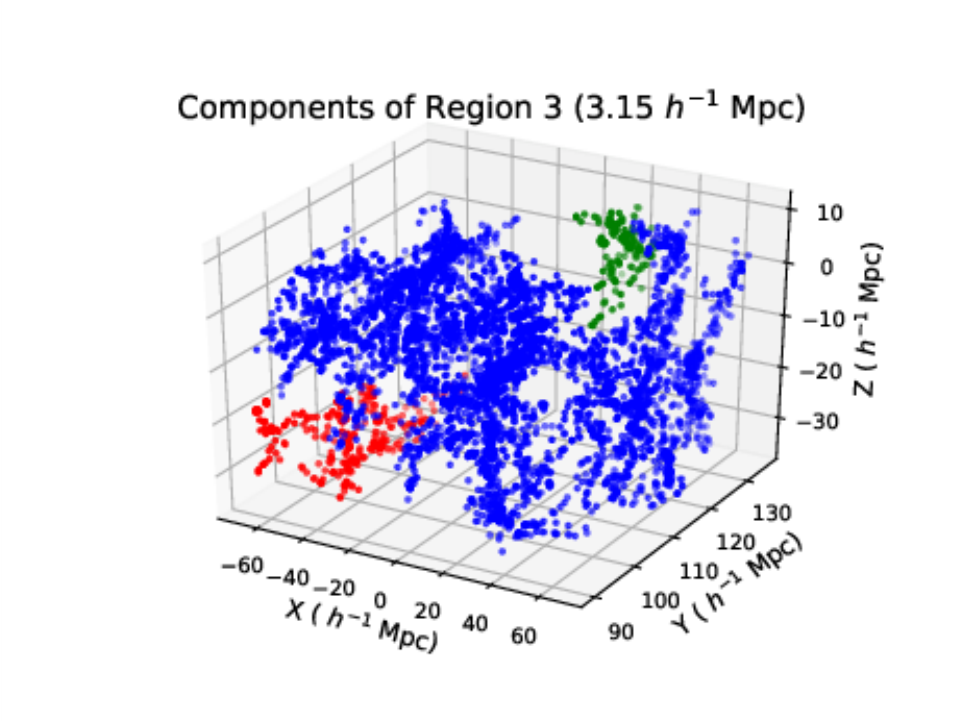}
\caption{Regions 3. The largest components with $r=2.21 h^{-1}$ Mpc  are: blue
points with 1592 galaxies; green points with 626 galaxies; magenta points with
499 galaxies. With $r_{1}=2.73 h^{-1}$ Mpc : blue points with 3571 galaxies and
green points with 1334 galaxies. $r_{1}=3.15 h^{-1} $Mpc : blue points with 5833
galaxies and red points with 354 galaxies. }
\label{fig:region3max_comp}
\end{figure}

We divide the Region 3 into three smaller samples, Region 3a, 3b, and 3c, to
observe how the network diameter as a function of $r$ changes with the sample
size (see Figure \ref{fig:diamregion3abc}). We observe that regions of a high
density of galaxies at the smaller samples will predominate in the Region 3. For example,
the giant component  at Region 3, formed with $r_{2}=2.21 h^{-1} $Mpc
corresponds to the one at Region 3b formed with
$r_{2}=2.21 h^{-1} $Mpc. At this $r$ the sample is in a
percolation phase, the giant component of Region 3b  connect
galaxies of Region 3a and 3c (compare Figure  \ref{fig:region3giant} with Figure
\ref{fig:region3max_comp}. 

\begin{figure}
\centering
 \includegraphics[width=0.32\textwidth]{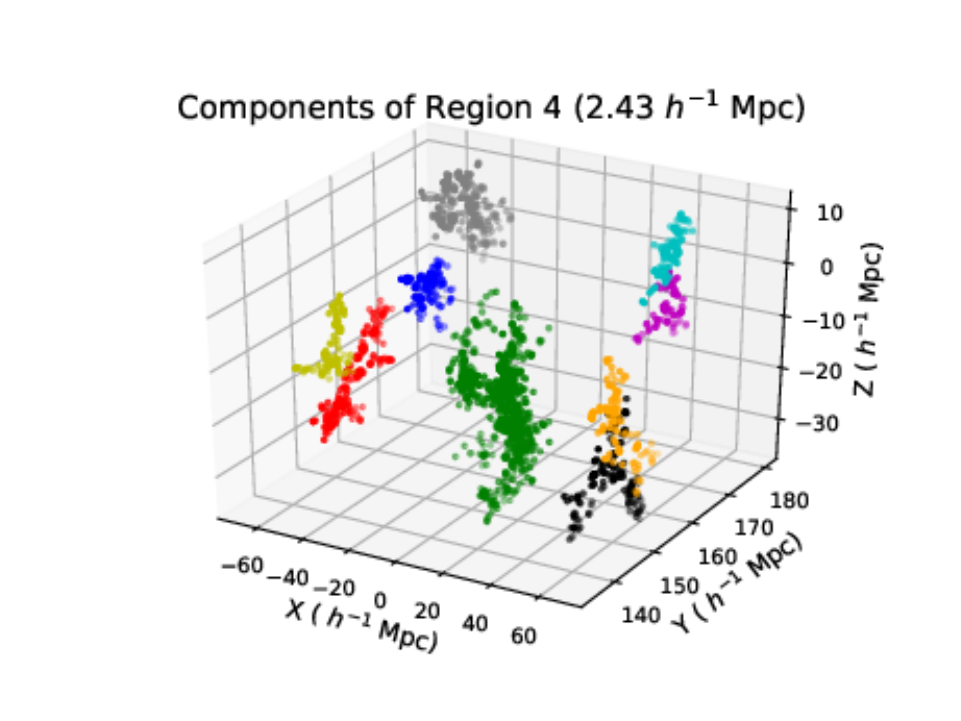}
 \includegraphics[width=0.32\textwidth]{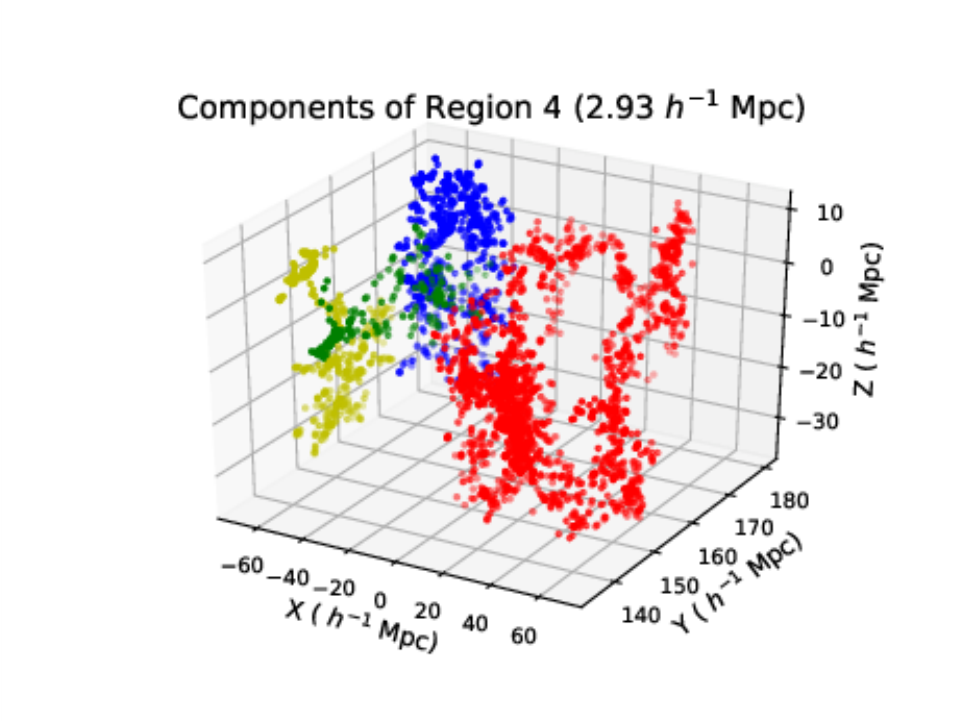}
 \includegraphics[width=0.32\textwidth]{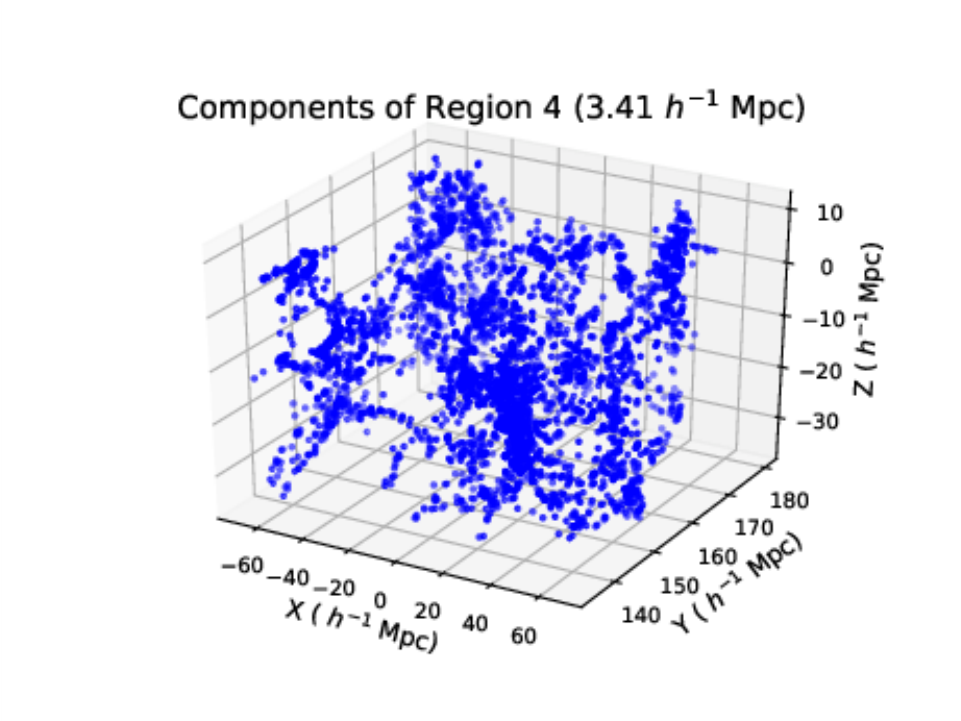}
\caption{Regions 4. The largest components with $r=2.43 h^{-1} $Mpc  are: green
points with 978 galaxies; gray points with 265 galaxies; red points with 232
galaxies and blue points with 208 galaxies. With $r_{1}=2.93 h^{-1} $Mpc : red
points with 2288 galaxies; blue points with 825 galaxies; yellow points with 528
galaxies and green points with 285 galaxies. $r_{1}=3.41 h^{-1} $Mpc : blue
points with 4851 galaxies.   }
\label{fig:region4max_comp}
\end{figure}

For completeness, we show the largest components of the Region 4 with the
connection radius corresponding to each peak in the network diameter as a
function of $r$. In Region 4, the giant component formed with  $r=2.43 h^{-1}
$Mpc (green points in Figure \ref{fig:region4max_comp}) is the structure with a
higher density of galaxies of the region. 

We observe that the method highlights the most statistically relevant region of
each sample. That is, according to the network definition used, the region of
the sample where the distance between galaxies connects the largest number of
galaxies of the sample, forming a giant component. The process
of percolation is dominated by the characteristics of the galaxy distribution
and not for the sample size.  The sample size interferes with
adding more galaxies to the structure, completing or making this structures
statistically relevant.

\begin{table}[ph]
\caption{Values of the network diameter ($diam$), APL ($\langle d \rangle$),
giant component ($Giant_{C}$), mean local clustering coefficient ($C$), and
transitivity ($T$), for the connection radii $r$ that maximizes the network
diameter in Regions 3, 3a, 3b, and 3c.}
\centering

\begin{tabular}{lrrrrrrr} 
\hline
   & diam & $\langle d \rangle$& Giant{C}& $ C $ & $T$\\
\hline 
 R3 &  &   &  &  &   \\
$r_{1}=0.37 h^{-1} $Mpc & 28  & 6.04 & 187 & 0.23 & 0.66 \\
$r_{2}=2.21 h^{-1} $Mpc & 67 & 14.53 & 1592  &0.67 &0.85 \\
$r_{3}=2.73 h^{-1} $Mpc & 89  & 25.29 & 3571 &0.70 &0.87  \\
$r_{4}=3.15 h^{-1} $Mpc & 102  & 32.24 & 5833 &0.71 &0.86 \\[.2cm]
R3a &  &   &  &  &  \\
$r_{1}=0.35 h^{-1} $Mpc & 13  & 3.00 & 46 & 0.23 & 0.72  \\
$r_{2}=2.67 h^{-1} $Mpc & 50 & 14.85 & 1162  & 0.70 & 0.87  \\
$r_{3}=3.71 h^{-1} $Mpc & 54  & 16.52 & 2173 & 0.72 & 0.81 \\[.2cm]
R3b &  &   &  &  & \\
$r_{1}=0.27 h^{-1} $Mpc & 16  & 3.84 & 54 & 0.13 & 0.64 \\
$r_{2}=2.21 h^{-1} $Mpc & 59 & 16.15 & 911  &0.66 &0.88 \\
$r_{3}=3.33 h^{-1} $Mpc & 65  & 15.85 & 1811 &0.72 &0.81 \\[.2cm]
R3c &  &   &  &  & \\
$r_{1}=0.37 h^{-1} $Mpc & 28  & 7.40 & 187 & 0.23 & 0.63 \\
$r_{3}=2.65 h^{-1} $Mpc & 49  & 13.48 & 1293 &0.71 &0.87 \\
$r_{4}=3.39 h^{-1} $Mpc & 68  & 18.43 & 2817 &0.73 &0.87 \\
\hline
\end{tabular}
\end{table}

\section*{Acknowledgements}

 E.G. thanks the CAPES-PNPD program: This study was financed in part by the
 Coordena\c{c}\~ao de Aperfei\c{c}oamento de Pessoal de N\'ivel Superior -
 Brasil (CAPES) - Finance Code 001.
 


\bibliography{references}

\end{document}